\documentclass[acmsmall,screen]{acmart}

\usepackage{xspace}
\usepackage{multirow}

\usepackage{algorithmic}
\usepackage{graphicx}
\usepackage{textcomp}
\usepackage{xcolor}
\usepackage{comment}
\usepackage{balance}
\usepackage{booktabs}
\usepackage{tabularray}
\usepackage[ruled, lined, linesnumbered, commentsnumbered, longend]{algorithm2e}
\usepackage{multirow,multicol}
\usepackage{tikz}
\usepackage{xcolor}
\usepackage{booktabs}
\usepackage{tabularx}
\usepackage{colortbl}
\usepackage{array}
\usepackage{rotating}
\usepackage{enumitem}
\usepackage{tcolorbox}
\usepackage{graphicx}
\usepackage{booktabs}
\usetikzlibrary{matrix,chains,positioning,decorations.pathreplacing,arrows}

\usepackage{mdframed}
\usepackage{framed}
\usepackage{tablefootnote}

\definecolor{mygreen}{RGB}{0, 128, 0}   % Green
\definecolor{myorange}{RGB}{255, 165, 0} % Orange
\definecolor{myteal}{RGB}{0, 128, 128}   % Teal

\newcommand{\prompt}[4]{
    \begin{figure}[htpb]
        \centering
        \begin{tcolorbox}[fontupper=\footnotesize,
            colframe=black, colback=white, coltitle=white, colbacktitle=black,
            title=#1 %, label=#3
        ]
            \textcolor{black}{
                #2
            }
        \end{tcolorbox}
        \caption{#4}
         \label{#3}
    \end{figure}
}

\newcommand{\rb}[1]{{ #1}} % placeholder for rebuttal to clean

\SetCommentSty{mycommfont}

\mdfdefinestyle{MyShadeQuoteStyle}{%
    leftmargin=0pt,
    rightmargin=0pt,
    backgroundcolor=gray!25,
    linewidth=0pt,
    skipbelow=\topskip,
    skipabove=\topskip
}

{\begin{snugshade}\begin{quote}}
{\hfill\end{quote}\end{snugshade}}
\definecolor{shadecolor}{rgb}{0.9,0.9,0.9}
\definecolor{darkyellow}{rgb}{0.5, 0.4, 0.0}

\usepackage{tcolorbox}

\newcommand{\ourTool}{VulScribeR\xspace}
\newcommand{\rqone}{\textit{RQ1: How effective is \ourTool compared to SOTA approaches?}\xspace}

\newcommand{\rqtwo}{\textit{RQ2: How does RAG contribute to \ourTool? }\xspace}

\newcommand{\rqthree}{\textit{RQ3: How does the quantity of the generated samples impact the effectiveness of vulnerability detection models?}\xspace}

\newcommand{\rqfour}{\textit{RQ4: Can \ourTool be effective on large, heavily-imbalanced datasets with complex real-world vulnerabilities? }\xspace}

\usepackage{listings}

\usepackage{xcolor}

\usepackage{array}

\usepackage{caption}
\usepackage{subcaption}

\definecolor{codegreen}{rgb}{0,0.6,0}
\definecolor{codegray}{rgb}{0.5,0.5,0.5}
\definecolor{codepurple}{rgb}{0.58,0,0.82}
\definecolor{backcolour}{rgb}{0.95,0.95,0.92}

\lstdefinestyle{javaStyle}{
    basicstyle=\ttfamily\footnotesize,
    keywordstyle=\color{blue},
    commentstyle=\color{gray},
    stringstyle=\color{red},
    showstringspaces=false,
    language=Java,
    frame=single,
    numbers=none,
    numberstyle=\tiny,
    stepnumber=1,
    numbersep=5pt,
    aboveskip=10pt,
    belowskip=10pt,
    escapeinside={(*@}{@*)},
}

\newtcbox{\errorbox}[1][red!20]{
    on line,
    arc=0pt,
    outer arc=0pt,
    colback=#1,
    colframe=#1,
    boxrule=0pt,
    boxsep=0pt,
    left=1pt,
    right=1pt,
    top=1pt,
    bottom=1pt,
}

\newtcbox{\fixbox}[1][green!20]{
    on line,
    arc=0pt,
    outer arc=0pt,
    colback=#1,
    colframe=#1,
    boxrule=0pt,
    boxsep=0pt,
    left=1pt,
    right=1pt,
    top=1pt,
    bottom=1pt,
}

\lstdefinestyle{styleformodificationdate}{
    backgroundcolor=\color{white}, % set background color to white
    basicstyle=\ttfamily,          % set font to typewriter
    numbers=none,                  % remove line numbers
    frame=none                     % remove frame
}

\lstdefinestyle{mystyle}{
  backgroundcolor=\color{backcolour}, commentstyle=\color{codegreen},
  keywordstyle=\color{magenta},
  numberstyle=\tiny\color{codegray},
  stringstyle=\color{codepurple},
  basicstyle=\ttfamily\footnotesize,
  breakatwhitespace=false,         
  breaklines=true,                 
  captionpos=b,                    
  keepspaces=true,                 
  numbers=left,                    
  numbersep=5pt,                  
  showspaces=false,                
  showstringspaces=false,
  showtabs=false,                  
  tabsize=2
}

\newcommand{\rqboxc}[1]{\begin{tcolorbox}[left=1pt,right=1pt,top=0pt,bottom=0pt,colback=gray!5,colframe=gray!40!black,before skip=5pt,after skip=5pt]#1\end{tcolorbox}}

\begin{document}

\title{VulScribeR: Exploring RAG-based Vulnerability Augmentation with LLMs} 


\author{Seyed Shayan Daneshvar}
\orcid{0000-0002-3463-7873}
\affiliation{%
  \institution{University of Manitoba}
  %\streetaddress{66 Chancellors Cir}
  \city{Winnipeg}
  \country{Canada}}
\email{daneshvs@myumanitoba.ca}

\author{Yu Nong}
\orcid{0000-0002-8598-5181}
\affiliation{%
  \institution{Washington State University}
  %\streetaddress{66 Chancellors Cir}
  \city{Pullman}
  \country{USA}}
\email{yu.nong@wsu.edu}

\author{Xu Yang}
\orcid{0000-0001-9963-6225}
\affiliation{%
 \institution{University of Manitoba}
  %\streetaddress{66 Chancellors Cir}
  \city{Winnipeg}
  \country{Canada}}
\email{yangx4@myumanitoba.ca}

\author{Shaowei Wang}
\orcid{0000-0003-3823-1771}
\affiliation{%
  \institution{University of Manitoba}
  %\streetaddress{66 Chancellors Cir}
  \city{Winnipeg}
  \country{Canada}}
\email{shaowei.wang@umanitoba.ca}

\author{Haipeng Cai}
\orcid{0000-0002-5224-9970}
\affiliation{%
  \institution{University at Buffalo}
  %\streetaddress{66 Chancellors Cir}
  \city{Buffalo}
  \country{USA}}
\email{haipengc@buffalo.edu}

\renewcommand{\shortauthors}{Daneshvar et al.}

%\input{todo}
\begin{abstract}

Detecting vulnerabilities is vital for software security, yet deep learning-based vulnerability detectors (DLVD) face a data shortage, which limits their effectiveness. Data augmentation can potentially alleviate the data shortage, but augmenting vulnerable code is challenging and requires a generative solution that maintains vulnerability. Previous works have only focused on generating samples that contain single statements or specific types of vulnerabilities. Recently, large language models (LLMs) have been used to solve various code generation and comprehension tasks with inspiring results, especially when fused with retrieval augmented generation (RAG). Therefore, we propose \textbf{\ourTool}, a novel LLM-based solution that leverages carefully curated prompt templates to augment vulnerable datasets. More specifically, we explore three strategies to augment both single and multi-statement vulnerabilities, with LLMs, namely Mutation, Injection, and Extension. Our extensive evaluation across four vulnerability datasets and DLVD models, using three LLMs, show that our approach beats two SOTA methods Vulgen and VGX, and Random Oversampling (ROS) by 27.48\%, 27.93\%, and 15.41\% in f1-score with 5K generated vulnerable samples on average, and 53.84\%, 54.10\%,  69.90\%, and 40.93\% with 15K generated vulnerable samples. Our approach demonstrates its feasibility for large-scale data augmentation by generating 1K samples at as cheap as US\$ 1.88.

%Detecting vulnerabilities is vital for software security, yet deep learning-based vulnerability detectors (DLVD) face a data shortage, limiting their effectiveness. Data augmentation could address this, but generating vulnerable code is complex, requiring a generative approach that preserves vulnerability characteristics. Prior works have mostly generated single-statement or specific vulnerabilities. Recently, large language models (LLMs) have excelled in code-related tasks, especially when combined with retrieval-augmented generation (RAG).

%In this study, we introduce VulScribeR, an LLM-based approach using tailored prompt templates to augment vulnerable code datasets. We explore three augmentation strategies with LLMs—Mutation, Injection, and Extension—to create both single- and multi-statement vulnerabilities. Our extensive evaluation across three vulnerability datasets and DLVD models, using two LLMs, shows that our injection-based clustering RAG approach outperforms baseline (NoAug), Vulgen, VGX, and Random Oversampling (ROS) in F1-score by up to 69.90% with 15K samples. VulScribeR also offers cost-effective scalability, generating 1,000 samples for as low as $1.88.

\end{abstract}


%
% Keywords. The author(s) should pick words that accurately describe
% the work being presented. Separate the keywords with commas.



\maketitle

\section{Introduction}\label{sec:intro}
Vulnerability detection is a crucial task in software engineering. Recent Deep Learning-based Vulnerability Detection (DLVD) models~\cite{vuldeepecker, LineVul, Devign, Reveal, IVDetect, DeepVD, grace2024} have drawn more attention from research and industry communities due to their promising performance. However, these models suffer greatly from 1) a lack of sizable datasets and 2) discrepancies between the distribution of training and testing datasets. Data Sampling techniques~\cite{yang2023does, ganz2023codegraphsmote} can alleviate the former problem to an extent by upsampling vulnerable samples to balance the dataset and optionally adding more clean samples, which are abundant. However, Data Augmentation presents a more promising solution as it can help with both problems by generating large diverse datasets. For instance, Zhang~\cite{ZhangNullPointerDerefrence} and Liu et al.~\cite{liu2024enhancing} focus on generating specific type(s) of bug/vulnerability. Similarly, Nong et al. proposed VulGen~\cite{VULGEN} and VGX~\cite{VGX} to mine single-statement vulnerability patterns and inject them into clean samples to generate vulnerable samples. Alternatively, Daneshvar et al.~\cite{daneshvar2025studymixupinspiredaugmentationmethods} evaluated mixup-inspired~\cite{mixup} augmentation techniques for token-based DLVD and proposed conditioned variants for augmenting vulnerabilities at the representation level. Despite their contributions, these methods have limitations. They either generate only specific types of vulnerabilities~\cite{ZhangNullPointerDerefrence, VGX, VULGEN} or require sizable datasets to learn models for generating vulnerable code or pinpointing the location to inject the vulnerability~\cite{Graph2EditforVulGen, VGX, VULGEN}, or they are limited to a certain class of DLVD models while yielding minimal performance gains~\cite{daneshvar2025studymixupinspiredaugmentationmethods}, making them impractical for adoption in a real-world scenario.

It is worth noting that recent works have also explored vulnerability detection using LLMs~\cite{vulrag2024, grace2024, ding2024primevul, comparativeLLM4VDEval2025NDSS}, which pose as competitors to DLVD models. However, adopting LLM-based vulnerability detection methods poses various challenges and limitations~\cite{abdali2025securinglargelanguagemodels, LLM4SecPrivacyGoodBadUgly} that make DLVD models still highly relevant: 1) Data Privacy Concerns: in the industry, there are privacy concerns regarding the use of LLMs for vulnerability detection and essentially exposing private commercial code to LLMs; 2) Limited Robustness to Domain-specifc Vulnerabilities: LLMs require robust datasets to be used alongside various techniques, thus they usually struggle with project and domain-specifc vulnerabilities not present in the training data; 3) High Computational Costs: LLMs are resource-intensive to run while VulScribeR uses LLMs to augment vulnerable samples once and the trained DLVD models can be deployed while requiring significantly less compute and memory, making them easier to be deployed on-premise or in a constrained environment. Consequently, it is important to augment vulnerable data to improve the performance of DLVD models.

Recently, Large Language Models (LLMs) have demonstrated promising results in code-related tasks, such as code understanding~\cite{white2024chatgpt,leinonen2023comparing}, generation~\cite{gu2023llm}, and vulnerability understanding~\cite{fang2024llm,islam2024llm}. Intuitively, vulnerable data augmentation is a task that requires the ability of code comprehension, typically in vulnerability understanding, and code generation. LLMs’ strong ability for code comprehension and code generation fits this task well.

Therefore, to tackle both of those problems and overcome the limitations of previous works, we propose \textbf{\ourTool}, a novel LLM-based solution that leverages carefully curated prompt templates to augment vulnerable datasets. More specifically, we design three strategies to generate vulnerable new code samples, namely \textbf{Mutation}, \textbf{Injection}, and \textbf{Extension}. \textbf{Mutation} refers to changing vulnerable code using code transformation that preserves the code's semantics and syntactical correctness by prompting LLMs. \textbf{Injection} prompts LLMs to inject vulnerable segments of the vulnerable samples into a clean sample to create a new vulnerable sample. \textbf{Extention} aims to extend the vulnerable sample by adding parts of the logic of the clean sample to the vulnerable sample to improve the diversity of the context where the vulnerability could occur.

To generate realistic code, we employ Retrieval-Augmented Generation (RAG) in \textbf{Injection} and \textbf{Extension}. More specifically, in \textbf{Injection}, we retrieve vulnerable samples that are similar to the clean sample into which the vulnerable segment will be injected. Additionally, to enhance the diversity of the generated vulnerable samples, we employ a clustering process to ensure that the data in all clusters have the choice to be retrieved during RAG. We employ RAG and clustering in \textbf{Extension} similarly.

%Our main objective is to generate new samples we consider injecting vulnerable segments to clean samples (i.e. \textbf{Injection} strategy). Additionally, We explore two other strategies for designing the prompts-- \textbf{Mutation} and \textbf{Extension}-- to determine whether vulnerability injection is the most effective strategy. For the \textbf{Injection} strategy, our prompt is tailored for the task of vulnerability detection. It includes 3 placeholders for clean and vulnerable samples, as well as the vulnerable lines of the vulnerable sample, allowing us to provide detailed instructions to the LLM. The detailed and clear instruction increases the compliance of the LLM, enabling \ourTool to work effectively with smaller LLMs (e.g. Qwen-7B~\cite{qwen}) and not be dependent on foundational models with hundreds of billions parameters. \textbf{Extension} strategy is similar to \textbf{Injection} but instead of injecting the vulnerable samples into a clean sample, we extend the vulnerable sample using segments from the clean sample. \textbf{Mutation} strategy generates new samples by applying code transformations on vulnerable samples that do not alter their semantics, thus eliminating the need for clean samples.

Eventually, we employ a fuzzy parser as a filtering mechanism to discard invalid responses such as empty code samples and those with severe syntactical errors, to reduce noise in the generated samples.

To assess \ourTool, we generated vulnerable samples for each strategy using two different LLMs, and evaluated the effectiveness of each data augmentation approach on three SOTA DLVD models (including both token-based and graph-based models) and three commonly used vulnerable datasets. We conducted various experiments in the form of the following Research Questions (RQs):

\begin{itemize}
    \item \rqone \\
    \textbf{Results:} Both \textbf{Injection} and \textbf{Extension} outperform the baselines and the \textbf{Mutation} by a large margin, while \textbf{Injection} provides a slightly higher performance compared to the \textbf{Extension}. 
    For instance, \textbf{Injection} outperforms NoAug, Vulgen, VGX, and ROS by 30.80\%, 27.48\%, 27.93\%, and 15.41\% on average F1-score.
 \item \rqtwo \\
\textbf{Results:} RAG component makes a significant contribution to \textbf{Injection} and \textbf{Extension} strategies, accounting for 4.99\% and 10.77\% of their respective improvements. 
 \item\rqthree \\
 \textbf{Results:} Augmenting more vulnerable data by using \textbf{Injection} helps improve the effectiveness of DLVD models, while VulGen, VGX, and random over-sampling fail to improve the performance of DLVD models by augmenting more than 5K vulnerable samples. Our LLM-based approach is more feasible for large-scale vulnerable data augmentation.
 \item \rb{ \rqfour \\
 \textbf{Results:} on PrimeVul, both \textbf{Injection} and \textbf{Extension} outperform the baselines and \textbf{Mutation} by a considerable margin, while \textbf{Extension} provides a slightly higher performance compared to \textbf{Injection}. 
    For instance, \textbf{Extension} outperforms NoAug, Vulgen, VGX, and ROS by 22.80\%, 30.83\%, 25.24\%, and 3.87\% on average F1-score.
 }
\end{itemize}

In summary, our contributions are as follows:
\begin{itemize}
    \item To the best of our knowledge, we are the first to explore vulnerability augmentation using LLMs. We carefully designed three novel prompt templates with different strategies and proposed a comprehensive pipeline for vulnerability augmentation that can be used for large-scale vulnerability augmentation (that is as cheap as US\$1.88 per 1K samples).
    \item  We performed an extensive evaluation using two different LLMs, three DLVD models, and three datasets, demonstrating the superiority of \ourTool compared to SOTA baselines, including the best latest techniques.
    %\item We contribute two sets of 15K high-quality augmented vulnerable samples generated with \textbf{Injection} strategy using two LLMs.  \sw{we cannot say ours are high-quality}
    \item We made our source code, experimental results, and the augmented datasets publicly available for future research~\cite{datarepo}. 
    % \url{https://github.com/VulScribeR/VulScribeR}. %{github.com/VulScribeR/VulScribeR}. %\sd{TODO: put the unclean codes and datasets before submission, clean in the following week}
\end{itemize}

\section{Background \& Related work}\label{sec:background}
\subsection{Vulnerability Augmentation}
%\sw{if we need more space, we can shrink on this \sd{Already shrunked, more will lead to losing info}}
Vulnerable code samples are scarce in practice, and various approaches have been developed~
\cite{Graph2EditforVulGen,VULGEN,VGX,liu2024enhancing,ganz2023codegraphsmote} to enhance the capability of vulnerability detection models by generating such data and expanding the datasets.
%Graph2Edit~\cite{Graph2Edit} is a general-purpose code editor that transforms the programs by introducing a set of predicted changes to the AST of the program. \sw{this is not for vulnerable data augmentation}
For instance, Nong et al.~\cite{Graph2EditforVulGen}  explored the feasibility of vulnerability injection through a neural code editing model~\cite{Graph2Edit}, which is a DL-based model trained to transform clean code into vulnerable code via introducing a set of predicted changes to the AST of the program. This approach requires high-quality datasets and hence suffers from a chicken-egg dilemma and has limited use.  
% They found that graph-based models are more effective than sequence-based models in generating realistic vulnerabilities. By adding the generated samples to their training sets, the performance of DL-based vulnerability detectors get improved. 
Ganz et al. ~\cite{ganz2023codegraphsmote} present CodeGraphSMOTE, a method to generate new vulnerable samples by porting SMOTE~\cite{SMOTE} to the graph domain. They convert code to graphs and use graph autoencoders to encode the graphs into their latent space. Then, it applies SMOTE (i.e. an interpolation-based sampling method) on the latent space to generate new vulnerable items. 

Nong et al. later proposed VulGen~\cite{VULGEN}, which mines vulnerabilities to collect single-statement vulnerable patterns. Then uses a modified transformer model to locate where injection should take place. VGX~\cite{VGX} is an improved version of VulGen where a significantly bigger vulnerability dataset was used to cover a wider range of 
 single-statement vulnerabilities in the mining phase and the localization model was replaced with a semantics-aware contextualization transformer to predict the injection contexts better. However, both VulGen and VGX only target single statement vulnerabilities and require a comprehensive pattern mining process. Single statement vulnerabilities only account for less than 40\% of all vulnerabilities in the Bigvul~\cite{BigVul} dataset.

% \sd{Removed, extra information already covered in introduction}
% Different from those approaches that only focus on single statement vulnerabilities and need a heavy-weight program analysis process, we propose novel strategies to augment vulnerable samples by leveraging LLMs with carefully designed prompts. 

\subsection{Source Code Augmentation with RAG and LLMs}
Leveraging LLMs for source code augmentation is getting popular and it has been used for data augmentation for Semantic Code Search~\cite{YouAugmentMe} and Code Generation~\cite{CodeSearchIsAllYouNeed}, both of which heavily rely on RAG. Wang et al.~\cite{YouAugmentMe} leverages RAG to retrieve similar $query-code$ pairs, then for each pair they instruct ChatGPT~\cite{OpenAI2022ChatGPT} to change the query and code in separate prompts using predefined rules included in the prompts. Then, they filter the augmented pairs using an encoder model that calculates their similarity score, removing the low-scoring pairs. Chen et al.~\cite{CodeSearchIsAllYouNeed} utilized Code Search to augment data for Code Generation. Specifically, they retrieve $context-function$ pairs, where the context can be either a function header or a comment, to populate prompt templates that instruct the LLM to generate code based on the given context and retrieved function. 
Their architecture uses three components, namely Retriever, Formulator, and Generator. Similar to previous studies, we also employ RAG to retrieve relevant context to enhance the code generation, while different from them, we focus on augmenting vulnerabilities and we employ a clustering process to ensure the diversity of generated data. % We adopt a similar scheme with components named alike with an extra module named Verifier, but our designs differ significantly.

\section{Methodology}\label{sec:method}

\begin{figure*}[t!]

\includegraphics[width=1\textwidth]{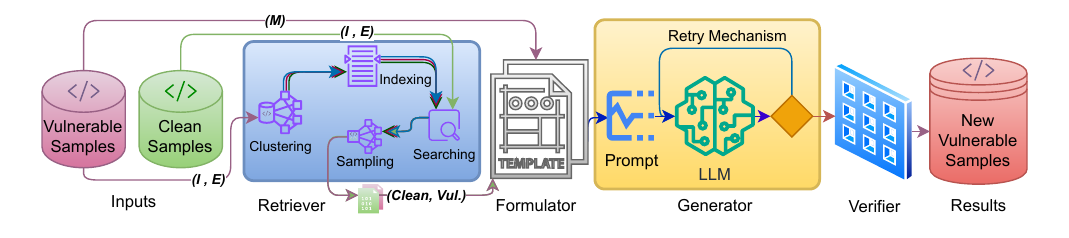}
	\caption{The components of \ourTool. All of the proposed strategies share the same generation and verifier components. The Formulator component of the mutation strategy only requires the vulnerable samples and receives input directly. \textit{(M, I, E)} tags distinguish the difference between the input data flows for \textbf{M}utation, \textbf{I}njection, and \textbf{E}xtension strategies, respectively.}
	\label{fig:appraoch_pipeline}
 \vspace{-0.2in}
\end{figure*}

% overview of our approach
In this section, we elaborate on the details of our methodology, \ourTool. 
We propose a RAG-based solution that leverages carefully curated prompt templates with a lenient filtering mechanism to generate vulnerable code snippets by utilizing LLMs, which allows us to generate diverse and realistic vulnerable code snippets. 

We specifically design three prompting strategies to generate vulnerable code samples for data augmentation.

Consequently, We carefully designed 3 prompt templates for each augmentation strategy, namely \textbf{Mutation}, \textbf{Injection}, and \textbf{Extension} templates, on which we will elaborate in Section~\ref{subsec:strategies}. %\sw{do we? I think we only need to introduce the best setting for injection and extension} \sd{Do we need what? Also, I don't get what celebrate means here?}Moreover, we have considered 2 strategies for each of the \textbf{Injection} and \textbf{Extension} prompts, and 1 strategy for the Mutation.

Figure~\ref{fig:appraoch_pipeline} presents the workflow for the three proposed strategies, and all of the three proposed augmentation strategies can be abstracted into at most four components if applicable as follows: 
\begin{itemize}

\item \textbf{\textit{Retriever}} Given input code samples (i.e., vulnerable and clean samples), Retriever is responsible for seeking suitable vulnerable and non-vulnerable pairs from the database and attaching them in the prompt to provide context for vulnerable sample generation if applicable. \textbf{Injection} and \textbf{Extension} strategies employ the Retriever component, while \textbf{Mutation} does not require a Retriever. 
   
\item \textbf{\textit{Formulator}} For different strategies, we employ their corresponding prompt template. The Formulator instantiates the corresponding prompt template by filling input code samples or the retrieved pairs from Retriever if applicable. For the \textbf{Mutation} prompt template, the input comes directly from the vulnerable samples dataset, while it comes in the form of pairs of clean and vulnerable samples from the Retriever for other templates. %It is worth noting that one can change our \textbf{Mutation} and \textbf{Extension} templates do not require the vulnerable lines, by relying only on the transforms that don't change the execution trace for the \textbf{Mutation} and enforcing a no change policy on the vulnerable code snippet as a whole.

\item \textbf{\textit{Generator}} In the Generator, we use the instantiated templates to prompt the LLM to generate vulnerable code samples. 
If the response does not contain a code snippet or we face an API error, a retry mechanism is activated to feed the prompt to the LLM up to three times to receive a response that contains a code snippet; otherwise, we move on to the next prompt.

\item \textbf{\textit{Verifier}}\rb{The goal of the Verifier\footnote{Verifier is a carefully chosen generic term; in this work, it is only a fuzzy parser that filters code with severe syntax errors.} is to have control over the quality of the generated code, as there’s no guarantee that the LLM produces compilable code. We use Joern’s~\cite{joern}, a fuzzy C language parser, which is based on Antlr’s~\cite{antlr} C Parser, to filter out generated codes that contain severe syntax errors. The rationale behind using a fuzzy parser instead of a strict parser is that 
in data augmentation for various machine learning tasks, the generated data does not need to conform to the original data’s strict standards~\cite{cutout, mixup, SenMixup, ExploringReprLevelAug, BoostSourceCodeAug, olsson2021classmix, yun2019cutmix, daneshvar2025studymixupinspiredaugmentationmethods}. Note that our goal in this study is to augment existing vulnerable datasets to help models capture the vulnerability patterns and generalize better, rather than generating high-quality vulnerable samples. As such, completely
correct code is not necessary for training models~\cite{wang2024natural}, and augmented code can improve the performance of the models even when they slightly break the source code syntax~\cite{ExploringReprLevelAug, BoostSourceCodeAug}; in fact, data with subtle noise helps with the generalization of models~\cite{NoisyDataTikhonovRegularization, ClassificationWithNoisyLabels, NoiseModelling1996}. We use a parser instead of a compiler as we wish to keep the post-processing lightweight, as compilation needs significant efforts, such as identifying the dependencies and resolving type errors, which are unavailable in the original datasets, as they only contain functions. Nevertheless,
we acknowledge that more sophisticated methods can be used for the verification phase, but starting from a simple module (i.e, a parser) was necessary for evaluating whether using LLMs for vulnerability detection is even feasible. As a result, the fuzzy C Parser filters out 2\%-13\% of
the generated data.}

%\sw{what do you mean by ambiguities, so our goal is to generate syntax-correct code?}\sd{Not really, we are using Joern to see if we're dealing with a code snippet in the first place and secondly, we try to produce code that doesn't have severe syntax problems, ambiguity is on the compilation phase, after mixing two pieces of code, some statement may be parsed with more than 1 grammar rule, in that case code is ambiguous and example: int a = 5; double b = 10.5; auto c = a + b; // Type of 'c' is ambiguous, and depends on the compiler}.

\end{itemize}

Note that all of our three strategies share the same \textbf{Generator} and  \textbf{Verifier} components, while different strategies have slightly different designs for \textbf{Retriever} and \textbf{Formulator} components. We discuss the details of the Retriever and Formulator components for each strategy in Section~\ref{subsec:strategies}. 
% introduce of each strategy 

\subsection{Augmentation Strategies} \label{subsec:strategies}
In this section, we break down the details of \textbf{Mutation}, \textbf{Injection}, and \textbf{Extension} strategies. Specifically, we introduce the prompt template and Retriever for each strategy. 
\newline

\subsubsection{\textbf{Mutation}}\label{sec:mutation}

The majority of previous works on program augmentation \cite{ProgramAugmnetaion_mutation, mutation_IPA, mutation_defect, semanticPerservingContrastiveCodeSearch, BoostSourceCodeAug, NatGen}, rely on code transforms that do not change the flow, semantics, and syntactical correctness of the program using program analysis. Variable name changing, replacing for with while loops and vice versa, and adding dead code are examples of this. However, program analysis is very time-consuming, and selecting the locations for transformation is challenging. Previous studies usually select the types and locations of transformation randomly without any program comprehension~\cite{ProgramAugmnetaion_mutation, mutation_IPA, mutation_defect, semanticPerservingContrastiveCodeSearch, BoostSourceCodeAug, NatGen}. In \textbf{Mutation} strategy, we aim to augment vulnerable code samples by utilizing the program comprehension capability of LLM to mutate existing vulnerable code samples and let the LLM choose both the type of transformation and the potential statements to transform. In this way, we get around the program analysis and rely on LLM's creativity to generate more suitable and diverse items. We design the prompt template as shown in Figure~\ref{box:mutation_prompt}.

%\sw{check out if it is possible to package the box as figure, so it can be organized by the compiler nicely. }\sd{DONE, why were the hrefs removed? don't remember where I referred to them to replace the fig}

\prompt{Prompt Template: Mutation}
{Here's a code snippet including a function. Except for the important lines mentioned below, mutate the rest of the code snippet so it is different from the original while keeping the original semantics.
To do so you can use one or more of the following rules:
\\
\\
Rules: 1- Replace the local variables’ identifiers with new non-repeated identifiers
\\
2- Replace the for statement with a semantic-equivalent while statement, and vice versa
\\
3-  \textit{\textbf{\{...remaining\_rules\_are\_hidden\_to\_save\_space\}}}
\\
\\
Code Snippet: 
\textit{\textbf{\{input\_vulnerable\_sample\}}}
\\
\\
The following lines are important and should be included in the final result, but they can still be changed using only the first 5 rules, the rest may be changed using any of the rules or can even be removed if they have no relation to these lines: (Lines are separated via \texttt{/\texttildelow/})
\textit{\textbf{\{input\_vulnerable\_lines\}}}
\\
\\
Put the generated code inside ```C ```. \\(Do not include comments)}{box:mutation_prompt}{The mutation prompt template.}

We instruct the LLM to use one of the 18 program transformation rules by following a recent study~\cite{ProgramAugmnetaion_mutation} and mutating vulnerable code samples while using at least one of the rules. When designing the template, we prioritize the following points. First, we aim to generate diverse code. For this, we mention all of the rules in a single prompt instead of having a prompt per rule to give the LLM the freedom to choose the more suitable rules and apply them to the vulnerable code snippet. Second, we seek to keep the existing vulnerability unchanged in the code. 
As some of the rules might change the execution trace of the code (e.g. transforming a ``switch-case'' to an ``if-else'' statement) and potentially change the status of the vulnerability, we instruct the LLM to use such rules only on the lines that are not important. Important lines are essentially the vulnerable lines of the code snippet. We refer to vulnerable lines as important lines in the prompt to prevent unwanted changes in the resulting code. However, these lines can still be transformed with the rules (i.e., the first five rules in the template) that do not change the execution trace without changing the vulnerable state of the code snippet.

As presented in Figure~\ref{fig:appraoch_pipeline}, the overall workflow for \textbf{Mutation} is straightforward. To augment $N$ vulnerable samples, $N$ vulnerable samples are randomly sampled from the input dataset, and then are directly fed into the Formulator to instantiate the template, as a result, $N$ prompts are instantiated by filling the templates. The prompts are then fed to the Generator to generate $N$ vulnerable samples. It's worth noting that on average 3\% of the generated samples will be filtered out in the Verifier component and so if one desires to end up with at least $N$ samples, a higher target should be selected in the generator phase, or else they should redo the generation after the verification to create more items to reach the target.
\newline

\subsubsection{\textbf{Injection}}\label{sec:injection}

Similar to recent works~\cite{VGX, VULGEN, Graph2EditforVulGen}, we also focus on injecting vulnerable code segments into a clean sample as our main strategy, but we aim to cover all types of vulnerabilities and not just single statement vulnerabilities. More specifically, we instruct the LLM to inject the logic of the vulnerable sample into a clean sample by prioritizing the injection of vulnerable segments. Using an LLM for injecting the vulnerable segments into a clean item gives the freedom to LLM to identify the best location. We present our \textbf{Injection} template prompt as shown in Figure~\ref{box:injection_prompt}.

\prompt{Prompt Template: Injection }{
Here are two code snippets specified below, modify code snippet 2 in a way that it includes the logic of code snippet 1:
\\
\\
Code Snippet 1: 
\textit{\textbf{\{input\_vulnerable\_sample\}}}
\\
Code Snippet 2: 
\textit{\textbf{\{input\_clean\_sample\}}}
\\
\\
Note that the following lines from the first code snippet have a high priority to be in the final modified code:
\\
Lines separated by \texttt{/\texttildelow/}:
\textit{\textbf{\{input\_vulnerable\_lines\}}}
\\
\\
Put the generated code inside ```C ```.
\\(Do not include comments)}
{box:injection_prompt}{The injection prompt template.}

The templates contain two or three placeholders to fill, namely \textit{input\_vulnerable\_sample}, \textit{input\_clean\_sample}, and \textit{input\_vulnerable\_lines}. \textit{input\_clean\_sample} is the input clean samples where we aim to inject vulnerability and is not present in the \textbf{Mutation} strategy. \textit{input\_vulnerable\_sample} is a vulnerable code example to be retrieved that is similar to the clean input sample. We retrieve similar vulnerable code because we believe that the logic from a similar vulnerable code could be integrated into the clean code more easily and naturally. \textit{input\_vulnerable\_lines} provides meta-information indicating the vulnerable lines in the retrieved vulnerable example. We instruct the LLM to prioritize including these vulnerable lines since our goal is to generate vulnerable code.

%\sw{the vulnerable lines is the whole slice containing the vul lines or just the vul lines? \sd{just vul lines, they are the same thing in this paper I have used these two interchangably}}\sw{could be issue attacked by reviewer}

To fill the template, we need to construct a dataset of pairs of clean samples and their corresponding retrieved similar vulnerable code samples (i.e., \textit{clean-vul} pairs). Note that we choose to provide a clean code sample as the input and search for similar vulnerable samples, rather than providing a vulnerable code sample and searching for similar non-vulnerable samples. This strategy is faster and more efficient due to the smaller number of vulnerable samples, which reduces the dataset size for the retrieval process. %This approach offers two key benefits: first, with a smaller set of documents, the precision of BM25's calculated document frequency (DF), term frequency (TF), and document length increases, enhancing the accuracy of relevance scoring; second, the indexing and querying processes become faster and more efficient due to the reduced dataset size.

Algorithm~\ref{alg:retriever} outlines our method for constructing a dataset of \textit{clean-vul} pairs through a retrieval process. Given a dataset containing both vulnerable samples ($V$) and clean samples ($C$), our goal is to retrieve $N$ \textit{clean-vul} pairs. A straightforward approach would be to retrieve the most similar vulnerable samples for each clean sample, sort them in descending order of similarity, and select the top $N$ pairs. However, focusing solely on similarity reduces the diversity of the retrieved samples, which is counterproductive for data augmentation. Previous studies have shown that higher diversity in the training dataset improves the generalizability of deep learning models \cite{diversity1, diversity2}. Therefore, to enhance diversity, we incorporate a clustering phase to group the vulnerable samples into $G$ clusters, ensuring that samples from all clusters are considered for selection.

First, we cluster the vulnerable samples into $G$ groups (line 2). To do this, we use CodeBERT~\cite{CodeBERT} to embed each code sample into a vector, specifically utilizing the [CLS] token's embedding for each sample. We then apply KMeans for clustering, using cosine similarity to measure the similarity of each pair of samples.

Following clustering, we create an index on each cluster using Lucene to facilitate efficient search (lines 7-10). For each clean sample, we search for the most similar vulnerable sample within each cluster using the cluster index, and store the resulting \textit{clean-vul} pairs for further processing (lines 12-17). To find similar vulnerable samples for each clean sample, we use the BM-25 algorithm~\cite{bm25}, as previous studies have shown that there is no significant difference between sparse methods like BM-25 and dense methods like CodeBERT when leveraging RAG~\cite{CodeSearchIsAllYouNeed, InContextLearningExamples, IRBasedPromptSelectionForCodeMesbah, EvaluatingDemonstraionRetrieversPengfei}. We use Lucene's implementation of BM-25 for this study.

Once the similarity score is calculated for each pair of clean-vul, we sort the results within each cluster based on the similarity score (lines 19-21). Then, we iteratively select the top pair from each cluster, ensuring a diverse yet relevant selection (lines 25-32). We also sort the groups based on size and start from the largest group ensuring a higher coverage for higher values of $G$ (line 23) by following previous study~\cite{ma2024llmparser}. We set $G=5$ in our study.

\begin{algorithm}[htpb]
\footnotesize
    \KwIn{$V$: Vulnerable samples}
    \KwIn{$C$: Clean samples}
    \KwIn{$N$: Number of pairs to sample}
    \KwIn{$G$: Number of groups into which the vulnerable samples will be clustered}
    \KwOut{$D$: Dataset with $N$ Clean-Vulnerable pairs}

    $\textit{vul\_embeddings} \gets \text{compute\_codebert\_embedding}(V)$ 
    
    $\textit{clustered\_vuls} \gets \text{KMeans}(G, V, \textit{vul\_embeddings})$ 
    
    $\textit{indices} \gets []$ 
    
    $\textit{clustered\_pairs} \gets [[], [], ..., []]$ \tcp{An array of $G$ arrays}
    
    $D \gets []$ 
    
    \tcp{Index vulnerable samples of each cluster separately}
    \For{$\textit{cluster\_id} = 0$ \KwTo $G - 1$}{ 
       
    $search\_engine\_index \gets \text{index}(\textit{clustered\_vuls}[\textit{cluster\_id}])$
    
    $\textit{indices}[\textit{cluster\_id}].\text{append}(search\_engine\_index)$

    }

    \tcp{each clean item is matched with the most similar vulnerable item of each cluster}
    \ForEach{$\textit{clean} \in C$}{
        \For{$\textit{cluster\_id} = 0$ \KwTo $G - 1$}{
            $\textit{vul} \gets \text{search}(\textit{indices}[\textit{cluster\_id}], \textit{clean})$ 
            
            $\textit{clustered\_pairs}[\textit{cluster\_id}].\text{append}((\textit{clean}, \textit{vul}))$ 
        }
    }
    \tcp{Sort each cluster's pairs based on their similarity score}
    \For{$\textit{cluster\_id} = 0$ \KwTo $G - 1$}{ 
     $sort\_by\_score(\textit{clustered\_pairs}[cluster\_id], "desc")$ 
        
    }
    
    \tcp{sort the arrays based on their size in descending order}
    $\textit{clustered\_pairs} \gets\text{sort\_arrays\_by\_size}(\textit{clustered\_pairs}, \text{"desc"})$ 
    
    $i \gets 0$  \tcp{Iterate between the clustered\_pairs and select the best pair each time}
    \While{\text{size}($D$) \textless $N$}{
        $\textit{pairs} \gets \textit{clustered\_pairs}[i \mod $\textit{G}$]$ 
        
        $\textit{index} \gets i // \textit{G}$ 
        
        \If{$\textit{index} < \text{size}(\textit{pairs})$}{
            $D.\text{append}(\textit{pairs}[\textit{index}])$ 
        }
        $i \gets i + 1$ 
    }
    
    \Return $D$ 
    \caption{(Clean, Vulnerable) Pair Retrieval}
    \label{alg:retriever}
\end{algorithm}

After the clean-vul pairs are retrieved, those pairs will be fed into Formulator to instantiate the \textbf{Injection} template and go through the rest of the components in the workflow to generate $N$ vulnerable new samples similar to \textbf{Mutation}. 
\newline

\subsubsection{\textbf{Extension}}\label{sec:extension}

While injecting vulnerable code segments into clean code is the defacto yet effective way of augmenting vulnerabilities, extending an already vulnerable code snippet by adding additional logic can also generate a new vulnerable sample, which enriches the context where vulnerable code happens. 
Adding certain parts from a clean sample to an already vulnerable sample has a higher chance of keeping the context of the original vulnerable code intact compared to injecting vulnerable segments into a clean item. By extending an already vulnerable sample, only a small section will be irrelevant to the vulnerable part of the code which is the important part for the models, while injecting vulnerable segments into a clean sample results in a code snippet where the vulnerable section has little connection to the original context and gives itself away. Therefore, we aim to explore it as an alternative to vulnerability injection. We refer to this strategy as \textbf{Extension}. We present the prompt template for \textbf{Extension} in Figure~\ref{box:extension_prompt}.\\

%\sw{did we try to prompt that asks the LLM to extend vulnerable freely without providing clean samples? if we tried, we can justify here, why we did not use this? } \sd{Yes, it didn't work, the LLM usually leaves it unchanged, or just works like mutation, they suck at generating content when they are not given clear instructions in general}\sw{sure, let's add this to justify our design. it shows we spent time on prompt design. }

\prompt{Prompt Template: Extension}{Here are two code snippets 1 and 2, each including a function. Add some parts of the logic of 1 to 2 in a way that the result would include important lines of 2 with some parts from 1. You can add variables to 2 to produce a more correct code.
\\
\\
Code Snippet 1: \textit{\textbf{\{input\_clean\_sample\}}}
\\
Code Snippet 2: \textit{\textbf{\{input\_vulnerable\_sample\}}}
\\\\                
The following lines from 2 are important and should be included in the final result, the rest may be changed or even removed if they have no relation to these lines: (Lines are separated via \text{/\texttildelow/})
\textit{\textbf{\{input\_vulnerable\_lines\}}}
\\
\\
Put the generated code inside ```C ``` and note that the final result should be a function that takes all the input args of 2 and more if required.
\\(Do not include comments)}
{box:extension_prompt}{The extension prompt template.}

Similar to the \textbf{Injection} strategy, we instruct the LLM to add sections of the logic of the clean sample to the vulnerable sample, while enforcing a no-change policy on the vulnerable lines. The only difference is that rather than injecting vulnerable samples into clean samples, the \textbf{Extension} strategy does the reverse. Therefore, we can exactly reuse the workflow of the \textbf{Injection} strategy by modifying the Retriever, where we aim to retrieve similar clean samples for given vulnerable samples and keep all other components the same. Note that we also tried instructing the LLM to extend vulnerable samples freely without providing clean samples, which offers the LLMs more freedom in extending the vulnerable samples. However, it does not work well, and tends to hallucinate in most cases. Specifically, The LLM usually leaves the vulnerable sample unchanged or behaves like a mutator (e.g., mutating variable names or adding dead code). The LLMs perform poorly at generating code when they are not given clear instructions, hence we designed our strategies in a way that we simply ask the LLM to follow clear instructions (i.e. mix two code pieces) without mentioning the word vulnerability and without asking the LLM anything vague or too general. This prevents the LLM from focusing on the vulnerability aspect of the codes and encourages following the given instructions, rather than hallucinating.

\section{Experimental Setting}\label{sec:experimentalsetting}
In this section, we present our research questions (RQs), the datasets, DLVD models, LLMs, evaluation metrics, our analysis approach for each of the RQs, and implementation details.

\subsection{Research Questions}
We evaluate \ourTool from different aspects to answer the following research questions.

\begin{itemize}
    \item\rqone
    \hfill
    \item \rqtwo
    \hfill
    \item \rqthree
    \hfill
    \item \rb{\rqfour}
    %\hfill
    %\item \rqfive
    
\end{itemize}

In RQ1, we evaluate the effectiveness of \ourTool for improving the performance of DLVD models by augmenting vulnerable data, by comparing it to current SOTA approaches. In RQ2, we aim to investigate the effectiveness of our design for the Retriever component. In RQ3, we explore how the number of generated samples by our approach and SOTA methods affects the performance of DLVD models. \rb{In RQ4, we investigate the usefulness of \ourTool on a new large, high-quality, heavily-imbalanced dataset with real-world vulnerabilities.}

% \begin{table*}[bt]
% \caption{The Studied Datasets}
%   \centering
%   \begin{tabular}{l!{\vrule width 0.5pt}|c!{\vrule width 0.5pt}|c !{\vrule width 0.5pt}|c| c|c}
%     \hline
    
%     \textbf{Dataset} & \textbf{Devign} & \textbf{Reveal} & \textbf{ BigVul Train} & \textbf{BigVul Validation} & \textbf{BigVul Test} \\
%     \hline
    
%     \textbf{Vulnerable Items} & 10768 & 2240 &  8,783 & 1038 & 1079 \\
%     % \quad Flaw Lines Metadata & X & X & 6610 & X & X \\
%     \hline
%     \textbf{Clean Items} & 12024& 20494& 142,125 & 17826 & 17785\\
%     \hline
%     \textbf{All Items} & 22792 & 22734 & 150,908 & 18864 & 18864\\
%     \hline
%     \textbf{Ratio} & 1:1.1& 1:9.1& 1:16.2 & 1:17.2 & 1:16.5 \\
%     \hline
%   \end{tabular}
%   \vspace{8pt}
%   \label{tab:dataset}
% \end{table*}
%  Transposed version of the table above
\begin{table}[h]
\caption{The Studied Datasets}
  \centering
  \begin{tabular}{|l|c|c |c|c|c}
    \hline
    
    \textbf{} & \textbf{Vul. Samples} & \textbf{Clean Samples} & \textbf{Total} & \textbf{Ratio} \\
    \hline
    
    \textbf{Devign} & 10768 & 12024 & 22792 & 1:1.1 \\
    \hline
    \textbf{Reveal} & 2240 & 20494 & 22734 & 1:9.1 \\
    \hline
    \(\textbf{BigVul}_\textit{Train}\) & 8783 & 142125 & 150908 & 1:16.2 \\
    \hline
    \(\textbf{BigVul}_\textit{Validation}\) & 1038 & 17826 & 18864 & 1:17.2 \\
    \hline
    \(\textbf{BigVul}_\textit{Test}\) & 1079 & 17785 & 18864 & 1:16.5 \\
    \hline
    \(\textbf{PrimeVul}_\textit{Train}\) & 4802 & 169345 & 174147 & 1:35.3 \\
    \hline
    \(\textbf{PrimeVul}_\textit{Validation}\) & 593 & 23347 & 23940 & 1:39.4 \\
    \hline
    \(\textbf{PrimeVul}_\textit{Test}\) & 549 & 24232 & 18864 & 1:44.1 \\
%     # train 0 :169345 to target 1:4802 =  35.27
% # test 0 :24232 to target 1:549 =  44.14
% # val 0 :23347 to target 1:593 =  39.37
    \hline
  \end{tabular}
  \vspace{1pt}
  \label{tab:dataset}
\end{table}

\subsection{Datasets}
In this study, \rb{we use three different widely used vulnerability detection datasets~\cite{VULGEN,VGX,yang2023does,LineVul, Devign, Reveal}, namely Devign~\cite{Devign}, Reveal~\cite{Reveal}, and BigVul~\cite{BigVul}, to evaluate our approach. Additionally, we evaluate our method with PrimeVul~\cite{ding2024primevul}, which is a new large dataset with less label noise, in RQ4.} Table~\ref{tab:dataset} shows the details of these datasets. All of these datasets include C/C++ functions gathered from real-world projects. \rb{For Devign, Reveal, BigVul, and the extra clean data~\cite{vulgen43ExtraCleans} used for retaining the ratio, we used the cleaned version from VGX~\cite{VGX}, which had duplicates removed between the datasets. For PrimeVul, we used the SHA2 hash to remove samples available in Devign and the extra cleans~\cite{vulgen43ExtraCleans} from all three sets of the PrimeVul dataset.}

\subsubsection{Devign}
Following previous studies~\cite{VGX, VULGEN}, we use Devign as our primary training dataset as it offers a better balance between the two categories, namely the vulnerable and clean samples (i.e., non-vulnerable samples) by following a previous study, VulGen~\cite{VULGEN}. We also use the clean samples of this dataset as the clean inputs for the \textbf{Retriever} component.
\\
\subsubsection{Bigvul}
We use all three sets (i.e., training, validation, and testing set) of BigVul~\cite{BigVul} for different purposes. We use BigVul's training set as the source for collecting vulnerable samples to feed it to the \textbf{Retriever} as the vulnerable samples. However, not all of these items can be used as they lack the vulnerable lines metadata that are required for the \textbf{Injection} and \textbf{Extension} prompt templates. Hence, we only use the 6,610 vulnerable samples from BigVuls's training set that include the metadata of vulnerable lines. We use the testing set of BigVul as one of the testing sets in our study.
\\
\subsubsection{Reveal}
Similar to VulGen~\cite{VULGEN}, we use Reveal as another testing set in our study. Reveal is a slightly larger dataset than BigVul's testing set and has twice the number of vulnerable samples.
\\
\subsubsection{PrimeVul}
\rb{
Unlike VGX~\cite{VGX} and VulGen~\cite{VULGEN}, which use Devign and BigVul datasets, which are considerably noisy~\cite{Croft2023VulDataQuality, Risse2024VDLimits, ding2024primevul} (i.e., some vulnerable samples in these datasets are clean), we also use PrimeVul~\cite{ding2024primevul} in RQ4 with the settings of RQ1 with some changes (see more details in Section~\ref{sec:rq4}). PrimeVul is the largest dataset used in our study, and one of the most imbalanced datasets available. We use all three sets of Primevul in RQ4, with the train set replacing Devign as the main training set as well as BigVul's train set for the retrieved vulnerable samples, and the validation and testing sets replacing those of BigVul.
}

\subsection{Deep learning-based vulnerability detection (DLVD) Models}
There are two families of DLVD models, token-based and graph-based. To represent both families, we select a SOTA token-based model and two SOTA graph-based models, namely LineVul~\cite{LineVul}, Devign~\cite{Devign}, and Reveal~\cite{Reveal} models. For Devign and Reveal models, we use the same setting as VulGen~\cite{VULGEN} and train both of these models 5 times with random seed values, test them and report the results with the highest F1-score achieved. For Linevul, we follow the settings of VGX~\cite{VGX}, train for 10 epochs, and select the checkpoint that achieved the highest F1 score on Bigvul's validation set. 
We specifically train and test the models with different datasets, following the settings of VulGen~\cite{VULGEN} and VGX~\cite{VGX}, to ensure there is no information leakage and the results can be trusted to apply to real-world unseen data.

\subsection{Evaluation metrics}
DLVD is inherently a classification task as a category is assigned to the code snippet.
Hence, to evaluate the effectiveness of the studied methods, we use the common classification metrics~\cite{EvaluationMetrics}, in line with previous related  studies~\cite{yang2023does, EvaluationMetrics, vuldeepecker, DeepVD, FeatureImportanceDefectClassifiers, NoiseClassifierSE, VULGEN, VGX}, namely Precision, Recall, and F1-Score. 

\subsection{Base LLMs}
In our study, we use two different LLMs to generate vulnerable code, namely ChatGPT3.5~\cite{OpenAI2022ChatGPT}, a commercial general-purpose LLM, and CodeQwen1.5~\cite{codeqwen1.5}, an open-source LLM specialized for code tasks. 

For ChatGPT, We use the ``gpt-3.5-turbo'', variant of OpenAI's ChatGPT~\cite{OpenAI2022ChatGPT} which has shown great potential in previous studies~\cite{YouAugmentMe, InContextLearningExamples, CodeSearchIsAllYouNeed, AutomaticBugFixChatGPT}. We set the temperature hyper-parameter to 0.5 since we wish to reduce the randomness of the response to a degree that ChatGPT would follow our instructions, but would not limit its freedom of creativity. We left all the other hyper-parameters to their default values.
For CodeQwen, we use the 7B variant of CodeQwen1.5~\cite{codeqwen1.5}, ``CodeQwen1.5-7B-Chat'', which is a variant of the general purpose Qwen~\cite{qwen}. CodeQwen1.5, similar to ChatGPT, is a decoder-only transformer-based model with the difference that it was pre-trained on code data, and supports a larger context length. We did not change any of the hyper-parameters for this model and used the defaults, except for the number of new tokens that we set to ChatGPT's default (i.e. 4096 tokens).

\rb{Additionally, we use GPT4o-mini~\cite{gpt4omini} for our RQ4, which is as capable as ChatGPT3.5, but more affordable. Similar to ChatGPT3.5, we set the temperature parameter to 0.5, but we did not change any other hyperparameters. }

\subsection{Baselines}
To compare our studies with previous work and demonstrate the effectiveness of \ourTool we consider the following baselines:
\begin{itemize}
\item \textbf{NoAug:} 
This is the most basic of our baselines, where we use the original dataset for training without introducing any changes.

\item \textbf{ROS:} As Yang et al. \cite{yang2023does} show that Random Oversampling (ROS) can have a considerable effect on improving the performance of DLVD, thus we also ROS as another baseline.
% , \sw{same here, we don't need to mentioned 5k here. we can mention 5k in RQ1}in which we sample 5K extra items from Devign and add the same extra clean items as the SOTA baselines to keep the ratio.

\item \textbf{{VulGen}}
Vulgen~\cite{VULGEN} is a SOTA vulnerability generation method that mines single-statement vulnerabilities and uses a transformer model for locating where a vulnerable pattern should be used for injection. We use this baseline as it can generate a significant amount of data that can be used to improve the performance of DLVD models.

\item \textbf{VGX:} VGX~\cite{VGX} is the improved version of VulGen that uses a larger dataset for mining single-statement patterns and employs a more sophisticated localization model. Hence, we use VGX as it is potentially a more powerful method and can generate more diverse high-quality data than VulGen.

\end{itemize}

\subsection{Approach for RQs}

\subsubsection{Approach of RQ1}
To demonstrate the effectiveness of \ourTool and examine which of the proposed strategies is superior to others, we first generate 5K vulnerable samples by employing different approaches (i.e., \ourTool and baselines). More specifically, to sample 5K items with any of our strategies, we first aim to generate up to 6K samples with the \textbf{Generator} which are then filtered by the \textbf{Verifier} to yield more than 5K generated samples, from which we sample randomly to get 5K items. For VGX and VulGen, we use the generated results in VGX~\cite{VGX} and sample 5K from their results randomly. For ROS, we randomly up-sample 5K samples from the vulnerable samples in the Devign dataset.
% We add the proportional number of clean items from ~\cite{vulgen43ExtraCleans} to keep the original ratio of the Devign dataset. 

We then train the three DLVD models using the augmented Devign dataset. To augment the Devign dataset, we add the 5K generated vulnerable samples to it and add the proportional number of clean samples from the vulnerability benchmark proposed in~\cite{vulgen43ExtraCleans} to maintain the original ratio of the dataset, by following previous studies~\cite{VULGEN, VGX}. We use BigVul's testing set and Reveal dataset to evaluate the performance of all of these models. We assess the performance of a vulnerable data augmentation approach by assessing the performance of a specific DLVD with the augmented dataset of the specified strategy. Hence, each strategy will be tested in 12 instances (i.e. 3 DLVD Models * 2 Testing Datasets * 2 LLMs).

%\sd{This next paragraph (commented) is mostly not needed, we didn't use the notation, but we can clarify that we have 12 settings, so I added that to the previous paragraph.} % To simplify our writing, we make the following definition. We consider the assessment of the data augmentation approach of one DLVD approach on one dataset using one specific base LLM as one \textit{experimental instance}. For simplicity, we use the format $\{DLVD\}+\{Dataset\}+\{LLM\}$ to denote one specific instance in the rest of the sections. Therefore, We have 12 experimental instances, i.e., 3 DLVD * 2 datasets * 2 LLMs. For example, suppose we observe that the model Devign with the augmentation approach Injection with ChatGPT 3.5 Turbo outperforms Devign with Extension using same base LLM on same dataset. In that case, we say Injection outperforms Extension on the experimental instance $\{Devign\}+\{Reveal\}+\{ChatGPT 3.5 Turbo\}$.

\subsubsection{Approach of RQ2}
In this RQ, we investigate the usefulness of our proposed RAG for the \textbf{Injection} and \textbf{Extension} strategies. Specifically, we eliminate the \textbf{Retriever} component, and instead, we match the clean and vulnerable samples randomly to examine the effect of RAG (i.e. \textbf{Injection w/o Retriever} and \textbf{Extension w/o Retriever}).
Moreover, we also conduct an ablation study on the clustering phase of the \textbf{Retriever} component to examine the effect of diversity on the quality of the generated samples. To do so, we remove the clustering phase from the Retriever component and search for similar samples for each clean input item within all the vulnerable items instead of within each cluster. Then we take the top 5 retrieved items to end up with a similar number of $clean-vul$ pairs. In the sampling phase, we sort all the pairs based on their relevance score and start from the top. We denote the \textbf{Injection} strategy without clustering phase as \textbf{Injection w/o Clustering} and the \textbf{Extension} strategy without clustering as \textbf{Extension w/o Clustering}. \\ \\
Similar to RQ1, we generate 5k vulnerable samples using the studied approaches and compare their effectiveness by assessing the DLVD models that are trained on the augmented datasets.

\subsubsection{Approach of RQ3}
In this RQ, we investigate how the number of augmented samples impacts the performance of DLVD models, and how much performance can be gained by providing more data.

To see the impact and find out whether \ourTool can be used for large-scale vulnerability augmentation, we use the \textbf{Injection} strategy, our best strategy as depicted in Section~\ref{sec:results}, and generate up to 16.5K samples so that after the filtering we end up with a set of at least 15K vulnerable samples which is about 40\% higher than the number of vulnerable samples in the original dataset. Then we repeat the experiments for the \textbf{Injection} using both LLMs. We sample 10K and 15K vulnerable items while adding the proportional number of clean items to maintain the ratio. We do the same for VGX~\cite{VGX}, VulGen~\cite{VULGEN}, and ROS.

\subsubsection{Approach of RQ4}
\rb{
Lastly, we propose a slightly different setting than RQ1 for our approach to work with heavily imbalanced large datasets, and evaluate the effectiveness of our strategies.
\\
For this RQ, we utilized GPT4-omini and used PrimeVul's trainset as the primary training set as well as the source for collecting vulnerable samples that are fed to the \textbf{Retriever}, and used Devign's cleans as the source of cleans. Since the PrimeVul dataset does not contain line-level information of vulnerabilities, we collected the vulnerable lines by computing the diff between paired clean and vulnerable samples and extracting the vulnerable lines.
\\
\\
It is worth mentioning that PrimeVul is a challenging dataset with a highly imbalanced ratio between vulnerable and clean samples (1:35) for training DLVD models~\cite{ding2024primevul}. To make training possible, we balanced the dataset by randomly removing clean samples to achieve a ratio of 1:15 (larger ratios made training less effective on some models). Then we followed the settings of RQ1 for each strategy, and collected 5K new samples. To conduct a fair comparison between the methods, we used ROS to randomly oversample the original vulnerable data to get to a balanced dataset (i.e., 1:1 ratio) for training DLVD. Note that we do not change the ratio of the validation and testing sets.
}
%samples as well as the new 5K samples with a ratio of 3:2 \footnote{The 3:2 ratio for oversampling the augmented dataset worked better than 1:1 and 2:1.
%(i.e., out of five vulnerable samples in the final dataset, two would be from the new ones and three from the original dataset) to get to a balanced dataset (i.e. 1:1 ratio); therefore, no extra clean samples were added since the dataset was heavily-imbalanced and for our ROS baseline in this RQ, we oversampled the vulnerable samples to get to the same ratio of 1:1. 
%Hence, in this RQ, the changes compared to RQ1 are: LLM utilized, the source of clean samples being from an external dataset, vulnerable samples coming from the same source, the added oversampling mechanism for all methods, and no extra clean samples.

\subsection{Implementation details}
\rb{We used a machine with four 24GB Nvidia RTX 3090s for our experiments. We conducted a total of 781 experiments for all our RQs, which cost close to 2000 GPU hours solely for training and testing DLVD models (excluding the time spent generating the samples).
}
% 616 : each of our strategy needs 44 experiments (5 * 8 + 4 line vul) = 44 and each baseline 22 so 44 * (RQ1(3+ 4/2) + 4 RQ2 + (6/2 + 2)) = 14 * 44
% plus 21 new experiments for RQ4, 84 for mixed, and 60 for ROS_clean totaling 165 
% total 165+ 616 = 781

\section{Results}\label{sec:results}
\subsection{\rqone }\label{sec:rq1}
%\sw{all our RQs only present numbers, looks boring, think about if why analysis or examples we can provide. now everything looks like a black box, no insights gained at all.}\sd{Added the diversity diagram}
\begin{table}[htbp]
  \centering
  \caption{Comparison of \ourTool's strategies with baselines using ChatGPT and CodeQwen when augmenting 5K samples across the studied DLVD models (i.e., Devign, Reveal, and Linevul). The cells with larger values (better performance) compared to NoAug are highlighted darker.}  \label{tab:rq1}
  \resizebox{13.5cm}{!}{
\begin{tabular}{rrrrrrrrrrrrrrrrrrrrrr|}
 &
   &
   &
   &
   &
   &
   &
   &
   &
   &
   &
   &
   &
   &
   &
   &
   &
   &
   &
   &
   &
  \multicolumn{1}{r}{}
  \\
 &
   &
   &
  \multicolumn{9}{c|}{\textit{\textbf{ChatGPT 3.5 Turbo}}} &
  \multicolumn{9}{c}{\textit{\textbf{CodeQwen1.5-7B-Chat}}} &
  \multicolumn{1}{r}{}
 \\
\cline{3-21} &
  \multicolumn{1}{r|}{} &
  
  \multicolumn{1}{c|}{\multirow{2}[4]{*}{{\raisebox{0.4cm}{\textbf{Strategy}}}}} &
  \multicolumn{3}{c|}{\textbf{Devign}} &
  \multicolumn{3}{c|}{\textbf{Reveal}} &
  \multicolumn{3}{c|}{\textbf{Linevul}} &
  \multicolumn{3}{c|}{\textbf{Devign}} &
  \multicolumn{3}{c|}{\textbf{Reveal}} &
  \multicolumn{3}{c|}{\textbf{Linevul}} &
  \multicolumn{1}{r}{}
  \\
\cline{4-21} &
  \multicolumn{1}{r|}{} &
  \multicolumn{1}{c|}{} &
  \multicolumn{1}{c}{P} &
  \multicolumn{1}{c}{R} &
  \multicolumn{1}{c|}{F1} &
  \multicolumn{1}{c}{P} &
  \multicolumn{1}{c}{R} &
  \multicolumn{1}{c|}{F1} &
  \multicolumn{1}{c}{P} &
  \multicolumn{1}{c}{R} &
  \multicolumn{1}{c|}{F1} &
  \multicolumn{1}{c}{P} &
  \multicolumn{1}{c}{R} &
  \multicolumn{1}{c|}{F1} &
  \multicolumn{1}{c}{P} &
  \multicolumn{1}{c}{R} &
  \multicolumn{1}{c|}{F1} &
  \multicolumn{1}{c}{P} &
  \multicolumn{1}{c}{R} &
  \multicolumn{1}{c|}{F1} &
  \multicolumn{1}{r}{}
  \\
\cline{3-21} &
  \multicolumn{1}{c|}{\multirow{7}[2]{*}{\begin{sideways}\textbf{Reveal}\end{sideways}}} &
  \multicolumn{1}{c|}{NoAug} &
  \multicolumn{1}{c}{10.59} &
  \multicolumn{1}{c}{47.23} &
  \multicolumn{1}{c|}{17.30} &
  \multicolumn{1}{c}{12.90} &
  \multicolumn{1}{c}{65.14} &
  \multicolumn{1}{c|}{21.54} &
  \multicolumn{1}{c}{5.57} &
  \multicolumn{1}{c}{16.98} &
  \multicolumn{1}{c|}{8.39} &
  \multicolumn{1}{c}{\cellcolor[rgb]{ .973,  .984,  .984} 10.59} &
  \multicolumn{1}{c}{47.23} &
  \multicolumn{1}{c|}{\cellcolor[rgb]{ .988,  .988,  1} 17.30} &
  \multicolumn{1}{c}{12.90} &
  \multicolumn{1}{c}{65.14} &
  \multicolumn{1}{c|}{21.54} &
  \multicolumn{1}{c}{\cellcolor[rgb]{ .98,  .984,  .992} 5.57} &
  \multicolumn{1}{c}{16.98} &
  \multicolumn{1}{c|}{\cellcolor[rgb]{ .988,  .988,  1} 8.39} &
  \multicolumn{1}{r}{}
  \\
 &
  \multicolumn{1}{c|}{} &
  \multicolumn{1}{c|}{VulGen} &
  \multicolumn{1}{c}{10.45} &
  \multicolumn{1}{c}{\cellcolor[rgb]{ .431,  .765,  .518} 57.90} &
  \multicolumn{1}{c|}{\cellcolor[rgb]{ .933,  .969,  .953} 17.70} &
  \multicolumn{1}{c}{12.46} &
  \multicolumn{1}{c}{64.60} &
  \multicolumn{1}{c|}{20.90} &
  \multicolumn{1}{c}{5.31} &
  \multicolumn{1}{c}{\cellcolor[rgb]{ .388,  .745,  .482} 98.09} &
  \multicolumn{1}{c|}{\cellcolor[rgb]{ .859,  .937,  .89} 10.08} &
  \multicolumn{1}{c}{\cellcolor[rgb]{ .988,  .988,  1} 10.45} &
  \multicolumn{1}{c}{\cellcolor[rgb]{ .486,  .784,  .569} 57.90} &
  \multicolumn{1}{c|}{\cellcolor[rgb]{ .937,  .969,  .957} 17.70} &
  \multicolumn{1}{c}{12.46} &
  \multicolumn{1}{c}{64.60} &
  \multicolumn{1}{c|}{20.90} &
  \multicolumn{1}{c}{\cellcolor[rgb]{ .988,  .988,  1} 5.31} &
  \multicolumn{1}{c}{\cellcolor[rgb]{ .388,  .745,  .482} 98.09} &
  \multicolumn{1}{c|}{\cellcolor[rgb]{ .851,  .933,  .882} 10.08} &
  \multicolumn{1}{r}{}
  \\
 &
  \multicolumn{1}{c|}{} &
  \multicolumn{1}{c|}{VGX} &
  \multicolumn{1}{c}{\cellcolor[rgb]{ .914,  .961,  .937} 11.02} &
  \multicolumn{1}{c}{\cellcolor[rgb]{ .388,  .745,  .482} 59.35} &
  \multicolumn{1}{c|}{\cellcolor[rgb]{ .804,  .914,  .843} 18.58} &
  \multicolumn{1}{c}{11.69} &
  \multicolumn{1}{c}{\cellcolor[rgb]{ .482,  .784,  .565} 65.62} &
  \multicolumn{1}{c|}{19.84} &
  \multicolumn{1}{c}{\cellcolor[rgb]{ .792,  .91,  .831} 9.51} &
  \multicolumn{1}{c}{\cellcolor[rgb]{ .467,  .776,  .549} 87.24} &
  \multicolumn{1}{c|}{\cellcolor[rgb]{ .749,  .894,  .796} 11.47} &
  \multicolumn{1}{c}{\cellcolor[rgb]{ .918,  .961,  .941} 11.02} &
  \multicolumn{1}{c}{\cellcolor[rgb]{ .447,  .773,  .533} 59.35} &
  \multicolumn{1}{c|}{\cellcolor[rgb]{ .82,  .922,  .855} 18.58} &
  \multicolumn{1}{c}{11.69} &
  \multicolumn{1}{c}{\cellcolor[rgb]{ .388,  .745,  .482} 65.62} &
  \multicolumn{1}{c|}{19.84} &
  \multicolumn{1}{c}{\cellcolor[rgb]{ .827,  .925,  .859} 9.51} &
  \multicolumn{1}{c}{\cellcolor[rgb]{ .467,  .776,  .549} 87.24} &
  \multicolumn{1}{c|}{\cellcolor[rgb]{ .737,  .886,  .784} 11.47} &
  \multicolumn{1}{r}{}
  \\
 &
  \multicolumn{1}{c|}{} &
  \multicolumn{1}{c|}{ROS} &
  \multicolumn{1}{c}{\cellcolor[rgb]{ .616,  .839,  .678} 13.28} &
  \multicolumn{1}{c}{39.93} &
  \multicolumn{1}{c|}{\cellcolor[rgb]{ .612,  .835,  .675} 19.93} &
  \multicolumn{1}{c}{\cellcolor[rgb]{ .565,  .816,  .635} 14.35} &
  \multicolumn{1}{c}{54.52} &
  \multicolumn{1}{c|}{\cellcolor[rgb]{ .631,  .843,  .694} 22.72} &
  \multicolumn{1}{c}{\cellcolor[rgb]{ .773,  .902,  .816} 9.88} &
  \multicolumn{1}{c}{\cellcolor[rgb]{ .922,  .961,  .941} 22.71} &
  \multicolumn{1}{c|}{\cellcolor[rgb]{ .573,  .82,  .639} 13.77} &
  \multicolumn{1}{c}{\cellcolor[rgb]{ .631,  .847,  .694} 13.28} &
  \multicolumn{1}{c}{39.93} &
  \multicolumn{1}{c|}{\cellcolor[rgb]{ .643,  .851,  .702} 19.93} &
  \multicolumn{1}{c}{\cellcolor[rgb]{ .51,  .796,  .588} 14.35} &
  \multicolumn{1}{c}{54.52} &
  \multicolumn{1}{c|}{\cellcolor[rgb]{ .561,  .816,  .631} 22.72} &
  \multicolumn{1}{c}{\cellcolor[rgb]{ .812,  .918,  .847} 9.88} &
  \multicolumn{1}{c}{\cellcolor[rgb]{ .918,  .961,  .941} 22.71} &
  \multicolumn{1}{c|}{\cellcolor[rgb]{ .545,  .812,  .62} 13.77} &
  \multicolumn{1}{r}{}
  \\
 &
  \multicolumn{1}{c|}{} &
  \multicolumn{1}{c|}{Mutation} &
  \multicolumn{1}{c}{\cellcolor[rgb]{ .918,  .961,  .941} 10.98} &
  \multicolumn{1}{c}{\cellcolor[rgb]{ .549,  .812,  .624} 53.44} &
  \multicolumn{1}{c|}{\cellcolor[rgb]{ .859,  .937,  .886} 18.22} &
  \multicolumn{1}{c}{12.23} &
  \multicolumn{1}{c}{\cellcolor[rgb]{ .388,  .745,  .482} 68.46} &
  \multicolumn{1}{c|}{20.75} &
  \multicolumn{1}{c}{\cellcolor[rgb]{ .875,  .941,  .902} 7.80} &
  \multicolumn{1}{c}{\cellcolor[rgb]{ .922,  .961,  .941} 22.62} &
  \multicolumn{1}{c|}{\cellcolor[rgb]{ .741,  .89,  .788} 11.60} &
  \multicolumn{1}{c}{\cellcolor[rgb]{ .925,  .965,  .945} 10.96} &
  \multicolumn{1}{c}{\cellcolor[rgb]{ .388,  .745,  .482} 61.64} &
  \multicolumn{1}{c|}{\cellcolor[rgb]{ .82,  .922,  .855} 18.60} &
  \multicolumn{1}{c}{\cellcolor[rgb]{ .388,  .745,  .482} 15.01} &
  \multicolumn{1}{c}{57.96} &
  \multicolumn{1}{c|}{\cellcolor[rgb]{ .388,  .745,  .482} \textbf{23.85}} &
  \multicolumn{1}{c}{\cellcolor[rgb]{ .827,  .925,  .863} 9.48} &
  \multicolumn{1}{c}{\cellcolor[rgb]{ .867,  .941,  .894} 30.43} &
  \multicolumn{1}{c|}{\cellcolor[rgb]{ .49,  .788,  .573} 14.45} &
  \multicolumn{1}{r}{}
  \\
 &
  \multicolumn{1}{c|}{} &
  \multicolumn{1}{c|}{Injection} &
  \multicolumn{1}{c}{\cellcolor[rgb]{ .592,  .827,  .659} 13.44} &
  \multicolumn{1}{c}{\cellcolor[rgb]{ .557,  .816,  .627} 53.26} &
  \multicolumn{1}{c|}{\cellcolor[rgb]{ .388,  .745,  .482} \textbf{21.46}} &
  \multicolumn{1}{c}{\cellcolor[rgb]{ .388,  .745,  .482} 15.43} &
  \multicolumn{1}{c}{61.46} &
  \multicolumn{1}{c|}{\cellcolor[rgb]{ .388,  .745,  .482} \textbf{24.66}} &
  \multicolumn{1}{c}{\cellcolor[rgb]{ .694,  .871,  .745} 11.57} &
  \multicolumn{1}{c}{\cellcolor[rgb]{ .894,  .953,  .918} 26.43} &
  \multicolumn{1}{c|}{\cellcolor[rgb]{ .388,  .745,  .482} \textbf{16.10}} &
  \multicolumn{1}{c}{\cellcolor[rgb]{ .584,  .827,  .651} 13.67} &
  \multicolumn{1}{c}{\cellcolor[rgb]{ .576,  .824,  .647} 54.28} &
  \multicolumn{1}{c|}{\cellcolor[rgb]{ .388,  .745,  .482} \textbf{21.83}} &
  \multicolumn{1}{c}{\cellcolor[rgb]{ .529,  .804,  .604} 14.23} &
  \multicolumn{1}{c}{64.48} &
  \multicolumn{1}{c|}{\cellcolor[rgb]{ .471,  .78,  .553} 23.32} &
  \multicolumn{1}{c}{\cellcolor[rgb]{ .808,  .918,  .843} 9.99} &
  \multicolumn{1}{c}{\cellcolor[rgb]{ .898,  .953,  .922} 25.79} &
  \multicolumn{1}{c|}{\cellcolor[rgb]{ .494,  .788,  .576} 14.40} &
  \multicolumn{1}{r}{}
  \\
 &
  \multicolumn{1}{c|}{} &
  \multicolumn{1}{c|}{Extension} &
  \multicolumn{1}{c}{\cellcolor[rgb]{ .388,  .745,  .482} 14.97} &
  \multicolumn{1}{c}{37.21} &
  \multicolumn{1}{c|}{\cellcolor[rgb]{ .408,  .753,  .498} 21.35} &
  \multicolumn{1}{c}{\cellcolor[rgb]{ .498,  .788,  .576} 14.76} &
  \multicolumn{1}{c}{50.00} &
  \multicolumn{1}{c|}{\cellcolor[rgb]{ .624,  .843,  .686} 22.80} &
  \multicolumn{1}{c}{\cellcolor[rgb]{ .388,  .745,  .482} 18.01} &
  \multicolumn{1}{c}{12.81} &
  \multicolumn{1}{c|}{\cellcolor[rgb]{ .478,  .784,  .561} 14.97} &
  \multicolumn{1}{c}{\cellcolor[rgb]{ .388,  .745,  .482} 15.20} &
  \multicolumn{1}{c}{38.18} &
  \multicolumn{1}{c|}{\cellcolor[rgb]{ .404,  .753,  .494} 21.75} &
  \multicolumn{1}{c}{\cellcolor[rgb]{ .424,  .761,  .514} 14.82} &
  \multicolumn{1}{c}{52.71} &
  \multicolumn{1}{c|}{\cellcolor[rgb]{ .498,  .792,  .576} 23.13} &
  \multicolumn{1}{c}{\cellcolor[rgb]{ .388,  .745,  .482} 20.65} &
  \multicolumn{1}{c}{12.62} &
  \multicolumn{1}{c|}{\cellcolor[rgb]{ .388,  .745,  .482} \textbf{15.67}} &
  \multicolumn{1}{r}{}
  \\
\cline{2-21} &
  \multicolumn{1}{c|}{\multirow{7}[2]{*}{\begin{sideways}\textbf{Bigvul}\end{sideways}}} &
  \multicolumn{1}{c|}{NoAug} &
  \multicolumn{1}{c}{6.27} &
  \multicolumn{1}{c}{43.24} &
  \multicolumn{1}{c|}{10.95} &
  \multicolumn{1}{c}{7.28} &
  \multicolumn{1}{c}{77.27} &
  \multicolumn{1}{c|}{13.31} &
  \multicolumn{1}{c}{7.90} &
  \multicolumn{1}{c}{29.19} &
  \multicolumn{1}{c|}{12.43} &
  \multicolumn{1}{c}{\cellcolor[rgb]{ .961,  .976,  .976} 6.27} &
  \multicolumn{1}{c}{43.24} &
  \multicolumn{1}{c|}{\cellcolor[rgb]{ .988,  .988,  1} 10.95} &
  \multicolumn{1}{c}{7.28} &
  \multicolumn{1}{c}{77.27} &
  \multicolumn{1}{c|}{13.31} &
  \multicolumn{1}{c}{7.90} &
  \multicolumn{1}{c}{29.19} &
  \multicolumn{1}{c|}{12.43} &
  \multicolumn{1}{r}{}
  \\
 &
  \multicolumn{1}{c|}{} &
  \multicolumn{1}{c|}{VulGen} &
  \multicolumn{1}{c}{\cellcolor[rgb]{ .933,  .969,  .953} 6.49} &
  \multicolumn{1}{c}{\cellcolor[rgb]{ .412,  .757,  .502} 62.32} &
  \multicolumn{1}{c|}{\cellcolor[rgb]{ .827,  .925,  .859} 11.75} &
  \multicolumn{1}{c}{6.36} &
  \multicolumn{1}{c}{70.59} &
  \multicolumn{1}{c|}{11.66} &
  \multicolumn{1}{c}{\cellcolor[rgb]{ .851,  .933,  .882} 8.51} &
  \multicolumn{1}{c}{22.150 } &
  \multicolumn{1}{c|}{12.29} &
  \multicolumn{1}{c}{\cellcolor[rgb]{ .929,  .965,  .949} 6.49} &
  \multicolumn{1}{c}{\cellcolor[rgb]{ .388,  .745,  .482} 62.32} &
  \multicolumn{1}{c|}{\cellcolor[rgb]{ .886,  .949,  .914} 11.75} &
  \multicolumn{1}{c}{6.36} &
  \multicolumn{1}{c}{70.59} &
  \multicolumn{1}{c|}{11.66} &
  \multicolumn{1}{c}{\cellcolor[rgb]{ .808,  .918,  .843} 8.51} &
  \multicolumn{1}{c}{22.15 } &
  \multicolumn{1}{c|}{12.29} &
   \multicolumn{1}{r}{}
  \\
 &
  \multicolumn{1}{c|}{} &
  \multicolumn{1}{c|}{VGX} &
  \multicolumn{1}{c}{6.06} &
  \multicolumn{1}{c}{\cellcolor[rgb]{ .451,  .773,  .537} 59.30} &
  \multicolumn{1}{c|}{\cellcolor[rgb]{ .98,  .988,  .996} 10.99} &
  \multicolumn{1}{c}{6.89} &
  \multicolumn{1}{c}{73.29} &
  \multicolumn{1}{c|}{12.60} &
  \multicolumn{1}{c}{5.35} &
  \multicolumn{1}{c}{\cellcolor[rgb]{ .388,  .745,  .482} 98.82} &
  \multicolumn{1}{c|}{10.15} &
  \multicolumn{1}{c}{\cellcolor[rgb]{ .988,  .988,  1} 6.06} &
  \multicolumn{1}{c}{\cellcolor[rgb]{ .439,  .769,  .525} 59.30} &
  \multicolumn{1}{c|}{\cellcolor[rgb]{ .984,  .988,  .996} 10.99} &
  \multicolumn{1}{c}{6.89} &
  \multicolumn{1}{c}{73.29} &
  \multicolumn{1}{c|}{12.60} &
  \multicolumn{1}{c}{5.35} &
  \multicolumn{1}{c}{\cellcolor[rgb]{ .388,  .745,  .482} 98.82} &
  \multicolumn{1}{c|}{10.15} &
   \multicolumn{1}{r}{}
  \\
 &
  \multicolumn{1}{c|}{} &
  \multicolumn{1}{c|}{ROS} &
  \multicolumn{1}{c}{\cellcolor[rgb]{ .796,  .91,  .835} 7.54} &
  \multicolumn{1}{c}{25.60} &
  \multicolumn{1}{c|}{\cellcolor[rgb]{ .847,  .933,  .878} 11.65} &
  \multicolumn{1}{c}{\cellcolor[rgb]{ .808,  .918,  .843} 7.96} &
  \multicolumn{1}{c}{33.39} &
  \multicolumn{1}{c|}{12.86} &
  \multicolumn{1}{c}{\cellcolor[rgb]{ .745,  .89,  .792} 10.91} &
  \multicolumn{1}{c}{13.16} &
  \multicolumn{1}{c|}{11.93} &
  \multicolumn{1}{c}{\cellcolor[rgb]{ .78,  .906,  .824} 7.54} &
  \multicolumn{1}{c}{25.60} &
  \multicolumn{1}{c|}{\cellcolor[rgb]{ .898,  .953,  .925} 11.65} &
  \multicolumn{1}{c}{7.96} &
  \multicolumn{1}{c}{33.39} &
  \multicolumn{1}{c|}{12.86} &
  \multicolumn{1}{c}{\cellcolor[rgb]{ .671,  .859,  .725} 10.91} &
  \multicolumn{1}{c}{13.16} &
  \multicolumn{1}{c|}{11.93} &
   \multicolumn{1}{r}{}
  \\
 &
  \multicolumn{1}{c|}{} &
  \multicolumn{1}{c|}{Mutation} &
  \multicolumn{1}{c}{\cellcolor[rgb]{ .761,  .898,  .804} 7.80} &
  \multicolumn{1}{c}{\cellcolor[rgb]{ .388,  .745,  .482} 63.91} &
  \multicolumn{1}{c|}{\cellcolor[rgb]{ .388,  .745,  .482} \textbf{13.90}} &
  \multicolumn{1}{c}{\cellcolor[rgb]{ .847,  .933,  .878} 7.63} &
  \multicolumn{1}{c}{67.73} &
  \multicolumn{1}{c|}{\cellcolor[rgb]{ .788,  .91,  .827} 13.72} &
  \multicolumn{1}{c}{6.78} &
  \multicolumn{1}{c}{\cellcolor[rgb]{ .561,  .816,  .631} 74.14} &
  \multicolumn{1}{c|}{12.42} &
  \multicolumn{1}{c}{\cellcolor[rgb]{ .78,  .906,  .82} 7.56} &
  \multicolumn{1}{c}{\cellcolor[rgb]{ .475,  .78,  .557} 57.23} &
  \multicolumn{1}{c|}{\cellcolor[rgb]{ .678,  .863,  .733} 13.36} &
  \multicolumn{1}{c}{\cellcolor[rgb]{ .522,  .8,  .596} 9.37} &
  \multicolumn{1}{c}{53.26} &
  \multicolumn{1}{c|}{\cellcolor[rgb]{ .388,  .745,  .482} \textbf{15.94}} &
  \multicolumn{1}{c}{6.65} &
  \multicolumn{1}{c}{\cellcolor[rgb]{ .533,  .804,  .608} 78.31} &
  \multicolumn{1}{c|}{12.25} &
  \multicolumn{1}{r}{}
  \\
 &
  \multicolumn{1}{c|}{} &
  \multicolumn{1}{c|}{Injection} &
  \multicolumn{1}{c}{\cellcolor[rgb]{ .631,  .843,  .69} 8.79} &
  \multicolumn{1}{c}{29.41} &
  \multicolumn{1}{c|}{\cellcolor[rgb]{ .463,  .776,  .549} 13.53} &
  \multicolumn{1}{c}{\cellcolor[rgb]{ .388,  .745,  .482} 11.64} &
  \multicolumn{1}{c}{38.00} &
  \multicolumn{1}{c|}{\cellcolor[rgb]{ .388,  .745,  .482} \textbf{17.82}} &
  \multicolumn{1}{c}{\cellcolor[rgb]{ .722,  .882,  .773} 11.43} &
  \multicolumn{1}{c}{18.81} &
  \multicolumn{1}{c|}{\cellcolor[rgb]{ .451,  .773,  .537} 14.22} &
  \multicolumn{1}{c}{\cellcolor[rgb]{ .388,  .745,  .482} 10.33} &
  \multicolumn{1}{c}{31.80} &
  \multicolumn{1}{c|}{\cellcolor[rgb]{ .388,  .745,  .482} \textbf{15.59}} &
  \multicolumn{1}{c}{\cellcolor[rgb]{ .435,  .765,  .525} 9.91} &
  \multicolumn{1}{c}{38.16} &
  \multicolumn{1}{c|}{\cellcolor[rgb]{ .42,  .757,  .51} 15.74} &
  \multicolumn{1}{c}{7.85} &
  \multicolumn{1}{c}{\cellcolor[rgb]{ .863,  .937,  .89} 30.49} &
  \multicolumn{1}{c|}{\cellcolor[rgb]{ .576,  .824,  .643} 12.49} &
   \multicolumn{1}{r}{}
  \\
 &
  \multicolumn{1}{c|}{} &
  \multicolumn{1}{c|}{Extension} &
  \multicolumn{1}{c}{\cellcolor[rgb]{ .388,  .745,  .482} 10.60} &
  \multicolumn{1}{c}{17.33} &
  \multicolumn{1}{c|}{\cellcolor[rgb]{ .541,  .808,  .616} 13.16} &
  \multicolumn{1}{c}{\cellcolor[rgb]{ .447,  .773,  .533} 11.13} &
  \multicolumn{1}{c}{34.02} &
  \multicolumn{1}{c|}{\cellcolor[rgb]{ .494,  .788,  .573} 16.77} &
  \multicolumn{1}{c}{\cellcolor[rgb]{ .388,  .745,  .482} 19.03} &
  \multicolumn{1}{c}{11.96} &
  \multicolumn{1}{c|}{\cellcolor[rgb]{ .388,  .745,  .482} \textbf{14.68}} &
  \multicolumn{1}{c}{\cellcolor[rgb]{ .451,  .773,  .537} 9.89} &
  \multicolumn{1}{c}{31.48} &
  \multicolumn{1}{c|}{\cellcolor[rgb]{ .459,  .776,  .545} 15.05} &
  \multicolumn{1}{c}{\cellcolor[rgb]{ .388,  .745,  .482} 10.22} &
  \multicolumn{1}{c}{33.86} &
  \multicolumn{1}{c|}{\cellcolor[rgb]{ .424,  .761,  .514} 15.70} &
  \multicolumn{1}{c}{\cellcolor[rgb]{ .388,  .745,  .482} 15.74} &
  \multicolumn{1}{c}{11.86} &
  \multicolumn{1}{c|}{\cellcolor[rgb]{ .388,  .745,  .482} \textbf{13.53}} &
   \multicolumn{1}{r}{}
  \\
\cline{3-21}

\end{tabular}%
    }
\end{table}%

% \sw{also can you measure the diversity of the augmented samples? maybe, calculate the average distance of each generated samples to the center on the vector using codebert, like if it is less diverse, all points will close to the center and average distance should be smaller. }

% \sd{TODO: 
% 2- Mutation slightly is better than ROS, so probably doesn't worth it
% \\ DONE
% 3- Injection and Extension outperform others and INjection is slightly better
% \\ DONE
% 4- Note that both ChatGPT and CodeQwen perform well, ChatGPT slightly better in some cases and in some CodeQwen (Put this in discussion instead)
% \\
% 1- Talk about how ROS outperforms VGX and VUlgen
% }
% \sw{focus on our approach first. move point 1 to the last.}

\textbf{Extension and Injection outperforms Mutation. Injection slightly outperforms Extension}. Table~\ref{tab:rq1} presents the results of \ourTool and baselines across different experimental instances. As observed, \textbf{Extension} and \textbf{Injection} beat \textbf{Mutation} in most (9 out of 12) instances. \textbf{Extension} and \textbf{Injection} perform close to each other, while \textbf{Injection} appears to have the advantage by a tiny margin in most (9 out of 12) settings. In terms of F1-score, \textbf{Injection} achieves an F1-score of 17.60\%, outperforming \textbf{Extension} (F1-score 17.38\%) and \textbf{Mutation} (F1-score  15.75\%) by 0.96\% and 12.44\% on average, respectively. One possible reason is that \textbf{Injection} and \textbf{Extension} enrich the context where vulnerabilities could occur while \textbf{Mutation} does not alter the semantics of vulnerable code. \textbf{Mutation} does not enrich the context of the semantics of the original vulnerable code. %Any reasons why injection and extension outperform mutation? is it because injection and extension enrich the context of vul code by adding additional logic? }\sd{yes and the fact that mutation doesn't change the code too much }

\begin{figure}[htpb]

\includegraphics[width=0.9\textwidth]{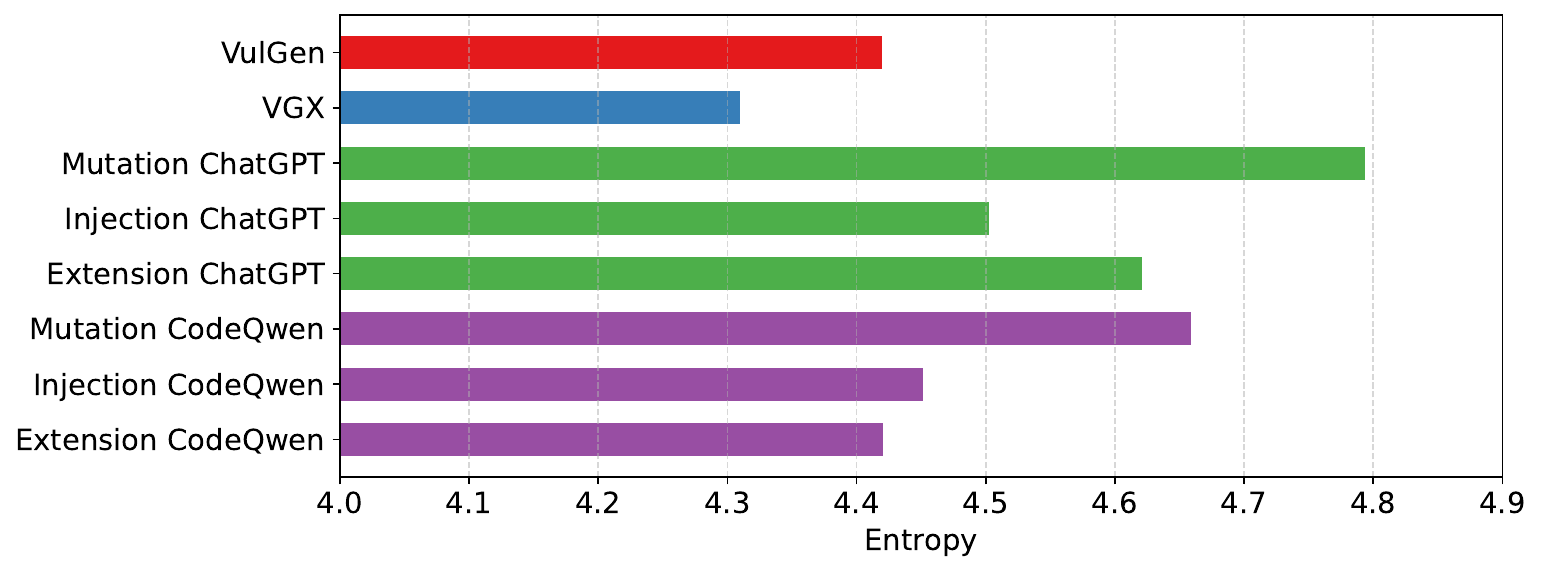}
	\caption{Entropy of the augmented vulnerable datasets by different approaches. Higher indicates more diversity.\\\\}
 % making things more compact, will add this if got accepted on Camera ready or we had more space at the end
 
 % We use PCA on CodeBERT's embeddings to reduce dimension to 3. Next, we calculate the histogram using 10 bins, from which we calculate the entropy of the vectors. Higher entropy values indicate greater diversity among the vulnerable samples, while lower values suggest a more concentrated distribution.}
 
	\label{fig:entropy}
 \vspace{-0.2in}
\end{figure}

\textbf{All our proposed LLM-based augmentation strategies outperform the baselines. Typically, \textbf{Injection} and \textbf{Extension} always outperform baselines in all instances. \textbf{Injection} outperforms NoAug, Vulgen, VGX, and ROS  by 30.80\%, 27.48\%, 27.93\%, and 15.41\% on average F1-score, respectively.} As shown in Table~\ref{tab:rq1}, \textbf{Injection} and \textbf{Extension} strategies outperform baselines (i.e., NoAug, VulGen, VGX, ROS) in all of the experimental instances in terms of F1-score. 
More specifically, \textbf{Extension} beats the baselines: NoAug, Vulgen, VGX, and ROS by 29.68\%, 26.27\%, 26.90\%, and 14.35\% on average F1-score, while the \textbf{Injection} strategy beats the baselines: NoAug, Vulgen, VGX, and ROS by 30.80\%, 27.48\%, 27.93\%, and 15.41\% on average F1-score respectively.\\
\\
As observed, \textbf{Mutation} beats most baselines by a large margin. However, \textbf{Mutation} does not have a dominating advantage over ROS, and ROS beats \textbf{Mutation} in some (4 out of 12) instances. It is worth noting that ROS beats the baseline, Vulgen, and VGX by 13.96\%, 10.72\%, and 11.21\% in terms of F1-score on average. 
%\sw{need revisit after get the results. for the entropy can we measure only the generated vulnerable samples? \sd{We can, and I did the results didn't show anything because the diversity of the extra samples don't matter, the diversity of the final dataset matters, I have sent the table for it in TEAMS}}
\\
As discussed in Section~\ref{sec:background}, SOTA approaches like VGX and VulGen only focus on single-statement that limits the diversity of generated vulnerabilities. However, our approaches do not have this limitation and are able to generate more diverse vulnerable samples. To examine this, we measure the diversity of the augmented vulnerable samples by our approaches and VGX and VulGen. By following previous studies, we first apply principle component analysis (PCA) on CodeBERT's embeddings of the generated vulnerable samples and reduce the dimension to three. Next, we calculate the histogram using 10 bins, from which we calculate the entropy of the vectors. Higher entropy values indicate greater diversity among the vulnerable samples, while lower values suggest a more concentrated distribution with similar samples. As depicted in Figure~\ref{fig:entropy} the entropy for \textbf{Mutation}, \textbf{Injection}, and \textbf{Extension} are 4.79, 4.5, and 4.62 for ChatGPT, and 4.66, 4.45, 4.42 for CodeQwen, while for Vulgen and VGX, it is 4.42 and 4.31 respectively. The results demonstrate that our approaches have a better potential for generating more diverse samples as we beat them while using only one of the datasets they used for vulnerability mining. \textbf{\ourTool produces more diverse vulnerable samples compared to SOTA approaches VGX and VulGen.}\\ 
% \sw{can you provide an example that our approach could generate better vulnerable code, for example, multiple line vulnerable injection, which vulgen cannot.\sd{I don't think this is needed, because vulgen specifcly generates one-line vuls and I don't really think we have much space left anyway}}\sw{I strongly recommend to provide a strong example if we have space. let's see if we have space. \sd{We actually don't have space left, but we can add a few to the appendix}}
%\textbf{The mutation strategy beats the baselines, VulGen, VGX, and ROS by 16.73\%, 13.92\%, 14.71\%, and 3.49\% on average F1-score across all settings.}. Table~\ref{tab:rq1} shows the studied strategies' performance (P, R, F1). 
\\
Lastly, it is worth noting that the results on both of the LLMs perform very close to each other. CodeQwen1.5-7B-Chat slightly outperforms ChatGPT3.5 Turbo by a tiny margin (i.e. 1.49\%) averaged across all three strategies, yet one cannot conclude that CodeQwen is a better LLM for vulnerability generation as we did not explore the hyper-parameters for the optimum setting for each of the LLMs. However, we can conclude that our strategies work with similar LLM to ChatGPT and CodeQwen (e.g. GPT4~\cite{openai2024gpt4technicalreport} and DeepSeek-Coder~\cite{deepseek-coder}) and do not require an LLM that is trained specifically on code.\\

\rqboxc{Both \textbf{Injection} and \textbf{Extension} outperforms the baselines and the \textbf{Mutation} by a large margin, while \textbf{Injection} provides a slightly higher performance compared to the \textbf{Extension}. For instance, \textbf{Injection}
outperforms NoAug, Vulgen, VGX, and ROS by 30.80\%,
27.48\%, 27.93\%, and 15.41\% on average F1-score.
}

%\sw{this is overclaim, we can say so.unless we do experiments and the results support so, }\sd{why is that? we have tested chatGPT which is a big model but isn't code specific, CodeQwen is code specific and very small, so if in terms of metrics people find similar LLMs then we should be fine, I did test DeepSeekCoder7B and the results it generated was pretty much the same for 8 out of 10 items}\sw{never say ANY if you haven't tested all...}

% REsize: 
% % replace -> \resizebox{18cm}{!}{TABULAR}
% 
% delete [$b] and [$t] remove the $$s

\subsection{\rqtwo}\label{sec:rq2}

\begin{table}[htbp]
  \centering
  \caption{Results of Ablation Studies on the Retriever and Clustering for Injection across the studied DLVD models (i.e., Devign, Reveal, and Linevul).}
  \resizebox{13.5cm}{!}{
% Table generated by Excel2LaTeX from sheet 'Sheet2'
\begin{tabular}{rrrrrrrrrrrrrrrrrrrrr}
 &
   &
  \multicolumn{9}{c|}{\textit{\textbf{ChatGPT 3.5 Turbo}}} &
  \multicolumn{9}{c}{\textit{\textbf{CodeQwen1.5-7B-Chat}}} &
  
  \\
\cline{2-20}\multicolumn{1}{r|}{} &
  \multicolumn{1}{c|}{\multirow{2}[4]{*}{{\raisebox{0.4cm}{\textbf{Strategy}}}}} &
  \multicolumn{3}{c|}{\textbf{Devign}} &
  \multicolumn{3}{c|}{\textbf{Reveal}} &
  \multicolumn{3}{c|}{\textbf{Linevul}} &
  \multicolumn{3}{c|}{\textbf{Devign}} &
  \multicolumn{3}{c|}{\textbf{Reveal}} &
  \multicolumn{3}{c|}{\textbf{Linevul}} &
  
  \\
\cline{3-20}\multicolumn{1}{r|}{} &
  \multicolumn{1}{c|}{} &
  \multicolumn{1}{c}{P} &
  \multicolumn{1}{c}{R} &
  \multicolumn{1}{c|}{F1} &
  \multicolumn{1}{c}{P} &
  \multicolumn{1}{c}{R} &
  \multicolumn{1}{c|}{F1} &
  \multicolumn{1}{c}{P} &
  \multicolumn{1}{c}{R} &
  \multicolumn{1}{c|}{F1} &
  \multicolumn{1}{c}{P} &
  \multicolumn{1}{c}{R} &
  \multicolumn{1}{c|}{F1} &
  \multicolumn{1}{c}{P} &
  \multicolumn{1}{c}{R} &
  \multicolumn{1}{c|}{F1} &
  \multicolumn{1}{c}{P} &
  \multicolumn{1}{c}{R} &
  \multicolumn{1}{c|}{F1} &
  
  \\
\cline{2-2}\multicolumn{1}{c|}{\multirow{3}[2]{*}{\begin{sideways}\textbf{ Reveal}\end{sideways}}} &
  \multicolumn{1}{c|}{w/o Retriever} &
  \multicolumn{1}{c}{\cellcolor[rgb]{ .773,  .902,  .816} 13.91} &
  \multicolumn{1}{c}{\cellcolor[rgb]{ .706,  .875,  .757} 44.33} &
  \multicolumn{1}{c|}{\cellcolor[rgb]{ .753,  .894,  .796} 21.18} &
  \multicolumn{1}{c}{\cellcolor[rgb]{ .988,  .988,  1} 14.17} &
  \multicolumn{1}{c}{\cellcolor[rgb]{ .388,  .745,  .482} 64.29} &
  \multicolumn{1}{c|}{\cellcolor[rgb]{ .988,  .988,  1} 23.22} &
  \multicolumn{1}{c}{\cellcolor[rgb]{ .988,  .988,  1} 8.11} &
  \multicolumn{1}{c}{\cellcolor[rgb]{ .388,  .745,  .482} 32.15} &
  \multicolumn{1}{c|}{\cellcolor[rgb]{ .988,  .988,  1} 12.95} &
  \multicolumn{1}{c}{\cellcolor[rgb]{ .988,  .988,  1} 11.94} &
  \multicolumn{1}{c}{\cellcolor[rgb]{ .388,  .745,  .482} 57.12} &
  \multicolumn{1}{c|}{\cellcolor[rgb]{ .988,  .988,  1} 19.75} &
  \multicolumn{1}{c}{\cellcolor[rgb]{ .988,  .988,  1} 12.52} &
  \multicolumn{1}{c}{\cellcolor[rgb]{ .388,  .745,  .482} 72.74} &
  \multicolumn{1}{c|}{\cellcolor[rgb]{ .988,  .988,  1} 21.37} &
  \multicolumn{1}{c}{\cellcolor[rgb]{ .388,  .745,  .482} 13.00} &
  \multicolumn{1}{c}{\cellcolor[rgb]{ .388,  .745,  .482} 32.06} &
  \multicolumn{1}{c|}{\cellcolor[rgb]{ .388,  .745,  .482} 18.50} &
  
  \\
\multicolumn{1}{c|}{} &
  \multicolumn{1}{c|}{ w/o Clustering} &
  \multicolumn{1}{c}{\cellcolor[rgb]{ .388,  .745,  .482} 14.75} &
  \multicolumn{1}{c}{\cellcolor[rgb]{ .988,  .988,  1} 36.37} &
  \multicolumn{1}{c|}{\cellcolor[rgb]{ .988,  .988,  1} 20.99} &
  \multicolumn{1}{c}{\cellcolor[rgb]{ .388,  .745,  .482} 15.59} &
  \multicolumn{1}{c}{\cellcolor[rgb]{ .988,  .988,  1} 59.89} &
  \multicolumn{1}{c|}{\cellcolor[rgb]{ .388,  .745,  .482} 24.74} &
  \multicolumn{1}{c}{\cellcolor[rgb]{ .678,  .863,  .733} 9.92} &
  \multicolumn{1}{c}{\cellcolor[rgb]{ .988,  .988,  1} 24.61} &
  \multicolumn{1}{c|}{\cellcolor[rgb]{ .765,  .898,  .808} 14.14} &
  \multicolumn{1}{c}{\cellcolor[rgb]{ .388,  .745,  .482} 13.86} &
  \multicolumn{1}{c}{\cellcolor[rgb]{ .988,  .988,  1} 43.37} &
  \multicolumn{1}{c|}{\cellcolor[rgb]{ .627,  .843,  .69} 21.01} &
  \multicolumn{1}{c}{\cellcolor[rgb]{ .388,  .745,  .482} 14.92} &
  \multicolumn{1}{c}{\cellcolor[rgb]{ .988,  .988,  1} 52.17} &
  \multicolumn{1}{c|}{\cellcolor[rgb]{ .427,  .761,  .514} 23.20} &
  \multicolumn{1}{c}{\cellcolor[rgb]{ .545,  .808,  .616} 12.23} &
  \multicolumn{1}{c}{\cellcolor[rgb]{ .988,  .988,  1} 24.61} &
  \multicolumn{1}{c|}{\cellcolor[rgb]{ .706,  .875,  .757} 16.34} &
  
  \\
\multicolumn{1}{c|}{} &
  \multicolumn{1}{c|}{Injection } &
  \multicolumn{1}{c}{\cellcolor[rgb]{ .988,  .988,  1} 13.44} &
  \multicolumn{1}{c}{\cellcolor[rgb]{ .388,  .745,  .482} 53.26} &
  \multicolumn{1}{c|}{\cellcolor[rgb]{ .388,  .745,  .482} \textbf{21.46}} &
  \multicolumn{1}{c}{\cellcolor[rgb]{ .459,  .773,  .541} 15.43} &
  \multicolumn{1}{c}{\cellcolor[rgb]{ .776,  .902,  .816} 61.46} &
  \multicolumn{1}{c|}{\cellcolor[rgb]{ .42,  .757,  .51} \textbf{24.66}} &
  \multicolumn{1}{c}{\cellcolor[rgb]{ .388,  .745,  .482} 11.57} &
  \multicolumn{1}{c}{\cellcolor[rgb]{ .847,  .933,  .878} 26.43} &
  \multicolumn{1}{c|}{\cellcolor[rgb]{ .388,  .745,  .482} \textbf{16.10}} &
  \multicolumn{1}{c}{\cellcolor[rgb]{ .451,  .773,  .537} 13.67} &
  \multicolumn{1}{c}{\cellcolor[rgb]{ .514,  .796,  .592} 54.28} &
  \multicolumn{1}{c|}{\cellcolor[rgb]{ .388,  .745,  .482} \textbf{21.83}} &
  \multicolumn{1}{c}{\cellcolor[rgb]{ .561,  .816,  .631} 14.23} &
  \multicolumn{1}{c}{\cellcolor[rgb]{ .631,  .843,  .694} 64.48} &
  \multicolumn{1}{c|}{\cellcolor[rgb]{ .388,  .745,  .482} \textbf{23.32}} &
  \multicolumn{1}{c}{\cellcolor[rgb]{ .988,  .988,  1} 9.99} &
  \multicolumn{1}{c}{\cellcolor[rgb]{ .894,  .953,  .922} 25.79} &
  \multicolumn{1}{c|}{\cellcolor[rgb]{ .988,  .988,  1} 14.40} &
  
  \\
\cline{1-20}\multicolumn{1}{c|}{\multirow{3}[2]{*}{\begin{sideways}\textbf{Bigvul}\end{sideways}}} &
  \multicolumn{1}{c|}{w/o Retriever} &
  \multicolumn{1}{c}{\cellcolor[rgb]{ .988,  .988,  1} 8.47} &
  \multicolumn{1}{c}{\cellcolor[rgb]{ .388,  .745,  .482} 36.09} &
  \multicolumn{1}{c|}{\cellcolor[rgb]{ .388,  .745,  .482} 13.72} &
  \multicolumn{1}{c}{\cellcolor[rgb]{ .988,  .988,  1} 8.96} &
  \multicolumn{1}{c}{\cellcolor[rgb]{ .388,  .745,  .482} 65.50} &
  \multicolumn{1}{c|}{\cellcolor[rgb]{ .988,  .988,  1} 15.76} &
  \multicolumn{1}{c}{\cellcolor[rgb]{ .988,  .988,  1} 6.67} &
  \multicolumn{1}{c}{\cellcolor[rgb]{ .388,  .745,  .482} 66.54} &
  \multicolumn{1}{c|}{\cellcolor[rgb]{ .988,  .988,  1} 12.12} &
  \multicolumn{1}{c}{\cellcolor[rgb]{ .988,  .988,  1} 8.87} &
  \multicolumn{1}{c}{\cellcolor[rgb]{ .388,  .745,  .482} 53.74} &
  \multicolumn{1}{c|}{\cellcolor[rgb]{ .796,  .91,  .835} 15.23} &
  \multicolumn{1}{c}{\cellcolor[rgb]{ .988,  .988,  1} 8.88} &
  \multicolumn{1}{c}{\cellcolor[rgb]{ .388,  .745,  .482} 64.39} &
  \multicolumn{1}{c|}{\cellcolor[rgb]{ .988,  .988,  1} 15.60} &
  \multicolumn{1}{c}{\cellcolor[rgb]{ .988,  .988,  1} 7.09} &
  \multicolumn{1}{c}{\cellcolor[rgb]{ .388,  .745,  .482} 61.72} &
  \multicolumn{1}{c|}{\cellcolor[rgb]{ .961,  .98,  .976} 12.72} &
  
  \\
\multicolumn{1}{c|}{} &
  \multicolumn{1}{c|}{w/o Clustering} &
  \multicolumn{1}{c}{\cellcolor[rgb]{ .878,  .945,  .902} 8.53} &
  \multicolumn{1}{c}{\cellcolor[rgb]{ .647,  .851,  .706} 33.23} &
  \multicolumn{1}{c|}{\cellcolor[rgb]{ .867,  .941,  .894} 13.57} &
  \multicolumn{1}{c}{\cellcolor[rgb]{ .49,  .788,  .573} 11.19} &
  \multicolumn{1}{c}{\cellcolor[rgb]{ .988,  .988,  1} 32.75} &
  \multicolumn{1}{c|}{\cellcolor[rgb]{ .722,  .882,  .773} 16.68} &
  \multicolumn{1}{c}{\cellcolor[rgb]{ .624,  .843,  .686} 9.56} &
  \multicolumn{1}{c}{\cellcolor[rgb]{ .988,  .988,  1} 16.68} &
  \multicolumn{1}{c|}{\cellcolor[rgb]{ .98,  .984,  .992} 12.16} &
  \multicolumn{1}{c}{\cellcolor[rgb]{ .62,  .839,  .682} 9.77} &
  \multicolumn{1}{c}{\cellcolor[rgb]{ .965,  .98,  .98} 32.75} &
  \multicolumn{1}{c|}{\cellcolor[rgb]{ .988,  .988,  1} 15.05} &
  \multicolumn{1}{c}{\cellcolor[rgb]{ .635,  .847,  .698} 9.49} &
  \multicolumn{1}{c}{\cellcolor[rgb]{ .749,  .894,  .796} 48.65} &
  \multicolumn{1}{c|}{\cellcolor[rgb]{ .388,  .745,  .482} 15.88} &
  \multicolumn{1}{c}{\cellcolor[rgb]{ .388,  .745,  .482} 14.61} &
  \multicolumn{1}{c}{\cellcolor[rgb]{ .988,  .988,  1} 21.78} &
  \multicolumn{1}{c|}{\cellcolor[rgb]{ .388,  .745,  .482} 17.49} &
  
  \\
\multicolumn{1}{c|}{} &
  \multicolumn{1}{c|}{Injection } &
  \multicolumn{1}{c}{\cellcolor[rgb]{ .388,  .745,  .482} 8.79} &
  \multicolumn{1}{c}{\cellcolor[rgb]{ .988,  .988,  1} 29.41} &
  \multicolumn{1}{c|}{\cellcolor[rgb]{ .988,  .988,  1} 13.53} &
  \multicolumn{1}{c}{\cellcolor[rgb]{ .388,  .745,  .482} 11.64} &
  \multicolumn{1}{c}{\cellcolor[rgb]{ .894,  .953,  .918} 38.00} &
  \multicolumn{1}{c|}{\cellcolor[rgb]{ .388,  .745,  .482} \textbf{17.82}} &
  \multicolumn{1}{c}{\cellcolor[rgb]{ .388,  .745,  .482} 11.43} &
  \multicolumn{1}{c}{\cellcolor[rgb]{ .965,  .98,  .98} 18.81} &
  \multicolumn{1}{c|}{\cellcolor[rgb]{ .388,  .745,  .482} \textbf{14.22}} &
  \multicolumn{1}{c}{\cellcolor[rgb]{ .388,  .745,  .482} 10.33} &
  \multicolumn{1}{c}{\cellcolor[rgb]{ .988,  .988,  1} 31.80} &
  \multicolumn{1}{c|}{\cellcolor[rgb]{ .388,  .745,  .482} \textbf{15.59}} &
  \multicolumn{1}{c}{\cellcolor[rgb]{ .388,  .745,  .482} 9.91} &
  \multicolumn{1}{c}{\cellcolor[rgb]{ .988,  .988,  1} 38.16} &
  \multicolumn{1}{c|}{\cellcolor[rgb]{ .694,  .871,  .749} 15.74} &
  \multicolumn{1}{c}{\cellcolor[rgb]{ .929,  .965,  .949} 7.85} &
  \multicolumn{1}{c}{\cellcolor[rgb]{ .859,  .937,  .89} 30.49} &
  \multicolumn{1}{c|}{\cellcolor[rgb]{ .988,  .988,  1} 12.49} &
  
  \\
\cline{2-20} &
   &
   &
   &
   &
   &
   &
   &
   &
   &
   &
   &
   &
   &
   &
   &
   &
   &
   &
   &
  
  \\
\end{tabular}%
    }
  \label{tab:rq2_injection}%
\end{table}%
\begin{table}[htbp]
  \centering
  \caption{Results of Ablation Studies on the Retriever and Clustering for Extension across the studied DLVD models (i.e., Devign, Reveal, and Linevul).}
  \resizebox{13.5cm}{!}{
% Table generated by Excel2LaTeX from sheet 'Sheet2'
\begin{tabular}{rrrrrrrrrrrrrrrrrrrrr}
 &
   &
  \multicolumn{9}{c|}{\textit{\textbf{ChatGPT 3.5 Turbo}}} &
  \multicolumn{9}{c}{\textit{\textbf{CodeQwen1.5-7B-Chat}}} &
  
  \\
\cline{2-20}\multicolumn{1}{r|}{} &
  \multicolumn{1}{c|}{\multirow{2}[4]{*}{{\raisebox{0.4cm}{\textbf{Strategy}}}}} &
  \multicolumn{3}{c|}{\textbf{Devign}} &
  \multicolumn{3}{c|}{\textbf{Reveal}} &
  \multicolumn{3}{c|}{\textbf{Linevul}} &
  \multicolumn{3}{c|}{\textbf{Devign}} &
  \multicolumn{3}{c|}{\textbf{Reveal}} &
  \multicolumn{3}{c|}{\textbf{Linevul}} &
  
  \\
\cline{3-20}\multicolumn{1}{r|}{} &
  \multicolumn{1}{c|}{} &
  \multicolumn{1}{c}{P} &
  \multicolumn{1}{c}{R} &
  \multicolumn{1}{c|}{F1} &
  \multicolumn{1}{c}{P} &
  \multicolumn{1}{c}{R} &
  \multicolumn{1}{c|}{F1} &
  \multicolumn{1}{c}{P} &
  \multicolumn{1}{c}{R} &
  \multicolumn{1}{c|}{F1} &
  \multicolumn{1}{c}{P} &
  \multicolumn{1}{c}{R} &
  \multicolumn{1}{c|}{F1} &
  \multicolumn{1}{c}{P} &
  \multicolumn{1}{c}{R} &
  \multicolumn{1}{c|}{F1} &
  \multicolumn{1}{c}{P} &
  \multicolumn{1}{c}{R} &
  \multicolumn{1}{c|}{F1} &
  
  \\
\cline{2-2}\multicolumn{1}{c|}{\multirow{3}[2]{*}{\begin{sideways}\textbf{ Reveal}\end{sideways}}} &
  \multicolumn{1}{c|}{w/o Retriever} &
  \multicolumn{1}{c}{\cellcolor[rgb]{ .988,  .988,  1} 13.96} &
  \multicolumn{1}{c}{\cellcolor[rgb]{ .988,  .988,  1} 35.59} &
  \multicolumn{1}{c|}{\cellcolor[rgb]{ .988,  .988,  1} 20.05} &
  \multicolumn{1}{c}{\cellcolor[rgb]{ .988,  .988,  1} 14.00} &
  \multicolumn{1}{c}{\cellcolor[rgb]{ .988,  .988,  1} 49.94} &
  \multicolumn{1}{c|}{\cellcolor[rgb]{ .988,  .988,  1} 21.87} &
  \multicolumn{1}{c}{\cellcolor[rgb]{ .941,  .973,  .961} 8.63} &
  \multicolumn{1}{c}{\cellcolor[rgb]{ .988,  .988,  1} 12.20} &
  \multicolumn{1}{c|}{\cellcolor[rgb]{ .988,  .988,  1} 12.20} &
  \multicolumn{1}{c}{\cellcolor[rgb]{ .988,  .988,  1} 11.55} &
  \multicolumn{1}{c}{\cellcolor[rgb]{ .388,  .745,  .482} 67.73} &
  \multicolumn{1}{c|}{\cellcolor[rgb]{ .988,  .988,  1} 19.74} &
  \multicolumn{1}{c}{\cellcolor[rgb]{ .988,  .988,  1} 13.21} &
  \multicolumn{1}{c}{\cellcolor[rgb]{ .388,  .745,  .482} 65.26} &
  \multicolumn{1}{c|}{\cellcolor[rgb]{ .988,  .988,  1} 21.98} &
  \multicolumn{1}{c}{\cellcolor[rgb]{ .894,  .953,  .918} 11.52} &
  \multicolumn{1}{c}{\cellcolor[rgb]{ .388,  .745,  .482} 32.15} &
  \multicolumn{1}{c|}{\cellcolor[rgb]{ .388,  .745,  .482} 16.96} &
  
  \\
\multicolumn{1}{c|}{} &
  \multicolumn{1}{c|}{ w/o Clustering} &
  \multicolumn{1}{c}{\cellcolor[rgb]{ .388,  .745,  .482} 15.14} &
  \multicolumn{1}{c}{\cellcolor[rgb]{ .835,  .925,  .867} 36.01} &
  \multicolumn{1}{c|}{\cellcolor[rgb]{ .404,  .753,  .494} 21.32} &
  \multicolumn{1}{c}{\cellcolor[rgb]{ .722,  .882,  .773} 14.34} &
  \multicolumn{1}{c}{\cellcolor[rgb]{ .388,  .745,  .482} 63.63} &
  \multicolumn{1}{c|}{\cellcolor[rgb]{ .388,  .745,  .482} 23.41} &
  \multicolumn{1}{c}{\cellcolor[rgb]{ .988,  .988,  1} 7.82} &
  \multicolumn{1}{c}{\cellcolor[rgb]{ .388,  .745,  .482} 48.38} &
  \multicolumn{1}{c|}{\cellcolor[rgb]{ .718,  .878,  .765} 13.46} &
  \multicolumn{1}{c}{\cellcolor[rgb]{ .388,  .745,  .482} 15.68} &
  \multicolumn{1}{c}{\cellcolor[rgb]{ .988,  .988,  1} 36.85} &
  \multicolumn{1}{c|}{\cellcolor[rgb]{ .388,  .745,  .482} 22.00} &
  \multicolumn{1}{c}{\cellcolor[rgb]{ .388,  .745,  .482} 14.93} &
  \multicolumn{1}{c}{\cellcolor[rgb]{ .988,  .988,  1} 50.84} &
  \multicolumn{1}{c|}{\cellcolor[rgb]{ .416,  .757,  .506} 23.09} &
  \multicolumn{1}{c}{\cellcolor[rgb]{ .988,  .988,  1} 9.77} &
  \multicolumn{1}{c}{\cellcolor[rgb]{ .722,  .882,  .773} 21.34} &
  \multicolumn{1}{c|}{\cellcolor[rgb]{ .988,  .988,  1} 13.40} &
  
  \\
\multicolumn{1}{c|}{} &
  \multicolumn{1}{c|}{Extention } &
  \multicolumn{1}{c}{\cellcolor[rgb]{ .478,  .784,  .561} 14.97} &
  \multicolumn{1}{c}{\cellcolor[rgb]{ .388,  .745,  .482} 37.21} &
  \multicolumn{1}{c|}{\cellcolor[rgb]{ .388,  .745,  .482} \textbf{21.35}} &
  \multicolumn{1}{c}{\cellcolor[rgb]{ .388,  .745,  .482} 14.76} &
  \multicolumn{1}{c}{\cellcolor[rgb]{ .988,  .988,  1} 50.00} &
  \multicolumn{1}{c|}{\cellcolor[rgb]{ .627,  .843,  .69} 22.80} &
  \multicolumn{1}{c}{\cellcolor[rgb]{ .388,  .745,  .482} 18.01} &
  \multicolumn{1}{c}{\cellcolor[rgb]{ .98,  .984,  .992} 12.81} &
  \multicolumn{1}{c|}{\cellcolor[rgb]{ .388,  .745,  .482} \textbf{14.97}} &
  \multicolumn{1}{c}{\cellcolor[rgb]{ .459,  .776,  .545} 15.20} &
  \multicolumn{1}{c}{\cellcolor[rgb]{ .965,  .98,  .98} 38.18} &
  \multicolumn{1}{c|}{\cellcolor[rgb]{ .459,  .773,  .541} 21.75} &
  \multicolumn{1}{c}{\cellcolor[rgb]{ .431,  .765,  .518} 14.82} &
  \multicolumn{1}{c}{\cellcolor[rgb]{ .914,  .957,  .933} 52.71} &
  \multicolumn{1}{c|}{\cellcolor[rgb]{ .388,  .745,  .482} \textbf{23.13}} &
  \multicolumn{1}{c}{\cellcolor[rgb]{ .388,  .745,  .482} 20.65} &
  \multicolumn{1}{c}{\cellcolor[rgb]{ .988,  .988,  1} 12.62} &
  \multicolumn{1}{c|}{\cellcolor[rgb]{ .608,  .835,  .671} 15.67} &
  
  \\
\cline{1-20}\multicolumn{1}{c|}{\multirow{3}[2]{*}{\begin{sideways}\textbf{Bigvul}\end{sideways}}} &
  \multicolumn{1}{c|}{w/o Retriever} &
  \multicolumn{1}{c}{\cellcolor[rgb]{ .988,  .988,  1} 7.88} &
  \multicolumn{1}{c}{\cellcolor[rgb]{ .988,  .988,  1} 15.42} &
  \multicolumn{1}{c|}{\cellcolor[rgb]{ .988,  .988,  1} 10.43} &
  \multicolumn{1}{c}{\cellcolor[rgb]{ .988,  .988,  1} 8.62} &
  \multicolumn{1}{c}{\cellcolor[rgb]{ .855,  .937,  .886} 25.12} &
  \multicolumn{1}{c|}{\cellcolor[rgb]{ .988,  .988,  1} 12.83} &
  \multicolumn{1}{c}{\cellcolor[rgb]{ .988,  .988,  1} 7.13} &
  \multicolumn{1}{c}{\cellcolor[rgb]{ .388,  .745,  .482} 72.75} &
  \multicolumn{1}{c|}{\cellcolor[rgb]{ .686,  .867,  .741} 12.99} &
  \multicolumn{1}{c}{\cellcolor[rgb]{ .988,  .988,  1} 7.66} &
  \multicolumn{1}{c}{\cellcolor[rgb]{ .388,  .745,  .482} 67.57} &
  \multicolumn{1}{c|}{\cellcolor[rgb]{ .988,  .988,  1} 13.77} &
  \multicolumn{1}{c}{\cellcolor[rgb]{ .988,  .988,  1} 8.91} &
  \multicolumn{1}{c}{\cellcolor[rgb]{ .388,  .745,  .482} 62.48} &
  \multicolumn{1}{c|}{\cellcolor[rgb]{ .988,  .988,  1} 15.60} &
  \multicolumn{1}{c}{\cellcolor[rgb]{ .988,  .988,  1} 6.89} &
  \multicolumn{1}{c}{\cellcolor[rgb]{ .388,  .745,  .482} 67.66} &
  \multicolumn{1}{c|}{\cellcolor[rgb]{ .988,  .988,  1} 12.50} &
  
  \\
\multicolumn{1}{c|}{} &
  \multicolumn{1}{c|}{w/o Clustering} &
  \multicolumn{1}{c}{\cellcolor[rgb]{ .388,  .745,  .482} 10.78} &
  \multicolumn{1}{c}{\cellcolor[rgb]{ .388,  .745,  .482} 33.70} &
  \multicolumn{1}{c|}{\cellcolor[rgb]{ .388,  .745,  .482} 16.33} &
  \multicolumn{1}{c}{\cellcolor[rgb]{ .388,  .745,  .482} 11.13} &
  \multicolumn{1}{c}{\cellcolor[rgb]{ .988,  .988,  1} 22.58} &
  \multicolumn{1}{c|}{\cellcolor[rgb]{ .675,  .863,  .729} 14.91} &
  \multicolumn{1}{c}{\cellcolor[rgb]{ .937,  .969,  .957} 8.20} &
  \multicolumn{1}{c}{\cellcolor[rgb]{ .929,  .965,  .949} 17.98} &
  \multicolumn{1}{c|}{\cellcolor[rgb]{ .988,  .988,  1} 11.26} &
  \multicolumn{1}{c}{\cellcolor[rgb]{ .388,  .745,  .482} 12.52} &
  \multicolumn{1}{c}{\cellcolor[rgb]{ .988,  .988,  1} 30.21} &
  \multicolumn{1}{c|}{\cellcolor[rgb]{ .388,  .745,  .482} 17.70} &
  \multicolumn{1}{c}{\cellcolor[rgb]{ .569,  .82,  .639} 9.83} &
  \multicolumn{1}{c}{\cellcolor[rgb]{ .867,  .941,  .894} 39.75} &
  \multicolumn{1}{c|}{\cellcolor[rgb]{ .388,  .745,  .482} 15.76} &
  \multicolumn{1}{c}{\cellcolor[rgb]{ .82,  .922,  .855} 9.38} &
  \multicolumn{1}{c}{\cellcolor[rgb]{ .835,  .925,  .867} 26.41} &
  \multicolumn{1}{c|}{\cellcolor[rgb]{ .388,  .745,  .482} 13.85} &
  
  \\
\multicolumn{1}{c|}{} &
  \multicolumn{1}{c|}{Extention } &
  \multicolumn{1}{c}{\cellcolor[rgb]{ .427,  .761,  .514} 10.60} &
  \multicolumn{1}{c}{\cellcolor[rgb]{ .929,  .965,  .949} 17.33} &
  \multicolumn{1}{c|}{\cellcolor[rgb]{ .714,  .878,  .765} 13.16} &
  \multicolumn{1}{c}{\cellcolor[rgb]{ .388,  .745,  .482} 11.13} &
  \multicolumn{1}{c}{\cellcolor[rgb]{ .388,  .745,  .482} 34.02} &
  \multicolumn{1}{c|}{\cellcolor[rgb]{ .388,  .745,  .482} \textbf{16.77}} &
  \multicolumn{1}{c}{\cellcolor[rgb]{ .388,  .745,  .482} 19.03} &
  \multicolumn{1}{c}{\cellcolor[rgb]{ .988,  .988,  1} 11.96} &
  \multicolumn{1}{c|}{\cellcolor[rgb]{ .388,  .745,  .482} \textbf{14.68}} &
  \multicolumn{1}{c}{\cellcolor[rgb]{ .714,  .878,  .765} 9.89} &
  \multicolumn{1}{c}{\cellcolor[rgb]{ .969,  .98,  .984} 31.48} &
  \multicolumn{1}{c|}{\cellcolor[rgb]{ .796,  .91,  .835} 15.05} &
  \multicolumn{1}{c}{\cellcolor[rgb]{ .388,  .745,  .482} 10.22} &
  \multicolumn{1}{c}{\cellcolor[rgb]{ .988,  .988,  1} 33.86} &
  \multicolumn{1}{c|}{\cellcolor[rgb]{ .616,  .839,  .678} \textbf{15.70}} &
  \multicolumn{1}{c}{\cellcolor[rgb]{ .388,  .745,  .482} 15.74} &
  \multicolumn{1}{c}{\cellcolor[rgb]{ .988,  .988,  1} 11.86} &
  \multicolumn{1}{c|}{\cellcolor[rgb]{ .533,  .804,  .608} 13.53} &
  
  \\
\cline{2-20} &
   &
   &
   &
   &
   &
   &
   &
   &
   &
   &
   &
   &
   &
   &
   &
   &
   &
   &
   &
  
  \\
\end{tabular}
    }
  \label{tab:rq2_extension}%
\end{table}%

\textbf{For Injection, the clustering phase of the Retriever adds 0.49\% to the F1-score averaged across all settings compared to w/o clustering, and w/o clustering improves the average f1-score by 6.34\% compared to w/o Retriever.} As shown in Table~\ref{tab:rq2_injection}, both similarity and diversity affect the effectiveness of the models positively with similarity having a more significant impact. When comparing \textbf{Injection} with \textbf{Injection} w/o Retriever, we observe that the complete Retriever component improves the effectiveness of the \textbf{Injection} by 4.99\% on the average F1-score and in 8 out of 12 instances. Specifically, when we compare \textbf{Injection} with w/o the clustering phase, we see that the clustering phase enhances the effectiveness of RAG by 0.49\% on the average F1-score and in 8 out of 12 instances, and by comparing w/o clustering with w/o Retriever, we see that RAG enhances the effectiveness of RAG by 6.34\% on the average F1-score and in 8 out of 12 instances.
%\sw{it seems removing retriever in injection, the recall increased a lot in most cases, any reasons?\sd{probably because it becomes too noisy and the noise pushes the model towards overfitting to noise and be deviated towards becoming the all-1 classfier, can't really say for sure to include in the paper} only reporting number is very boring, readers usually would like to know why.\sd{well we can guess, but can't really say for sure} } \sw{why on linevul with codeQwen, without retriever the performance seems the best.\sd{Linevul is a bigger model compared to the others, so it needs more data to be trained efficiently hence if we train all these models on a big dataset like bigvul, linevul will probably outperform others. Here, if it performs better, it is just an exceptional case, can't really say anything}}
\\ \\
\textbf{For Extension strategy, the clustering phase of The Retriever provides 2.54\%  of improvement to the F1-score averaged across all settings compared to w/o clustering, and w/o clustering improves the average f1-score by 5.61\% compared to w/o Retriever}. As shown in Table~\ref{tab:rq2_extension}, both similarity and diversity affect the effectiveness of the models positively with similarity having a more significant impact. Overall, our complete \textbf{Retriever} module improves the effectiveness of the \textbf{Extension} strategy by 10.77\% on the average F1-score and beats the other settings in 6 out of 12 settings.
Specifically, by comparing \textbf{Extension} with w/o clustering we see that the clustering phase enhances the effectiveness of RAG by by 2.54\% on the average F1-score and in 7 out of 12 instances, and by comparing w/o clustering with w/o Retriever, we see that RAG enhances the effectiveness of RAG by by 5.61\% on the average F1-score and in 10 out of 12 instances.\\

\rqboxc{Retriever component makes significant contribution for \textbf{Injection} and \textbf{Extension}. \textbf{Extension} gains more than twice the increase from the \textbf{Retriever} component, implying that it is more sensitive to both the similarity and diversity of the retrieved pairs.}

\subsection{\rqthree}\label{sec:rq3}

\begin{table}[htbp!]
  \centering
  \caption{Impact of Generated Samples on Improving DLVD models' Performance at 5K, 10K, and 15K}
  \resizebox{13.5cm}{!}{
% Table generated by Excel2LaTeX from sheet 'Sheet3'
\begin{tabular}{rrrrrrrrrrrrrrrrrrrrrr}
 &
   &
   &
   &
   &
   &
   &
   &
   &
   &
   &
   &
   &
   &
   &
   &
   &
   &
   &
   &
   &
  
  \\
 &
   &
   &
  \multicolumn{9}{c|}{\textit{\textbf{ChatGPT 3.5 Turbo}}} &
  \multicolumn{9}{c}{\textit{\textbf{CodeQwen1.5-7B-Chat}}} &
  
  \\
\cline{3-21} &
  \multicolumn{1}{r|}{} &
  
  \multicolumn{1}{c|}{\multirow{2}[4]{*}{{\raisebox{0.4cm}{\textbf{Strategy}}}}} &
  \multicolumn{3}{c|}{\textbf{Devign}} &
  \multicolumn{3}{c|}{\textbf{Reveal}} &
  \multicolumn{3}{c|}{\textbf{Linevul}} &
  \multicolumn{3}{c|}{\textbf{Devign}} &
  \multicolumn{3}{c|}{\textbf{Reveal}} &
  \multicolumn{3}{c|}{\textbf{Linevul}} &
  
  \\
\cline{4-21} &
  \multicolumn{1}{r|}{} &
  \multicolumn{1}{c|}{} &
  \multicolumn{1}{c}{P} &
  \multicolumn{1}{c}{R} &
  \multicolumn{1}{c|}{F1} &
  \multicolumn{1}{c}{P} &
  \multicolumn{1}{c}{R} &
  \multicolumn{1}{c|}{F1} &
  \multicolumn{1}{c}{P} &
  \multicolumn{1}{c}{R} &
  \multicolumn{1}{c|}{F1} &
  \multicolumn{1}{c}{P} &
  \multicolumn{1}{c}{R} &
  \multicolumn{1}{c|}{F1} &
  \multicolumn{1}{c}{P} &
  \multicolumn{1}{c}{R} &
  \multicolumn{1}{c|}{F1} &
  \multicolumn{1}{c}{P} &
  \multicolumn{1}{c}{R} &
  \multicolumn{1}{c|}{F1} &
  
  \\
\cline{3-21} &
  \multicolumn{1}{c|}{\multirow{13}[4]{*}{\begin{sideways}\textbf{Reveal}\end{sideways}}} &
  \multicolumn{1}{c|}{NoAug} &
  \multicolumn{1}{c}{10.59} &
  \multicolumn{1}{c}{47.23} &
  \multicolumn{1}{c|}{17.30} &
  \multicolumn{1}{c}{12.90} &
  \multicolumn{1}{c}{65.14} &
  \multicolumn{1}{c|}{21.54} &
  \multicolumn{1}{c}{5.57} &
  \multicolumn{1}{c}{\cellcolor[rgb]{ .988,  .988,  1} 16.98} &
  \multicolumn{1}{c|}{\cellcolor[rgb]{ .988,  .988,  1} 8.39} &
  \multicolumn{1}{c}{10.59} &
  \multicolumn{1}{c}{47.23} &
  \multicolumn{1}{c|}{17.30} &
  \multicolumn{1}{c}{12.90} &
  \multicolumn{1}{c}{65.14} &
  \multicolumn{1}{c|}{21.54} &
  \multicolumn{1}{c}{5.57} &
  \multicolumn{1}{c}{16.98} &
  \multicolumn{1}{c|}{8.39} &
  
  \\
 &
  \multicolumn{1}{c|}{} &
  \multicolumn{1}{c|}{VulGen 5K} &
  \multicolumn{1}{c}{10.45} &
  \multicolumn{1}{c}{\cellcolor[rgb]{ .424,  .761,  .514} 57.90} &
  \multicolumn{1}{c|}{\cellcolor[rgb]{ .812,  .918,  .847} 17.70} &
  \multicolumn{1}{c}{12.46} &
  \multicolumn{1}{c}{64.60} &
  \multicolumn{1}{c|}{20.90} &
  \multicolumn{1}{c}{5.31} &
  \multicolumn{1}{c}{\cellcolor[rgb]{ .404,  .753,  .498} 98.09} &
  \multicolumn{1}{c|}{\cellcolor[rgb]{ .898,  .953,  .922} 10.08} &
  \multicolumn{1}{c}{10.45} &
  \multicolumn{1}{c}{\cellcolor[rgb]{ .424,  .761,  .514} 57.90} &
  \multicolumn{1}{c|}{\cellcolor[rgb]{ .78,  .906,  .824} 17.70} &
  \multicolumn{1}{c}{12.46} &
  \multicolumn{1}{c}{64.60} &
  \multicolumn{1}{c|}{20.90} &
  \multicolumn{1}{c}{5.31} &
  \multicolumn{1}{c}{\cellcolor[rgb]{ .404,  .753,  .498} 98.09} &
  \multicolumn{1}{c|}{\cellcolor[rgb]{ .898,  .953,  .922} 10.08} &
  
  \\
 &
  \multicolumn{1}{c|}{} &
  \multicolumn{1}{c|}{VulGen 10K} &
  \multicolumn{1}{c}{10.21} &
  \multicolumn{1}{c}{46.20} &
  \multicolumn{1}{c|}{16.72} &
  \multicolumn{1}{c}{11.31} &
  \multicolumn{1}{c}{66.77} &
  \multicolumn{1}{c|}{19.34} &
  \multicolumn{1}{c}{5.37} &
  \multicolumn{1}{c}{\cellcolor[rgb]{ .388,  .745,  .482} 100.00} &
  \multicolumn{1}{c|}{\cellcolor[rgb]{ .89,  .949,  .914} 10.20} &
  \multicolumn{1}{c}{10.21} &
  \multicolumn{1}{c}{46.20} &
  \multicolumn{1}{c|}{16.72} &
  \multicolumn{1}{c}{11.31} &
  \multicolumn{1}{c}{\cellcolor[rgb]{ .557,  .816,  .627} 66.77} &
  \multicolumn{1}{c|}{19.34} &
  \multicolumn{1}{c}{5.37} &
  \multicolumn{1}{c}{\cellcolor[rgb]{ .388,  .745,  .482} 100.00} &
  \multicolumn{1}{c|}{\cellcolor[rgb]{ .89,  .949,  .914} 10.20} &
  
  \\
 &
  \multicolumn{1}{c|}{} &
  \multicolumn{1}{c|}{Vulgen 15K} &
  \multicolumn{1}{c}{10.30} &
  \multicolumn{1}{c}{\cellcolor[rgb]{ .49,  .788,  .569} 55.25} &
  \multicolumn{1}{c|}{\cellcolor[rgb]{ .827,  .925,  .863} 17.37} &
  \multicolumn{1}{c}{11.23} &
  \multicolumn{1}{c}{75.51} &
  \multicolumn{1}{c|}{19.55} &
  \multicolumn{1}{c}{5.35} &
  \multicolumn{1}{c}{\cellcolor[rgb]{ .396,  .749,  .49} 99.09} &
  \multicolumn{1}{c|}{\cellcolor[rgb]{ .894,  .949,  .918} 10.15} &
  \multicolumn{1}{c}{10.30} &
  \multicolumn{1}{c}{\cellcolor[rgb]{ .49,  .788,  .569} 55.25} &
  \multicolumn{1}{c|}{\cellcolor[rgb]{ .8,  .914,  .839} 17.37} &
  \multicolumn{1}{c}{11.23} &
  \multicolumn{1}{c}{\cellcolor[rgb]{ .388,  .745,  .482} 75.51} &
  \multicolumn{1}{c|}{19.55} &
  \multicolumn{1}{c}{5.35} &
  \multicolumn{1}{c}{\cellcolor[rgb]{ .396,  .749,  .49} 99.09} &
  \multicolumn{1}{c|}{\cellcolor[rgb]{ .89,  .949,  .918} 10.15} &
  
  \\
 &
  \multicolumn{1}{c|}{} &
  \multicolumn{1}{c|}{VGX 5K} &
  \multicolumn{1}{c}{\cellcolor[rgb]{ .847,  .933,  .878} 11.02} &
  \multicolumn{1}{c}{\cellcolor[rgb]{ .388,  .745,  .482} 59.35} &
  \multicolumn{1}{c|}{\cellcolor[rgb]{ .765,  .898,  .808} 18.58} &
  \multicolumn{1}{c}{11.69} &
  \multicolumn{1}{c}{\cellcolor[rgb]{ .388,  .745,  .482} 65.62} &
  \multicolumn{1}{c|}{19.84} &
  \multicolumn{1}{c}{\cellcolor[rgb]{ .773,  .902,  .812} 9.51} &
  \multicolumn{1}{c}{\cellcolor[rgb]{ .482,  .784,  .565} 87.24} &
  \multicolumn{1}{c|}{\cellcolor[rgb]{ .82,  .922,  .855} 11.47} &
  \multicolumn{1}{c}{\cellcolor[rgb]{ .816,  .918,  .851} 11.02} &
  \multicolumn{1}{c}{\cellcolor[rgb]{ .388,  .745,  .482} 59.35} &
  \multicolumn{1}{c|}{\cellcolor[rgb]{ .725,  .882,  .776} 18.58} &
  \multicolumn{1}{c}{11.69} &
  \multicolumn{1}{c}{\cellcolor[rgb]{ .576,  .824,  .647} 65.62} &
  \multicolumn{1}{c|}{19.84} &
  \multicolumn{1}{c}{\cellcolor[rgb]{ .753,  .894,  .796} 9.51} &
  \multicolumn{1}{c}{\cellcolor[rgb]{ .482,  .784,  .565} 87.24} &
  \multicolumn{1}{c|}{\cellcolor[rgb]{ .82,  .922,  .855} 11.47} &
  
  \\
 &
  \multicolumn{1}{c|}{} &
  \multicolumn{1}{c|}{VGX 10K} &
  \multicolumn{1}{c}{10.31} &
  \multicolumn{1}{c}{\cellcolor[rgb]{ .424,  .761,  .514} 57.90} &
  \multicolumn{1}{c|}{\cellcolor[rgb]{ .82,  .922,  .855} 17.50} &
  \multicolumn{1}{c}{11.53} &
  \multicolumn{1}{c}{56.45} &
  \multicolumn{1}{c|}{19.15} &
  \multicolumn{1}{c}{5.24} &
  \multicolumn{1}{c}{\cellcolor[rgb]{ .42,  .761,  .51} 95.91} &
  \multicolumn{1}{c|}{\cellcolor[rgb]{ .906,  .957,  .929} 9.93} &
  \multicolumn{1}{c}{10.31} &
  \multicolumn{1}{c}{\cellcolor[rgb]{ .424,  .761,  .514} 57.90} &
  \multicolumn{1}{c|}{\cellcolor[rgb]{ .792,  .91,  .831} 17.50} &
  \multicolumn{1}{c}{11.53} &
  \multicolumn{1}{c}{56.45} &
  \multicolumn{1}{c|}{19.15} &
  \multicolumn{1}{c}{5.24} &
  \multicolumn{1}{c}{\cellcolor[rgb]{ .42,  .761,  .51} 95.91} &
  \multicolumn{1}{c|}{\cellcolor[rgb]{ .906,  .957,  .929} 9.93} &
  
  \\
 &
  \multicolumn{1}{c|}{} &
  \multicolumn{1}{c|}{VGX 15K} &
  \multicolumn{1}{c}{9.01} &
  \multicolumn{1}{c}{34.68} &
  \multicolumn{1}{c|}{14.30} &
  \multicolumn{1}{c}{11.95} &
  \multicolumn{1}{c}{63.75} &
  \multicolumn{1}{c|}{20.13} &
  \multicolumn{1}{c}{5.38} &
  \multicolumn{1}{c}{\cellcolor[rgb]{ .392,  .749,  .486} 99.91} &
  \multicolumn{1}{c|}{\cellcolor[rgb]{ .89,  .949,  .914} 10.21} &
  \multicolumn{1}{c}{9.01} &
  \multicolumn{1}{c}{34.68} &
  \multicolumn{1}{c|}{14.30} &
  \multicolumn{1}{c}{11.95} &
  \multicolumn{1}{c}{63.75} &
  \multicolumn{1}{c|}{20.13} &
  \multicolumn{1}{c}{5.38} &
  \multicolumn{1}{c}{\cellcolor[rgb]{ .392,  .749,  .486} 99.91} &
  \multicolumn{1}{c|}{\cellcolor[rgb]{ .89,  .949,  .914} 10.21} &
  
  \\
 &
  \multicolumn{1}{c|}{} &
  \multicolumn{1}{c|}{ROS 5K} &
  \multicolumn{1}{c}{\cellcolor[rgb]{ .686,  .867,  .737} 13.28} &
  \multicolumn{1}{c}{39.93} &
  \multicolumn{1}{c|}{\cellcolor[rgb]{ .69,  .871,  .745} 19.93} &
  \multicolumn{1}{c}{\cellcolor[rgb]{ .667,  .859,  .722} 14.35} &
  \multicolumn{1}{c}{54.52} &
  \multicolumn{1}{c|}{\cellcolor[rgb]{ .686,  .867,  .741} 22.72} &
  \multicolumn{1}{c}{\cellcolor[rgb]{ .753,  .894,  .796} 9.88} &
  \multicolumn{1}{c}{\cellcolor[rgb]{ .949,  .973,  .965} 22.71} &
  \multicolumn{1}{c|}{\cellcolor[rgb]{ .69,  .871,  .745} 13.77} &
  \multicolumn{1}{c}{\cellcolor[rgb]{ .616,  .839,  .678} 13.28} &
  \multicolumn{1}{c}{39.93} &
  \multicolumn{1}{c|}{\cellcolor[rgb]{ .643,  .851,  .702} 19.93} &
  \multicolumn{1}{c}{\cellcolor[rgb]{ .62,  .839,  .682} 14.35} &
  \multicolumn{1}{c}{54.52} &
  \multicolumn{1}{c|}{\cellcolor[rgb]{ .643,  .851,  .702} 22.72} &
  \multicolumn{1}{c}{\cellcolor[rgb]{ .733,  .886,  .78} 9.88} &
  \multicolumn{1}{c}{\cellcolor[rgb]{ .949,  .973,  .965} 22.71} &
  \multicolumn{1}{c|}{\cellcolor[rgb]{ .69,  .871,  .745} 13.77} &
  
  \\
 &
  \multicolumn{1}{c|}{} &
  \multicolumn{1}{c|}{ROS 10K} &
  \multicolumn{1}{c}{\cellcolor[rgb]{ .71,  .878,  .761} 12.92} &
  \multicolumn{1}{c}{\cellcolor[rgb]{ .573,  .82,  .643} 51.81} &
  \multicolumn{1}{c|}{\cellcolor[rgb]{ .651,  .855,  .71} 20.68} &
  \multicolumn{1}{c}{\cellcolor[rgb]{ .702,  .875,  .753} 14.00} &
  \multicolumn{1}{c}{46.20} &
  \multicolumn{1}{c|}{21.49} &
  \multicolumn{1}{c}{\cellcolor[rgb]{ .776,  .906,  .82} 9.39} &
  \multicolumn{1}{c}{\cellcolor[rgb]{ .894,  .953,  .922} 30.06} &
  \multicolumn{1}{c|}{\cellcolor[rgb]{ .663,  .859,  .718} 14.31} &
  \multicolumn{1}{c}{\cellcolor[rgb]{ .647,  .851,  .706} 12.92} &
  \multicolumn{1}{c}{\cellcolor[rgb]{ .573,  .82,  .643} 51.81} &
  \multicolumn{1}{c|}{\cellcolor[rgb]{ .6,  .831,  .663} 20.68} &
  \multicolumn{1}{c}{\cellcolor[rgb]{ .659,  .855,  .718} 14.00} &
  \multicolumn{1}{c}{46.20} &
  \multicolumn{1}{c|}{21.49} &
  \multicolumn{1}{c}{\cellcolor[rgb]{ .761,  .898,  .804} 9.39} &
  \multicolumn{1}{c}{\cellcolor[rgb]{ .894,  .953,  .922} 30.06} &
  \multicolumn{1}{c|}{\cellcolor[rgb]{ .663,  .855,  .718} 14.31} &
  
  \\
 &
  \multicolumn{1}{c|}{} &
  \multicolumn{1}{c|}{ROS 15K} &
  \multicolumn{1}{c}{\cellcolor[rgb]{ .686,  .867,  .741} 13.25} &
  \multicolumn{1}{c}{\cellcolor[rgb]{ .682,  .867,  .737} 47.29} &
  \multicolumn{1}{c|}{\cellcolor[rgb]{ .651,  .851,  .71} 20.71} &
  \multicolumn{1}{c}{\cellcolor[rgb]{ .725,  .882,  .773} 13.77} &
  \multicolumn{1}{c}{43.85} &
  \multicolumn{1}{c|}{20.96} &
  \multicolumn{1}{c}{\cellcolor[rgb]{ .839,  .929,  .871} 8.18} &
  \multicolumn{1}{c}{\cellcolor[rgb]{ .969,  .98,  .984} 19.71} &
  \multicolumn{1}{c|}{\cellcolor[rgb]{ .816,  .918,  .851} 11.56} &
  \multicolumn{1}{c}{\cellcolor[rgb]{ .62,  .839,  .682} 13.25} &
  \multicolumn{1}{c}{\cellcolor[rgb]{ .682,  .867,  .737} 47.29} &
  \multicolumn{1}{c|}{\cellcolor[rgb]{ .596,  .831,  .663} 20.71} &
  \multicolumn{1}{c}{\cellcolor[rgb]{ .686,  .867,  .741} 13.77} &
  \multicolumn{1}{c}{43.85} &
  \multicolumn{1}{c|}{20.96} &
  \multicolumn{1}{c}{\cellcolor[rgb]{ .827,  .925,  .859} 8.18} &
  \multicolumn{1}{c}{\cellcolor[rgb]{ .969,  .98,  .984} 19.71} &
  \multicolumn{1}{c|}{\cellcolor[rgb]{ .812,  .918,  .851} 11.56} &
  
  \\
 &
  \multicolumn{1}{c|}{} &
  \multicolumn{1}{c|}{Injection 5K} &
  \multicolumn{1}{c}{\cellcolor[rgb]{ .675,  .863,  .729} 13.44} &
  \multicolumn{1}{c}{\cellcolor[rgb]{ .537,  .808,  .612} 53.26} &
  \multicolumn{1}{c|}{\cellcolor[rgb]{ .612,  .835,  .675} 21.46} &
  \multicolumn{1}{c}{\cellcolor[rgb]{ .553,  .812,  .624} 15.43} &
  \multicolumn{1}{c}{61.46} &
  \multicolumn{1}{c|}{\cellcolor[rgb]{ .522,  .8,  .6} 24.66} &
  \multicolumn{1}{c}{\cellcolor[rgb]{ .667,  .859,  .722} 11.57} &
  \multicolumn{1}{c}{\cellcolor[rgb]{ .922,  .961,  .941} 26.43} &
  \multicolumn{1}{c|}{\cellcolor[rgb]{ .565,  .816,  .631} 16.10} &
  \multicolumn{1}{c}{\cellcolor[rgb]{ .584,  .824,  .651} 13.67} &
  \multicolumn{1}{c}{\cellcolor[rgb]{ .514,  .796,  .592} 54.28} &
  \multicolumn{1}{c|}{\cellcolor[rgb]{ .525,  .804,  .604} 21.83} &
  \multicolumn{1}{c}{\cellcolor[rgb]{ .631,  .847,  .694} 14.23} &
  \multicolumn{1}{c}{64.48} &
  \multicolumn{1}{c|}{\cellcolor[rgb]{ .584,  .827,  .655} 23.32} &
  \multicolumn{1}{c}{\cellcolor[rgb]{ .725,  .882,  .773} 9.99} &
  \multicolumn{1}{c}{\cellcolor[rgb]{ .925,  .965,  .945} 25.79} &
  \multicolumn{1}{c|}{\cellcolor[rgb]{ .655,  .855,  .714} 14.40} &
  
  \\
 &
  \multicolumn{1}{c|}{} &
  \multicolumn{1}{c|}{Injection 10K} &
  \multicolumn{1}{c}{\cellcolor[rgb]{ .388,  .745,  .482} 17.39} &
  \multicolumn{1}{c}{42.70} &
  \multicolumn{1}{c|}{\cellcolor[rgb]{ .439,  .769,  .525} 24.71} &
  \multicolumn{1}{c}{\cellcolor[rgb]{ .455,  .773,  .541} 16.37} &
  \multicolumn{1}{c}{63.21} &
  \multicolumn{1}{c|}{\cellcolor[rgb]{ .408,  .753,  .502} 26.00} &
  \multicolumn{1}{c}{\cellcolor[rgb]{ .686,  .867,  .741} 11.19} &
  \multicolumn{1}{c}{\cellcolor[rgb]{ .882,  .945,  .91} 31.97} &
  \multicolumn{1}{c|}{\cellcolor[rgb]{ .537,  .808,  .612} 16.57} &
  \multicolumn{1}{c}{\cellcolor[rgb]{ .388,  .745,  .482} 15.87} &
  \multicolumn{1}{c}{\cellcolor[rgb]{ .659,  .855,  .718} 48.31} &
  \multicolumn{1}{c|}{\cellcolor[rgb]{ .4,  .753,  .494} 23.90} &
  \multicolumn{1}{c}{\cellcolor[rgb]{ .545,  .812,  .62} 14.96} &
  \multicolumn{1}{c}{63.03} &
  \multicolumn{1}{c|}{\cellcolor[rgb]{ .502,  .792,  .58} 24.18} &
  \multicolumn{1}{c}{\cellcolor[rgb]{ .443,  .769,  .529} 15.05} &
  \multicolumn{1}{c}{\cellcolor[rgb]{ .973,  .984,  .984} 19.53} &
  \multicolumn{1}{c|}{\cellcolor[rgb]{ .51,  .796,  .588} 17.00} &
  
  \\
 &
  \multicolumn{1}{c|}{} &
  \multicolumn{1}{c|}{Injection 15K} &
  \multicolumn{1}{c}{\cellcolor[rgb]{ .424,  .761,  .514} 16.94} &
  \multicolumn{1}{c}{\cellcolor[rgb]{ .553,  .812,  .624} 52.65} &
  \multicolumn{1}{c|}{\cellcolor[rgb]{ .388,  .745,  .482} 25.63} &
  \multicolumn{1}{c}{\cellcolor[rgb]{ .388,  .745,  .482} 17.00} &
  \multicolumn{1}{c}{57.36} &
  \multicolumn{1}{c|}{\cellcolor[rgb]{ .388,  .745,  .482} 26.22} &
  \multicolumn{1}{c}{\cellcolor[rgb]{ .592,  .827,  .659} 13.02} &
  \multicolumn{1}{c}{\cellcolor[rgb]{ .847,  .933,  .878} 36.61} &
  \multicolumn{1}{c|}{\cellcolor[rgb]{ .388,  .745,  .482} 19.21} &
  \multicolumn{1}{c}{\cellcolor[rgb]{ .416,  .757,  .506} 15.60} &
  \multicolumn{1}{c}{\cellcolor[rgb]{ .557,  .816,  .627} 52.53} &
  \multicolumn{1}{c|}{\cellcolor[rgb]{ .388,  .745,  .482} 24.06} &
  \multicolumn{1}{c}{\cellcolor[rgb]{ .388,  .745,  .482} 16.27} &
  \multicolumn{1}{c}{57.18} &
  \multicolumn{1}{c|}{\cellcolor[rgb]{ .388,  .745,  .482} 25.33} &
  \multicolumn{1}{c}{\cellcolor[rgb]{ .388,  .745,  .482} 16.01} &
  \multicolumn{1}{c}{\cellcolor[rgb]{ .941,  .969,  .961} 23.89} &
  \multicolumn{1}{c|}{\cellcolor[rgb]{ .388,  .745,  .482} 19.17} &
  
%   \\
% \cline{3-21} &
%   \multicolumn{1}{c|}{} &
%   \multicolumn{1}{c|}{Injection 20K} &
%   \multicolumn{1}{c}{\cellcolor[rgb]{ .435,  .765,  .525} 16.74} &
%   \multicolumn{1}{c}{\cellcolor[rgb]{ .635,  .847,  .694} 49.28} &
%   \multicolumn{1}{c|}{\cellcolor[rgb]{ .424,  .761,  .514} 24.99} &
%   \multicolumn{1}{c}{\cellcolor[rgb]{ .4,  .749,  .49} 16.91} &
%   \multicolumn{1}{c}{57.90} &
%   \multicolumn{1}{c|}{\cellcolor[rgb]{ .392,  .749,  .486} 26.18} &
%   \multicolumn{1}{c}{\cellcolor[rgb]{ .388,  .745,  .482} 16.97} &
%   \multicolumn{1}{c}{\cellcolor[rgb]{ .984,  .988,  .996} 17.80} &
%   \multicolumn{1}{c|}{\cellcolor[rgb]{ .49,  .788,  .573} 17.38} &
%   \multicolumn{1}{c}{\cellcolor[rgb]{ .455,  .773,  .541} 15.11} &
%   \multicolumn{1}{c}{\cellcolor[rgb]{ .471,  .78,  .557} 55.97} &
%   \multicolumn{1}{c|}{\cellcolor[rgb]{ .408,  .753,  .498} 23.79} &
%   \multicolumn{1}{c}{\cellcolor[rgb]{ .506,  .792,  .584} 15.31} &
%   \multicolumn{1}{c}{58.69} &
%   \multicolumn{1}{c|}{\cellcolor[rgb]{ .494,  .788,  .573} 24.28} &
%   \multicolumn{1}{c}{\cellcolor[rgb]{ .506,  .796,  .584} 13.92} &
%   \multicolumn{1}{c}{\cellcolor[rgb]{ .988,  .988,  1} 17.26} &
%   \multicolumn{1}{c|}{\cellcolor[rgb]{ .6,  .831,  .667} 15.41} &
  
  \\
\cline{2-21} &
  \multicolumn{1}{c|}{\multirow{13}[4]{*}{\begin{sideways}\textbf{Bigvul}\end{sideways}}} &
  \multicolumn{1}{c|}{NoAug} &
  \multicolumn{1}{c}{6.27} &
  \multicolumn{1}{c}{43.24} &
  \multicolumn{1}{c|}{10.95} &
  \multicolumn{1}{c}{7.28} &
  \multicolumn{1}{c}{77.27} &
  \multicolumn{1}{c|}{13.31} &
  \multicolumn{1}{c}{7.90} &
  \multicolumn{1}{c}{29.19} &
  \multicolumn{1}{c|}{12.43} &
  \multicolumn{1}{c}{6.27} &
  \multicolumn{1}{c}{43.24} &
  \multicolumn{1}{c|}{10.95} &
  \multicolumn{1}{c}{7.28} &
  \multicolumn{1}{c}{77.27} &
  \multicolumn{1}{c|}{13.31} &
  \multicolumn{1}{c}{7.90} &
  \multicolumn{1}{c}{29.19} &
  \multicolumn{1}{c|}{12.43} &
  
  \\
 &
  \multicolumn{1}{c|}{} &
  \multicolumn{1}{c|}{VulGen 5K} &
  \multicolumn{1}{c}{\cellcolor[rgb]{ .941,  .973,  .961} 6.49} &
  \multicolumn{1}{c}{\cellcolor[rgb]{ .4,  .749,  .49} 62.32} &
  \multicolumn{1}{c|}{\cellcolor[rgb]{ .906,  .957,  .929} 11.75} &
  \multicolumn{1}{c}{6.36} &
  \multicolumn{1}{c}{70.59} &
  \multicolumn{1}{c|}{11.66} &
  \multicolumn{1}{c}{\cellcolor[rgb]{ .706,  .875,  .757} 8.51} &
  \multicolumn{1}{c}{22.150 } &
  \multicolumn{1}{c|}{12.29} &
  \multicolumn{1}{c}{\cellcolor[rgb]{ .929,  .965,  .949} 6.49} &
  \multicolumn{1}{c}{\cellcolor[rgb]{ .4,  .749,  .494} 62.32} &
  \multicolumn{1}{c|}{\cellcolor[rgb]{ .898,  .953,  .922} 11.75} &
  \multicolumn{1}{c}{6.36} &
  \multicolumn{1}{c}{70.59} &
  \multicolumn{1}{c|}{11.66} &
  \multicolumn{1}{c}{\cellcolor[rgb]{ .706,  .875,  .757} 8.51} &
  \multicolumn{1}{c}{22.150 } &
  \multicolumn{1}{c|}{12.29} &
  
  \\
 &
  \multicolumn{1}{c|}{} &
  \multicolumn{1}{c|}{VulGen 10K} &
  \multicolumn{1}{c}{\cellcolor[rgb]{ .933,  .969,  .953} 6.59} &
  \multicolumn{1}{c}{\cellcolor[rgb]{ .525,  .804,  .604} 52.62} &
  \multicolumn{1}{c|}{\cellcolor[rgb]{ .91,  .957,  .933} 11.72} &
  \multicolumn{1}{c}{6.21} &
  \multicolumn{1}{c}{76.79} &
  \multicolumn{1}{c|}{11.50} &
  \multicolumn{1}{c}{\cellcolor[rgb]{ .388,  .745,  .482} 12.04} &
  \multicolumn{1}{c}{12.70} &
  \multicolumn{1}{c|}{12.36} &
  \multicolumn{1}{c}{\cellcolor[rgb]{ .918,  .961,  .941} 6.59} &
  \multicolumn{1}{c}{\cellcolor[rgb]{ .537,  .808,  .612} 52.62} &
  \multicolumn{1}{c|}{\cellcolor[rgb]{ .902,  .953,  .925} 11.72} &
  \multicolumn{1}{c}{6.21} &
  \multicolumn{1}{c}{76.79} &
  \multicolumn{1}{c|}{11.50} &
  \multicolumn{1}{c}{\cellcolor[rgb]{ .388,  .745,  .482} 12.04} &
  \multicolumn{1}{c}{12.70} &
  \multicolumn{1}{c|}{12.36} &
  
  \\
 &
  \multicolumn{1}{c|}{} &
  \multicolumn{1}{c|}{Vulgen 15K} &
  \multicolumn{1}{c}{\cellcolor[rgb]{ .953,  .976,  .969} 6.35} &
  \multicolumn{1}{c}{\cellcolor[rgb]{ .388,  .745,  .482} 62.96} &
  \multicolumn{1}{c|}{\cellcolor[rgb]{ .925,  .965,  .945} 11.54} &
  \multicolumn{1}{c}{6.48} &
  \multicolumn{1}{c}{69.95} &
  \multicolumn{1}{c|}{11.86} &
  \multicolumn{1}{c}{\cellcolor[rgb]{ .722,  .882,  .773} 8.32} &
  \multicolumn{1}{c}{14.74} &
  \multicolumn{1}{c|}{10.64} &
  \multicolumn{1}{c}{\cellcolor[rgb]{ .941,  .969,  .961} 6.35} &
  \multicolumn{1}{c}{\cellcolor[rgb]{ .388,  .745,  .482} 62.96} &
  \multicolumn{1}{c|}{\cellcolor[rgb]{ .918,  .961,  .937} 11.54} &
  \multicolumn{1}{c}{6.48} &
  \multicolumn{1}{c}{69.95} &
  \multicolumn{1}{c|}{11.86} &
  \multicolumn{1}{c}{\cellcolor[rgb]{ .722,  .882,  .773} 8.32} &
  \multicolumn{1}{c}{14.74} &
  \multicolumn{1}{c|}{10.64} &
  
  \\
 &
  \multicolumn{1}{c|}{} &
  \multicolumn{1}{c|}{VGX 5K} &
  \multicolumn{1}{c}{6.06} &
  \multicolumn{1}{c}{\cellcolor[rgb]{ .439,  .765,  .525} 59.30} &
  \multicolumn{1}{c|}{\cellcolor[rgb]{ .965,  .98,  .98} 10.99} &
  \multicolumn{1}{c}{6.89} &
  \multicolumn{1}{c}{73.29} &
  \multicolumn{1}{c|}{12.60} &
  \multicolumn{1}{c}{5.35} &
  \multicolumn{1}{c}{\cellcolor[rgb]{ .388,  .745,  .482} 98.82} &
  \multicolumn{1}{c|}{10.15} &
  \multicolumn{1}{c}{6.06} &
  \multicolumn{1}{c}{\cellcolor[rgb]{ .443,  .769,  .529} 59.30} &
  \multicolumn{1}{c|}{\cellcolor[rgb]{ .961,  .98,  .976} 10.99} &
  \multicolumn{1}{c}{6.89} &
  \multicolumn{1}{c}{73.29} &
  \multicolumn{1}{c|}{12.60} &
  \multicolumn{1}{c}{5.35} &
  \multicolumn{1}{c}{\cellcolor[rgb]{ .388,  .745,  .482} 98.82} &
  \multicolumn{1}{c|}{10.15} &
  
  \\
 &
  \multicolumn{1}{c|}{} &
  \multicolumn{1}{c|}{VGX 10K} &
  \multicolumn{1}{c}{5.84} &
  \multicolumn{1}{c}{\cellcolor[rgb]{ .412,  .757,  .506} 61.21} &
  \multicolumn{1}{c|}{10.65} &
  \multicolumn{1}{c}{6.36} &
  \multicolumn{1}{c}{63.28} &
  \multicolumn{1}{c|}{11.56} &
  \multicolumn{1}{c}{7.14} &
  \multicolumn{1}{c}{10.10} &
  \multicolumn{1}{c|}{8.37} &
  \multicolumn{1}{c}{5.84} &
  \multicolumn{1}{c}{\cellcolor[rgb]{ .416,  .757,  .506} 61.21} &
  \multicolumn{1}{c|}{10.65} &
  \multicolumn{1}{c}{6.36} &
  \multicolumn{1}{c}{63.28} &
  \multicolumn{1}{c|}{11.56} &
  \multicolumn{1}{c}{7.14} &
  \multicolumn{1}{c}{10.10} &
  \multicolumn{1}{c|}{8.37} &
  
  \\
 &
  \multicolumn{1}{c|}{} &
  \multicolumn{1}{c|}{VGX 15K} &
  \multicolumn{1}{c}{6.11} &
  \multicolumn{1}{c}{\cellcolor[rgb]{ .557,  .816,  .627} 50.40} &
  \multicolumn{1}{c|}{10.90} &
  \multicolumn{1}{c}{6.42} &
  \multicolumn{1}{c}{66.14} &
  \multicolumn{1}{c|}{11.70} &
  \multicolumn{1}{c}{\cellcolor[rgb]{ .631,  .843,  .69} 9.36} &
  \multicolumn{1}{c}{6.49} &
  \multicolumn{1}{c|}{7.66} &
  \multicolumn{1}{c}{6.11} &
  \multicolumn{1}{c}{\cellcolor[rgb]{ .569,  .82,  .639} 50.40} &
  \multicolumn{1}{c|}{10.90} &
  \multicolumn{1}{c}{6.42} &
  \multicolumn{1}{c}{66.14} &
  \multicolumn{1}{c|}{11.70} &
  \multicolumn{1}{c}{\cellcolor[rgb]{ .631,  .843,  .69} 9.36} &
  \multicolumn{1}{c}{6.49} &
  \multicolumn{1}{c|}{7.66} &
  
  \\
 &
  \multicolumn{1}{c|}{} &
  \multicolumn{1}{c|}{ROS 5K} &
  \multicolumn{1}{c}{\cellcolor[rgb]{ .863,  .937,  .894} 7.54} &
  \multicolumn{1}{c}{25.60} &
  \multicolumn{1}{c|}{\cellcolor[rgb]{ .914,  .961,  .937} 11.65} &
  \multicolumn{1}{c}{\cellcolor[rgb]{ .824,  .922,  .859} 7.96} &
  \multicolumn{1}{c}{33.39} &
  \multicolumn{1}{c|}{12.86} &
  \multicolumn{1}{c}{\cellcolor[rgb]{ .49,  .788,  .573} 10.91} &
  \multicolumn{1}{c}{13.16} &
  \multicolumn{1}{c|}{11.93} &
  \multicolumn{1}{c}{\cellcolor[rgb]{ .827,  .925,  .863} 7.54} &
  \multicolumn{1}{c}{25.60} &
  \multicolumn{1}{c|}{\cellcolor[rgb]{ .906,  .957,  .929} 11.65} &
  \multicolumn{1}{c}{\cellcolor[rgb]{ .796,  .91,  .835} 7.96} &
  \multicolumn{1}{c}{33.39} &
  \multicolumn{1}{c|}{12.86} &
  \multicolumn{1}{c}{\cellcolor[rgb]{ .49,  .788,  .573} 10.91} &
  \multicolumn{1}{c}{13.16} &
  \multicolumn{1}{c|}{11.93} &
  
  \\
 &
  \multicolumn{1}{c|}{} &
  \multicolumn{1}{c|}{ROS 10K} &
  \multicolumn{1}{c}{\cellcolor[rgb]{ .855,  .937,  .886} 7.64} &
  \multicolumn{1}{c}{32.43} &
  \multicolumn{1}{c|}{\cellcolor[rgb]{ .859,  .937,  .89} 12.37} &
  \multicolumn{1}{c}{\cellcolor[rgb]{ .694,  .871,  .749} 9.32} &
  \multicolumn{1}{c}{33.55} &
  \multicolumn{1}{c|}{\cellcolor[rgb]{ .741,  .89,  .788} 14.59} &
  \multicolumn{1}{c}{\cellcolor[rgb]{ .741,  .89,  .788} 8.13} &
  \multicolumn{1}{c}{18.44} &
  \multicolumn{1}{c|}{11.28} &
  \multicolumn{1}{c}{\cellcolor[rgb]{ .816,  .922,  .855} 7.64} &
  \multicolumn{1}{c}{32.43} &
  \multicolumn{1}{c|}{\cellcolor[rgb]{ .847,  .933,  .878} 12.37} &
  \multicolumn{1}{c}{\cellcolor[rgb]{ .643,  .851,  .702} 9.32} &
  \multicolumn{1}{c}{33.55} &
  \multicolumn{1}{c|}{\cellcolor[rgb]{ .71,  .875,  .761} 14.59} &
  \multicolumn{1}{c}{\cellcolor[rgb]{ .741,  .89,  .788} 8.13} &
  \multicolumn{1}{c}{18.44} &
  \multicolumn{1}{c|}{11.28} &
  
  \\
 &
  \multicolumn{1}{c|}{} &
  \multicolumn{1}{c|}{ROS 15K} &
  \multicolumn{1}{c}{\cellcolor[rgb]{ .733,  .886,  .78} 9.31} &
  \multicolumn{1}{c}{21.14} &
  \multicolumn{1}{c|}{\cellcolor[rgb]{ .82,  .922,  .855} 12.93} &
  \multicolumn{1}{c}{\cellcolor[rgb]{ .8,  .914,  .839} 8.22} &
  \multicolumn{1}{c}{26.87} &
  \multicolumn{1}{c|}{12.59} &
  \multicolumn{1}{c}{\cellcolor[rgb]{ .682,  .867,  .737} 8.77} &
  \multicolumn{1}{c}{13.07} &
  \multicolumn{1}{c|}{10.49} &
  \multicolumn{1}{c}{\cellcolor[rgb]{ .659,  .855,  .714} 9.31} &
  \multicolumn{1}{c}{21.14} &
  \multicolumn{1}{c|}{\cellcolor[rgb]{ .8,  .914,  .839} 12.93} &
  \multicolumn{1}{c}{\cellcolor[rgb]{ .765,  .898,  .808} 8.22} &
  \multicolumn{1}{c}{26.87} &
  \multicolumn{1}{c|}{12.59} &
  \multicolumn{1}{c}{\cellcolor[rgb]{ .682,  .867,  .737} 8.77} &
  \multicolumn{1}{c}{13.07} &
  \multicolumn{1}{c|}{10.49} &
  
  \\
 &
  \multicolumn{1}{c|}{} &
  \multicolumn{1}{c|}{Injection 5K} &
  \multicolumn{1}{c}{\cellcolor[rgb]{ .773,  .902,  .812} 8.79} &
  \multicolumn{1}{c}{29.41} &
  \multicolumn{1}{c|}{\cellcolor[rgb]{ .773,  .902,  .816} 13.53} &
  \multicolumn{1}{c}{\cellcolor[rgb]{ .475,  .78,  .557} 11.64} &
  \multicolumn{1}{c}{38.00} &
  \multicolumn{1}{c|}{\cellcolor[rgb]{ .482,  .784,  .565} 17.82} &
  \multicolumn{1}{c}{\cellcolor[rgb]{ .443,  .769,  .529} 11.43} &
  \multicolumn{1}{c}{18.81} &
  \multicolumn{1}{c|}{\cellcolor[rgb]{ .565,  .82,  .635} 14.22} &
  \multicolumn{1}{c}{\cellcolor[rgb]{ .561,  .816,  .631} 10.33} &
  \multicolumn{1}{c}{31.80} &
  \multicolumn{1}{c|}{\cellcolor[rgb]{ .58,  .824,  .647} 15.59} &
  \multicolumn{1}{c}{\cellcolor[rgb]{ .576,  .824,  .647} 9.91} &
  \multicolumn{1}{c}{38.16} &
  \multicolumn{1}{c|}{\cellcolor[rgb]{ .604,  .835,  .671} 15.74} &
  \multicolumn{1}{c}{7.85} &
  \multicolumn{1}{c}{\cellcolor[rgb]{ .835,  .925,  .867} 30.49} &
  \multicolumn{1}{c|}{\cellcolor[rgb]{ .573,  .82,  .643} 12.49} &
  
  \\
 &
  \multicolumn{1}{c|}{} &
  \multicolumn{1}{c|}{Injection 10K} &
  \multicolumn{1}{c}{\cellcolor[rgb]{ .388,  .745,  .482} 13.94} &
  \multicolumn{1}{c}{17.60} &
  \multicolumn{1}{c|}{\cellcolor[rgb]{ .467,  .776,  .549} 17.60} &
  \multicolumn{1}{c}{\cellcolor[rgb]{ .431,  .765,  .522} 12.09} &
  \multicolumn{1}{c}{43.56} &
  \multicolumn{1}{c|}{\cellcolor[rgb]{ .396,  .749,  .49} 18.92} &
  \multicolumn{1}{c}{\cellcolor[rgb]{ .678,  .863,  .733} 8.84} &
  \multicolumn{1}{c}{\cellcolor[rgb]{ .816,  .922,  .851} 33.09} &
  \multicolumn{1}{c|}{\cellcolor[rgb]{ .584,  .824,  .651} 13.96} &
  \multicolumn{1}{c}{\cellcolor[rgb]{ .451,  .773,  .537} 11.49} &
  \multicolumn{1}{c}{30.37} &
  \multicolumn{1}{c|}{\cellcolor[rgb]{ .49,  .788,  .569} 16.67} &
  \multicolumn{1}{c}{\cellcolor[rgb]{ .396,  .749,  .49} 11.54} &
  \multicolumn{1}{c}{41.81} &
  \multicolumn{1}{c|}{\cellcolor[rgb]{ .392,  .749,  .486} 18.08} &
  \multicolumn{1}{c}{\cellcolor[rgb]{ .584,  .827,  .655} 9.86} &
  \multicolumn{1}{c}{19.83} &
  \multicolumn{1}{c|}{\cellcolor[rgb]{ .514,  .796,  .592} 13.17} &
  
  \\
 &
  \multicolumn{1}{c|}{} &
  \multicolumn{1}{c|}{Injection 15K} &
  \multicolumn{1}{c}{\cellcolor[rgb]{ .482,  .784,  .565} 12.70} &
  \multicolumn{1}{c}{34.66} &
  \multicolumn{1}{c|}{\cellcolor[rgb]{ .388,  .745,  .482} 18.59} &
  \multicolumn{1}{c}{\cellcolor[rgb]{ .447,  .773,  .533} 11.92} &
  \multicolumn{1}{c}{41.81} &
  \multicolumn{1}{c|}{\cellcolor[rgb]{ .424,  .761,  .514} 18.55} &
  \multicolumn{1}{c}{\cellcolor[rgb]{ .412,  .757,  .506} 11.78} &
  \multicolumn{1}{c}{\cellcolor[rgb]{ .835,  .929,  .871} 30.15} &
  \multicolumn{1}{c|}{\cellcolor[rgb]{ .388,  .745,  .482} 16.94} &
  \multicolumn{1}{c}{\cellcolor[rgb]{ .388,  .745,  .482} 12.12} &
  \multicolumn{1}{c}{33.86} &
  \multicolumn{1}{c|}{\cellcolor[rgb]{ .388,  .745,  .482} 17.85} &
  \multicolumn{1}{c}{\cellcolor[rgb]{ .396,  .749,  .49} 11.54} &
  \multicolumn{1}{c}{41.81} &
  \multicolumn{1}{c|}{\cellcolor[rgb]{ .388,  .745,  .482} 18.09} &
  \multicolumn{1}{c}{\cellcolor[rgb]{ .596,  .831,  .663} 9.74} &
  \multicolumn{1}{c}{29.19} &
  \multicolumn{1}{c|}{\cellcolor[rgb]{ .388,  .745,  .482} 14.60} &
  
  % \\
% \cline{3-21} &
%   \multicolumn{1}{c|}{} &
%   \multicolumn{1}{c|}{Injection 20K} &
%   \multicolumn{1}{c}{\cellcolor[rgb]{ .431,  .765,  .518} 13.41} &
%   \multicolumn{1}{c}{29.89} &
%   \multicolumn{1}{c|}{\cellcolor[rgb]{ .396,  .749,  .49} 18.51} &
%   \multicolumn{1}{c}{\cellcolor[rgb]{ .388,  .745,  .482} 12.54} &
%   \multicolumn{1}{c}{38.95} &
%   \multicolumn{1}{c|}{\cellcolor[rgb]{ .388,  .745,  .482} 18.98} &
%   \multicolumn{1}{c}{\cellcolor[rgb]{ .4,  .753,  .494} 11.91} &
%   \multicolumn{1}{c}{25.49} &
%   \multicolumn{1}{c|}{\cellcolor[rgb]{ .435,  .765,  .525} 16.23} &
%   \multicolumn{1}{c}{\cellcolor[rgb]{ .635,  .847,  .694} 9.55} &
%   \multicolumn{1}{c}{34.02} &
%   \multicolumn{1}{c|}{\cellcolor[rgb]{ .635,  .847,  .694} 14.92} &
%   \multicolumn{1}{c}{\cellcolor[rgb]{ .388,  .745,  .482} 11.60} &
%   \multicolumn{1}{c}{39.59} &
%   \multicolumn{1}{c|}{\cellcolor[rgb]{ .404,  .753,  .494} 17.94} &
%   \multicolumn{1}{c}{7.87} &
%   \multicolumn{1}{c}{\cellcolor[rgb]{ .784,  .906,  .824} 38.00} &
%   \multicolumn{1}{c|}{\cellcolor[rgb]{ .525,  .8,  .6} 13.05} &
  
  \\
\cline{3-21} &
   &
   &
   &
   &
   &
   &
   &
   &
   &
   &
   &
   &
   &
   &
   &
   &
   &
   &
   &
   &
  
  \\
\end{tabular}%

    }
  \label{tab:rq3}%
\end{table}%

\begin{figure*}[htbp]
\includegraphics[width=1\textwidth]{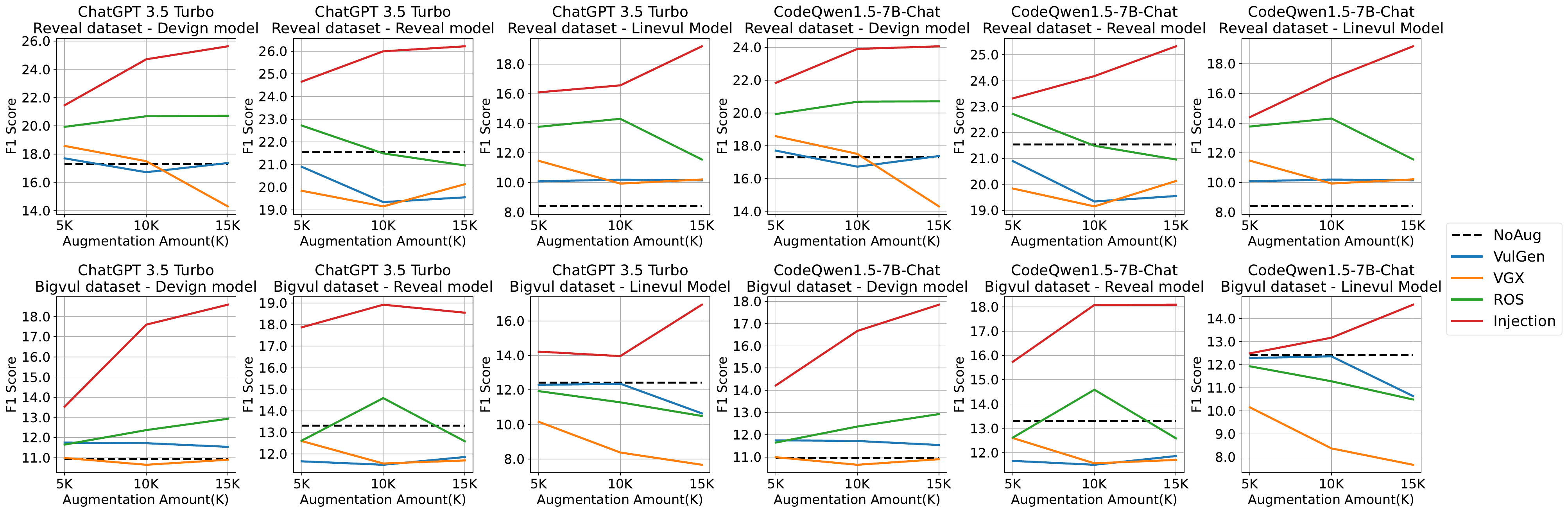}
	\caption{The impact of the number of augmented samples on the effectiveness of DLVD models across the studied datasets and LLMs across the studied DLVD models (i.e., Devign, Reveal, and Linevul). }
	\label{fig:chatgpt_plots}
 % \vspace{-0.2in}
\end{figure*}

\textbf{Augmenting more vulnerable data by using Injection helps improve the effectiveness of DLVD models.} Table~\ref{tab:rq3} and Figure~\ref{fig:chatgpt_plots} present the performance (in terms of F-1 score) of DLVD models when being trained on a dataset that is augmented with different vulnerable samples generated by \textbf{Injection} and other baselines (e.g., VulGen, VGX, ROS, and NoAug). As we can see, The performance of DLVD models increases as more augmented vulnerable samples are added to the training data in almost all experimental instances (11 out of 12), which indicates that adding more data augmented by \textbf{Injection} provides more useful information for the model to capture vulnerabilities.  

%\textbf{Our Injection Strategy achieves a considerably higher F1-score in different settings at 10K, and keeps on improving at 15K, beating all the baselines.} As it can be observed, \textbf{Injection} contributes to the performance of DLVD at 10K and 15K. This brings two notes to light: First, this augmentation method produces samples that are not noisy as the DLVD models do not degrade after 10K, losing performance due to overfitting to noise, and Secondly, it has enough diversity that the model can still learn from the extra samples that are added after 5K.

\textbf{In general, VulGen, VGX, and ROS fail to improve the performance of DLVD models by augmenting more vulnerable samples.} When looking at the baselines in Figures~\ref{fig:chatgpt_plots}, we notice that Vulgen, VGX, and ROS struggle with improving the performance of DLVD models by augmenting more data. For instance, if we compare the performance of DLVD models when augmenting 5K and 15K vulnerable samples, VGX, VulGen, and ROS decrease the performance in 5 out of 6, 4 out of 6, and 4 out of 6 instances, respectively (Note that the baselines are not dependent on the LLMs, but they are shown for easier comparison). This finding aligns with the results reported in Yang et al.~\cite{yang2023does}, indicating that excessive random over-sampling does not improve the effectiveness of DLVD models and can diminish their performance. One possible explanation is that since random over-sampling does not introduce any new information, and excessive over-sampling probably leads to overfitting of the oversampled data in the models.
For VGX and VulGen, this is probably due to the nature of their method, which solely focuses on single-statement vulnerabilities. This heavily limits the usefulness of such methods and makes them unfeasible to be used for large-scale data augmentation. 

In summary, \textbf{Injection} outperforms all baselines (i.e., VulGen, VGX, and ROS) in all scales, typically when used for large-scale data augmentation. The studied baselines fail to improve the effectiveness of DLVD models by augmenting more than 5K samples. Our LLM-based approach is more feasible for large-scale vulnerable data augmentation. 
%On the other hand, ROS beats VGX and VulGen and it keeps improving and beats them at all scales, making it a feasible cheap solution for dealing with a shortage of vulnerable items, but one should note that ROS's performance increase will be limited greatly as the number of sampled items increases.

\rqboxc{Augmenting more vulnerable data by using \textbf{Injection} helps improve the effectiveness of DLVD models, while VulGen, VGX, and ROS fail to improve the performance of DLVD models by augmenting more than 5K vulnerable samples. In contrast, our LLM-based approach is more feasible for large-scall vulnerable data augmentation. Explicitly, \textbf{Injection} at 15K beats NoAug, Vulgen at 15K, VGX at 15K, and ROS at 15K by 53.84\%, 54.10\%,  69.90\%, and 40.93\%.}

\subsection{\rb{\rqfour}}\label{sec:rq4}

\begin{table}[htbp!]
  \centering
  \caption{Comparison of VulScribeR’s strategies on the PrimeVul dataset with baselines using GPT4o-mini when augmenting 5K Samples across the studied DLVD models (i.e., Devign, Reveal, and Linevul). The cells with larger values (better performance) compared to NoAug are highlighted darker.}
   \label{tab:rq4}
  % \resizebox{13.5cm}{!}{ # too big
  \resizebox{10cm}{!}{
\begin{tabular}{rrrrrrrrrrrr}
  
  \\
\cline{2-11}\multicolumn{1}{r|}{} &
  \multicolumn{1}{c|}{\multirow{2}[4]{*}{{\raisebox{0.4cm}{\textbf{Strategy}}}}} &
  \multicolumn{3}{c|}{\textbf{Devign}} &
  \multicolumn{3}{c|}{\textbf{Reveal}} &
  \multicolumn{3}{c|}{\textbf{Linevul}} &
  
  \\
\cline{3-11}\multicolumn{1}{r|}{} &
  \multicolumn{1}{c|}{} &
  \multicolumn{1}{c}{P} &
  \multicolumn{1}{c}{R} &
  \multicolumn{1}{c|}{F1} &
  \multicolumn{1}{c}{P} &
  \multicolumn{1}{c}{R} &
  \multicolumn{1}{c|}{F1} &
  \multicolumn{1}{c}{P} &
  \multicolumn{1}{c}{R} &
  \multicolumn{1}{c|}{F1} &
  
  \\
\cline{2-11}\multicolumn{1}{r|}{} &
  \multicolumn{1}{c|}{NoAug} &
  \multicolumn{1}{c}{25.93} &
  \multicolumn{1}{c}{\cellcolor[rgb]{ .988,  .988,  1} 5.12} &
  \multicolumn{1}{c|}{\cellcolor[rgb]{ .988,  .988,  1} 8.55} &
  \multicolumn{1}{c}{7.35} &
  \multicolumn{1}{c}{44.63} &
  \multicolumn{1}{c|}{12.62} &
  \multicolumn{1}{c}{18.31} &
  \multicolumn{1}{c}{31.15} &
  \multicolumn{1}{c|}{23.06} &
  
  \\
\multicolumn{1}{r|}{} &
  \multicolumn{1}{c|}{VulGen} &
  \multicolumn{1}{c}{4.94} &
  \multicolumn{1}{c}{\cellcolor[rgb]{ .388,  .745,  .482} 59.76} &
  \multicolumn{1}{c|}{\cellcolor[rgb]{ .922,  .961,  .941} 9.11} &
  \multicolumn{1}{c}{5.29} &
  \multicolumn{1}{c}{\cellcolor[rgb]{ .388,  .745,  .482} 59.27} &
  \multicolumn{1}{c|}{9.72} &
  \multicolumn{1}{c}{\cellcolor[rgb]{ .447,  .769,  .533} 19.39} &
  \multicolumn{1}{c}{28.23} &
  \multicolumn{1}{c|}{22.99} &
  
  \\
\multicolumn{1}{r|}{} &
  \multicolumn{1}{c|}{VGX} &
  \multicolumn{1}{c}{5.61} &
  \multicolumn{1}{c}{\cellcolor[rgb]{ .459,  .773,  .541} 53.66} &
  \multicolumn{1}{c|}{\cellcolor[rgb]{ .788,  .91,  .827} 10.15} &
  \multicolumn{1}{c}{5.44} &
  \multicolumn{1}{c}{\cellcolor[rgb]{ .675,  .863,  .729} 50.73} &
  \multicolumn{1}{c|}{9.82} &
  \multicolumn{1}{c}{\cellcolor[rgb]{ .388,  .745,  .482} 19.77} &
  \multicolumn{1}{c}{27.69} &
  \multicolumn{1}{c|}{23.06} &
  
  \\
\multicolumn{1}{r|}{} &
  \multicolumn{1}{c|}{ROS} &
  \multicolumn{1}{c}{7.45} &
  \multicolumn{1}{c}{\cellcolor[rgb]{ .471,  .78,  .553} 52.44} &
  \multicolumn{1}{c|}{\cellcolor[rgb]{ .427,  .761,  .518} 13.04} &
  \multicolumn{1}{c}{\cellcolor[rgb]{ .565,  .816,  .635} 7.58} &
  \multicolumn{1}{c}{43.66} &
  \multicolumn{1}{c|}{\cellcolor[rgb]{ .533,  .804,  .608} 13.07} &
  \multicolumn{1}{c}{\cellcolor[rgb]{ .58,  .824,  .647} 18.48} &
  \multicolumn{1}{c}{30.05} &
  \multicolumn{1}{c|}{22.89} &
  
  \\
\multicolumn{1}{r|}{} &
  \multicolumn{1}{c|}{Mutation} &
  \multicolumn{1}{c}{7.17} &
  \multicolumn{1}{c}{\cellcolor[rgb]{ .467,  .776,  .549} 52.93} &
  \multicolumn{1}{c|}{\cellcolor[rgb]{ .478,  .784,  .561} 12.63} &
  \multicolumn{1}{c}{\cellcolor[rgb]{ .522,  .8,  .596} 7.81} &
  \multicolumn{1}{c}{43.66} &
  \multicolumn{1}{c|}{\cellcolor[rgb]{ .506,  .796,  .584} 13.25} &
  \multicolumn{1}{c}{15.71} &
  \multicolumn{1}{c}{26.96} &
  \multicolumn{1}{c|}{19.85} &
  
  \\
\multicolumn{1}{r|}{} &
  \multicolumn{1}{c|}{Injection} &
  \multicolumn{1}{c}{7.57} &
  \multicolumn{1}{c}{\cellcolor[rgb]{ .439,  .765,  .525} 55.37} &
  \multicolumn{1}{c|}{\cellcolor[rgb]{ .392,  .749,  .486} 13.31} &
  \multicolumn{1}{c}{\cellcolor[rgb]{ .522,  .8,  .596} 7.81} &
  \multicolumn{1}{c}{\cellcolor[rgb]{ .706,  .875,  .757} 49.76} &
  \multicolumn{1}{c|}{\cellcolor[rgb]{ .471,  .78,  .557} 13.51} &
  \multicolumn{1}{c}{17.29} &
  \multicolumn{1}{c}{\cellcolor[rgb]{ .388,  .745,  .482} 31.88} &
  \multicolumn{1}{c|}{22.42} &
  
  \\
\multicolumn{1}{r|}{} &
  \multicolumn{1}{c|}{Extension} &
  \multicolumn{1}{c}{7.61} &
  \multicolumn{1}{c}{\cellcolor[rgb]{ .459,  .776,  .545} 53.41} &
  \multicolumn{1}{c|}{\cellcolor[rgb]{ .388,  .745,  .482} \textbf{13.33}} &
  \multicolumn{1}{c}{\cellcolor[rgb]{ .388,  .745,  .482} 8.51} &
  \multicolumn{1}{c}{41.22} &
  \multicolumn{1}{c|}{\cellcolor[rgb]{ .388,  .745,  .482} \textbf{14.11}} &
  \multicolumn{1}{c}{\cellcolor[rgb]{ .576,  .824,  .643} 18.51} &
  \multicolumn{1}{c}{31.15} &
  \multicolumn{1}{c|}{\cellcolor[rgb]{ .388,  .745,  .482} \textbf{23.22}} &
  
  \\
\cline{2-11} &

\end{tabular}%
    }
  \label{tab:rq4}%
\end{table}%

\textbf{Extension and Injection outperform Mutation. Extension slightly outperforms Injection. Mutation continues to struggle against ROS.} Table~\ref{tab:rq4} presents the results of \ourTool and baselines across different DLVDs. As observed, \textbf{Extension} and \textbf{Injection} beat \textbf{Mutation} in all cases. \textbf{Extension} and \textbf{Injection} perform close to each other, while \textbf{Extension} appears to be more effective. In terms of F1-score, \textbf{Extension} achieves an F1-score of 16.89\%, outperforming \textbf{Injection} (F1-score 16.41\%) and \textbf{Mutation} (F1-score  15.24\%) by 2.88\% and 10.78\% on average, respectively. One possible reason is that the \textbf{Extension} strategy creates samples that are closer to the decision boundaries compared to \textbf{Injection}, with the reason lying in the source of the vulnerable samples and cleans. In RQ1~\ref{sec:rq1}, for the \textbf{Injection}, most of the context, being the clean sections, comes from the same dataset, while in this RQ, this is true for \textbf{Extension}.
\\
\\
\textbf{\textbf{Extension} appears to be useful and always outperforms baselines in all instances, with \textbf{Injection} performing close to \textbf{Extension} and outperforming all baselines with two of the three DLVDs. \textbf{Extension} outperforms NoAug, Vulgen, VGX, and ROS  by 14.54\%, 21.14\%, 17.73\%, and 3.39\% on average F1-score, respectively.} As shown in Table~\ref{tab:rq4}, \textbf{Extension} strategy outperforms baselines (i.e., NoAug, VulGen, VGX, ROS) in all of the experimental instances in terms of F1-score. The \textbf{Injection} fails to increase the performance for LineVul compared to the baselines. 
More specifically, \textbf{Extension} beats the baselines: NoAug, Vulgen, VGX, and ROS by 14.54\%, 21.14\%, 17.73\%, and 3.39\% on average F1-score, while the \textbf{Injection} strategy beats the baselines: NoAug, Vulgen, VGX, and ROS by 11.33\%, 17.74\%, 14.43\%, and 0.49\% on average F1-score, respectively.\\
\\
As observed, \textbf{Mutation} only beats all the baselines with Reveal. Similar to RQ1~\ref{sec:rq1}, \textbf{Mutation} does not have a dominating advantage over ROS, and ROS beats \textbf{Mutation} in the two other settings. ROS beats the baseline, Vulgen, VGX, and Mutation by 10.78\%, 17.17\%, 13.97\%, and 7.15\% in terms of F1-score on average.

\rqboxc{\textbf{Extension} outperforms the baselines and the \textbf{Mutation} by a large margin, while \textbf{Injection} provides a slightly lower performance compared to the \textbf{Extension}, but beats all the baselines with two DLVDs. \textbf{Extension}
outperforms NoAug, Vulgen, VGX, and ROS by 14.54\%, 21.14\%, 17.73\%, and 3.39\% on average F1-score. 
}

\section{Discussion}\label{sec:dis}

\begin{table}[htbp!]
  \centering
  \caption{Impact of mixing our strategies' generated samples on Improving DLVD models' performance with a total of 5K augmented samples. Each involved strategy provides an equal number of samples. The cells with larger values (better performance) compared to NoAug are highlighted darker.}
  \resizebox{13.5cm}{!}{
  % Table generated by Excel2LaTeX from sheet 'discussions - mixed'
\begin{tabular}{rrrrrrrrrrrrrrrrrrrrr}
 &
   &
   &
   &
   &
   &
   &
   &
   &
   &
   &
   &
   &
   &
   &
   &
   &
   &
   &
   &
  
  \\
 &
   &
  \multicolumn{9}{c|}{\textit{\textbf{ChatGPT 3.5 Turbo}}} &
  \multicolumn{9}{c}{\textit{\textbf{CodeQwen1.5-7B-Chat}}} &
  
  \\
\cline{2-20}\multicolumn{1}{r|}{} &
  \multicolumn{1}{c|}{\multirow{2}[4]{*}{{\raisebox{0.4cm}{\textbf{Strategy}}}}} &
  \multicolumn{3}{c|}{\textbf{Devign}} &
  \multicolumn{3}{c|}{\textbf{Reveal}} &
  \multicolumn{3}{c|}{\textbf{Linevul}} &
  \multicolumn{3}{c|}{\textbf{Devign}} &
  \multicolumn{3}{c|}{\textbf{Reveal}} &
  \multicolumn{3}{c|}{\textbf{Linevul}} &
  
  \\
\cline{3-20}\multicolumn{1}{r|}{} &
  \multicolumn{1}{c|}{} &
  \multicolumn{1}{c}{P} &
  \multicolumn{1}{c}{R} &
  \multicolumn{1}{c|}{F1} &
  \multicolumn{1}{c}{P} &
  \multicolumn{1}{c}{R} &
  \multicolumn{1}{c|}{F1} &
  \multicolumn{1}{c}{P} &
  \multicolumn{1}{c}{R} &
  \multicolumn{1}{c|}{F1} &
  \multicolumn{1}{c}{P} &
  \multicolumn{1}{c}{R} &
  \multicolumn{1}{c|}{F1} &
  \multicolumn{1}{c}{P} &
  \multicolumn{1}{c}{R} &
  \multicolumn{1}{c|}{F1} &
  \multicolumn{1}{c}{P} &
  \multicolumn{1}{c}{R} &
  \multicolumn{1}{c|}{F1} &
  
  \\
\cline{2-20}\multicolumn{1}{c|}{\multirow{7}[2]{*}{\begin{sideways}\textbf{Reveal}\end{sideways}}} &
  \multicolumn{1}{c|}{Mutation} &
  \multicolumn{1}{c}{10.98} &
  \multicolumn{1}{c}{53.44} &
  \multicolumn{1}{c|}{18.22} &
  \multicolumn{1}{c}{12.23} &
  \multicolumn{1}{c}{68.46} &
  \multicolumn{1}{c|}{20.75} &
  \multicolumn{1}{c}{\cellcolor[rgb]{ .988,  .988,  1} 7.80} &
  \multicolumn{1}{c}{22.62} &
  \multicolumn{1}{c|}{\cellcolor[rgb]{ .988,  .988,  1} 11.60} &
  \multicolumn{1}{c}{\cellcolor[rgb]{ .988,  .988,  1} 10.96} &
  \multicolumn{1}{c}{61.64} &
  \multicolumn{1}{c|}{18.60} &
  \multicolumn{1}{c}{15.01} &
  \multicolumn{1}{c}{57.96} &
  \multicolumn{1}{c|}{23.85} &
  \multicolumn{1}{c}{9.48} &
  \multicolumn{1}{c}{30.43} &
  \multicolumn{1}{c|}{14.45} &
  
  \\
\multicolumn{1}{c|}{} &
  \multicolumn{1}{c|}{Injection} &
  \multicolumn{1}{c}{\cellcolor[rgb]{ .62,  .839,  .682} 13.44} &
  \multicolumn{1}{c}{53.26} &
  \multicolumn{1}{c|}{\cellcolor[rgb]{ .388,  .745,  .482} \textbf{21.46}} &
  \multicolumn{1}{c}{\cellcolor[rgb]{ .388,  .745,  .482} 15.43} &
  \multicolumn{1}{c}{61.46} &
  \multicolumn{1}{c|}{\cellcolor[rgb]{ .388,  .745,  .482} \textbf{24.66}} &
  \multicolumn{1}{c}{\cellcolor[rgb]{ .831,  .925,  .867} 11.57} &
  \multicolumn{1}{c}{\cellcolor[rgb]{ .718,  .878,  .765} 26.43} &
  \multicolumn{1}{c|}{\cellcolor[rgb]{ .596,  .831,  .663} 16.10} &
  \multicolumn{1}{c}{\cellcolor[rgb]{ .608,  .835,  .671} 13.67} &
  \multicolumn{1}{c}{54.28} &
  \multicolumn{1}{c|}{\cellcolor[rgb]{ .388,  .745,  .482} \textbf{21.83}} &
  \multicolumn{1}{c}{14.23} &
  \multicolumn{1}{c}{\cellcolor[rgb]{ .675,  .863,  .729} 64.48} &
  \multicolumn{1}{c|}{23.32} &
  \multicolumn{1}{c}{\cellcolor[rgb]{ .949,  .973,  .969} 9.99} &
  \multicolumn{1}{c}{25.79} &
  \multicolumn{1}{c|}{14.40} &
  
  \\
\multicolumn{1}{c|}{} &
  \multicolumn{1}{c|}{Extension} &
  \multicolumn{1}{c}{\cellcolor[rgb]{ .388,  .745,  .482} 14.97} &
  \multicolumn{1}{c}{37.21} &
  \multicolumn{1}{c|}{\cellcolor[rgb]{ .412,  .757,  .502} 21.35} &
  \multicolumn{1}{c}{\cellcolor[rgb]{ .459,  .776,  .545} 14.76} &
  \multicolumn{1}{c}{50.00} &
  \multicolumn{1}{c|}{\cellcolor[rgb]{ .525,  .8,  .6} 22.80} &
  \multicolumn{1}{c}{\cellcolor[rgb]{ .561,  .816,  .631} 18.01} &
  \multicolumn{1}{c}{12.81} &
  \multicolumn{1}{c|}{\cellcolor[rgb]{ .694,  .871,  .745} 14.97} &
  \multicolumn{1}{c}{\cellcolor[rgb]{ .388,  .745,  .482} 15.20} &
  \multicolumn{1}{c}{38.18} &
  \multicolumn{1}{c|}{\cellcolor[rgb]{ .408,  .753,  .498} 21.75} &
  \multicolumn{1}{c}{14.82} &
  \multicolumn{1}{c}{52.71} &
  \multicolumn{1}{c|}{23.13} &
  \multicolumn{1}{c}{\cellcolor[rgb]{ .753,  .894,  .796} 20.65} &
  \multicolumn{1}{c}{12.62} &
  \multicolumn{1}{c|}{\cellcolor[rgb]{ .424,  .761,  .514} 15.67} &
  
  \\
\multicolumn{1}{c|}{} &
  \multicolumn{1}{c|}{M + I} &
  \multicolumn{1}{c}{\cellcolor[rgb]{ .855,  .933,  .882} 11.89} &
  \multicolumn{1}{c}{\cellcolor[rgb]{ .388,  .745,  .482} 66.41} &
  \multicolumn{1}{c|}{\cellcolor[rgb]{ .612,  .839,  .678} 20.17} &
  \multicolumn{1}{c}{\cellcolor[rgb]{ .565,  .82,  .635} 13.74} &
  \multicolumn{1}{c}{\cellcolor[rgb]{ .388,  .745,  .482} 70.39} &
  \multicolumn{1}{c|}{\cellcolor[rgb]{ .51,  .796,  .588} 22.99} &
  \multicolumn{1}{c}{\cellcolor[rgb]{ .835,  .929,  .871} 11.46} &
  \multicolumn{1}{c}{\cellcolor[rgb]{ .388,  .745,  .482} 47.14} &
  \multicolumn{1}{c|}{\cellcolor[rgb]{ .388,  .745,  .482} \textbf{18.43}} &
  \multicolumn{1}{c}{\cellcolor[rgb]{ .875,  .941,  .902} 11.78} &
  \multicolumn{1}{c}{58.02} &
  \multicolumn{1}{c|}{\cellcolor[rgb]{ .804,  .914,  .843} 19.58} &
  \multicolumn{1}{c}{13.14} &
  \multicolumn{1}{c}{\cellcolor[rgb]{ .455,  .773,  .537} 72.74} &
  \multicolumn{1}{c|}{22.25} &
  \multicolumn{1}{c}{9.35} &
  \multicolumn{1}{c}{\cellcolor[rgb]{ .388,  .745,  .482} 51.41} &
  \multicolumn{1}{c|}{\cellcolor[rgb]{ .388,  .745,  .482} \textbf{15.82}} &
  
  \\
\multicolumn{1}{c|}{} &
  \multicolumn{1}{c|}{M + E} &
  \multicolumn{1}{c}{\cellcolor[rgb]{ .953,  .976,  .969} 11.23} &
  \multicolumn{1}{c}{\cellcolor[rgb]{ .561,  .816,  .631} 58.02} &
  \multicolumn{1}{c|}{\cellcolor[rgb]{ .847,  .933,  .878} 18.82} &
  \multicolumn{1}{c}{\cellcolor[rgb]{ .525,  .8,  .6} 14.14} &
  \multicolumn{1}{c}{59.89} &
  \multicolumn{1}{c|}{\cellcolor[rgb]{ .522,  .8,  .596} 22.87} &
  \multicolumn{1}{c}{\cellcolor[rgb]{ .922,  .961,  .941} 9.45} &
  \multicolumn{1}{c}{\cellcolor[rgb]{ .451,  .773,  .537} 43.23} &
  \multicolumn{1}{c|}{\cellcolor[rgb]{ .647,  .851,  .706} 15.50} &
  \multicolumn{1}{c}{\cellcolor[rgb]{ .855,  .933,  .886} 11.92} &
  \multicolumn{1}{c}{55.19} &
  \multicolumn{1}{c|}{\cellcolor[rgb]{ .8,  .914,  .839} 19.60} &
  \multicolumn{1}{c}{13.20} &
  \multicolumn{1}{c}{\cellcolor[rgb]{ .635,  .847,  .694} 65.98} &
  \multicolumn{1}{c|}{22.00} &
  \multicolumn{1}{c}{7.85} &
  \multicolumn{1}{c}{\cellcolor[rgb]{ .518,  .8,  .596} 42.51} &
  \multicolumn{1}{c|}{13.26} &
  
  \\
\multicolumn{1}{c|}{} &
  \multicolumn{1}{c|}{I + E} &
  \multicolumn{1}{c}{\cellcolor[rgb]{ .627,  .843,  .69} 13.39} &
  \multicolumn{1}{c}{47.71} &
  \multicolumn{1}{c|}{\cellcolor[rgb]{ .486,  .784,  .565} 20.91} &
  \multicolumn{1}{c}{\cellcolor[rgb]{ .408,  .753,  .498} 15.26} &
  \multicolumn{1}{c}{59.71} &
  \multicolumn{1}{c|}{\cellcolor[rgb]{ .416,  .757,  .506} 24.31} &
  \multicolumn{1}{c}{\cellcolor[rgb]{ .388,  .745,  .482} 22.07} &
  \multicolumn{1}{c}{9.01} &
  \multicolumn{1}{c|}{\cellcolor[rgb]{ .886,  .949,  .91} 12.80} &
  \multicolumn{1}{c}{\cellcolor[rgb]{ .537,  .808,  .612} 14.15} &
  \multicolumn{1}{c}{46.98} &
  \multicolumn{1}{c|}{\cellcolor[rgb]{ .404,  .753,  .498} 21.75} &
  \multicolumn{1}{c}{14.92} &
  \multicolumn{1}{c}{\cellcolor[rgb]{ .733,  .886,  .78} 62.24} &
  \multicolumn{1}{c|}{\cellcolor[rgb]{ .388,  .745,  .482} \textbf{24.06}} &
  \multicolumn{1}{c}{\cellcolor[rgb]{ .945,  .973,  .961} 10.34} &
  \multicolumn{1}{c}{18.80} &
  \multicolumn{1}{c|}{13.35} &
  
  \\
\multicolumn{1}{c|}{} &
  \multicolumn{1}{c|}{M + I + E} &
  \multicolumn{1}{c}{\cellcolor[rgb]{ .969,  .98,  .984} 11.12} &
  \multicolumn{1}{c}{46.98} &
  \multicolumn{1}{c|}{17.98} &
  \multicolumn{1}{c}{9.67} &
  \multicolumn{1}{c}{53.26} &
  \multicolumn{1}{c|}{16.36} &
  \multicolumn{1}{c}{\cellcolor[rgb]{ .875,  .945,  .902} 10.55} &
  \multicolumn{1}{c}{\cellcolor[rgb]{ .388,  .745,  .482} 47.14} &
  \multicolumn{1}{c|}{\cellcolor[rgb]{ .494,  .788,  .573} 17.25} &
  \multicolumn{1}{c}{\cellcolor[rgb]{ .973,  .984,  .988} 11.07} &
  \multicolumn{1}{c}{57.60} &
  \multicolumn{1}{c|}{18.57} &
  \multicolumn{1}{c}{13.34} &
  \multicolumn{1}{c}{\cellcolor[rgb]{ .388,  .745,  .482} 75.09} &
  \multicolumn{1}{c|}{22.65} &
  \multicolumn{1}{c}{\cellcolor[rgb]{ .388,  .745,  .482} 40.33} &
  \multicolumn{1}{c}{9.47} &
  \multicolumn{1}{c|}{\cellcolor[rgb]{ .502,  .792,  .58} 15.34} &
  
  \\
\cline{1-20}\multicolumn{1}{c|}{\multirow{7}[2]{*}{\begin{sideways}\textbf{Bigvul}\end{sideways}}} &
  \multicolumn{1}{c|}{Mutation} &
  \multicolumn{1}{c}{7.80} &
  \multicolumn{1}{c}{63.91} &
  \multicolumn{1}{c|}{13.90} &
  \multicolumn{1}{c}{\cellcolor[rgb]{ .988,  .988,  1} 7.63} &
  \multicolumn{1}{c}{67.73} &
  \multicolumn{1}{c|}{\cellcolor[rgb]{ .988,  .988,  1} 13.72} &
  \multicolumn{1}{c}{\cellcolor[rgb]{ .988,  .988,  1} 6.78} &
  \multicolumn{1}{c}{74.14} &
  \multicolumn{1}{c|}{\cellcolor[rgb]{ .988,  .988,  1} 12.42} &
  \multicolumn{1}{c}{7.56} &
  \multicolumn{1}{c}{57.23} &
  \multicolumn{1}{c|}{13.36} &
  \multicolumn{1}{c}{9.37} &
  \multicolumn{1}{c}{53.26} &
  \multicolumn{1}{c|}{15.94} &
  \multicolumn{1}{c}{6.65} &
  \multicolumn{1}{c}{78.31} &
  \multicolumn{1}{c|}{12.25} &
  
  \\
\multicolumn{1}{c|}{} &
  \multicolumn{1}{c|}{Injection} &
  \multicolumn{1}{c}{\cellcolor[rgb]{ .706,  .875,  .757} 8.79} &
  \multicolumn{1}{c}{29.41} &
  \multicolumn{1}{c|}{13.53} &
  \multicolumn{1}{c}{\cellcolor[rgb]{ .388,  .745,  .482} 11.64} &
  \multicolumn{1}{c}{38.00} &
  \multicolumn{1}{c|}{\cellcolor[rgb]{ .388,  .745,  .482} \textbf{17.82}} &
  \multicolumn{1}{c}{\cellcolor[rgb]{ .761,  .898,  .804} 11.43} &
  \multicolumn{1}{c}{18.81} &
  \multicolumn{1}{c|}{\cellcolor[rgb]{ .514,  .796,  .592} 14.22} &
  \multicolumn{1}{c}{\cellcolor[rgb]{ .514,  .796,  .592} 10.33} &
  \multicolumn{1}{c}{31.80} &
  \multicolumn{1}{c|}{\cellcolor[rgb]{ .545,  .808,  .62} 15.59} &
  \multicolumn{1}{c}{\cellcolor[rgb]{ .533,  .804,  .608} 9.91} &
  \multicolumn{1}{c}{38.16} &
  \multicolumn{1}{c|}{15.74} &
  \multicolumn{1}{c}{\cellcolor[rgb]{ .91,  .957,  .933} 7.85} &
  \multicolumn{1}{c}{30.49} &
  \multicolumn{1}{c|}{\cellcolor[rgb]{ .875,  .945,  .902} 12.49} &
  
  \\
\multicolumn{1}{c|}{} &
  \multicolumn{1}{c|}{Extension} &
  \multicolumn{1}{c}{\cellcolor[rgb]{ .388,  .745,  .482} 10.60} &
  \multicolumn{1}{c}{17.33} &
  \multicolumn{1}{c|}{13.16} &
  \multicolumn{1}{c}{\cellcolor[rgb]{ .467,  .776,  .549} 11.13} &
  \multicolumn{1}{c}{34.02} &
  \multicolumn{1}{c|}{\cellcolor[rgb]{ .545,  .808,  .616} 16.77} &
  \multicolumn{1}{c}{\cellcolor[rgb]{ .388,  .745,  .482} 19.03} &
  \multicolumn{1}{c}{11.96} &
  \multicolumn{1}{c|}{\cellcolor[rgb]{ .388,  .745,  .482} \textbf{14.68}} &
  \multicolumn{1}{c}{\cellcolor[rgb]{ .584,  .824,  .651} 9.89} &
  \multicolumn{1}{c}{31.48} &
  \multicolumn{1}{c|}{\cellcolor[rgb]{ .639,  .847,  .698} 15.05} &
  \multicolumn{1}{c}{\cellcolor[rgb]{ .447,  .769,  .533} 10.22} &
  \multicolumn{1}{c}{33.86} &
  \multicolumn{1}{c|}{15.70} &
  \multicolumn{1}{c}{\cellcolor[rgb]{ .388,  .745,  .482} 15.74} &
  \multicolumn{1}{c}{11.86} &
  \multicolumn{1}{c|}{\cellcolor[rgb]{ .388,  .745,  .482} \textbf{13.53}} &
  
  \\
\multicolumn{1}{c|}{} &
  \multicolumn{1}{c|}{M + I} &
  \multicolumn{1}{c}{7.28} &
  \multicolumn{1}{c}{\cellcolor[rgb]{ .388,  .745,  .482} 66.93} &
  \multicolumn{1}{c|}{13.14} &
  \multicolumn{1}{c}{\cellcolor[rgb]{ .816,  .922,  .851} 8.79} &
  \multicolumn{1}{c}{66.45} &
  \multicolumn{1}{c|}{\cellcolor[rgb]{ .725,  .882,  .773} 15.53} &
  \multicolumn{1}{c}{\cellcolor[rgb]{ .961,  .976,  .976} 7.41} &
  \multicolumn{1}{c}{72.75} &
  \multicolumn{1}{c|}{\cellcolor[rgb]{ .718,  .878,  .765} 13.45} &
  \multicolumn{1}{c}{7.30} &
  \multicolumn{1}{c}{\cellcolor[rgb]{ .451,  .773,  .537} 58.82} &
  \multicolumn{1}{c|}{12.99} &
  \multicolumn{1}{c}{8.31} &
  \multicolumn{1}{c}{\cellcolor[rgb]{ .388,  .745,  .482} 64.71} &
  \multicolumn{1}{c|}{14.72} &
  \multicolumn{1}{c}{\cellcolor[rgb]{ .965,  .98,  .98} 7.03} &
  \multicolumn{1}{c}{\cellcolor[rgb]{ .388,  .745,  .482} 83.60} &
  \multicolumn{1}{c|}{\cellcolor[rgb]{ .651,  .851,  .71} 12.97} &
  
  \\
\multicolumn{1}{c|}{} &
  \multicolumn{1}{c|}{M + E} &
  \multicolumn{1}{c}{7.24} &
  \multicolumn{1}{c}{61.21} &
  \multicolumn{1}{c|}{12.94} &
  \multicolumn{1}{c}{\cellcolor[rgb]{ .835,  .925,  .867} 8.68} &
  \multicolumn{1}{c}{57.55} &
  \multicolumn{1}{c|}{\cellcolor[rgb]{ .788,  .91,  .827} 15.09} &
  \multicolumn{1}{c}{\cellcolor[rgb]{ .98,  .988,  .996} 6.94} &
  \multicolumn{1}{c}{\cellcolor[rgb]{ .388,  .745,  .482} 81.65} &
  \multicolumn{1}{c|}{\cellcolor[rgb]{ .894,  .953,  .922} 12.78} &
  \multicolumn{1}{c}{7.45} &
  \multicolumn{1}{c}{\cellcolor[rgb]{ .388,  .745,  .482} 61.84} &
  \multicolumn{1}{c|}{13.30} &
  \multicolumn{1}{c}{8.27} &
  \multicolumn{1}{c}{\cellcolor[rgb]{ .51,  .796,  .588} 58.51} &
  \multicolumn{1}{c|}{14.49} &
  \multicolumn{1}{c}{6.64} &
  \multicolumn{1}{c}{\cellcolor[rgb]{ .396,  .749,  .49} 82.76} &
  \multicolumn{1}{c|}{\cellcolor[rgb]{ .961,  .98,  .976} 12.30} &
  
  \\
\multicolumn{1}{c|}{} &
  \multicolumn{1}{c|}{I + E} &
  \multicolumn{1}{c}{\cellcolor[rgb]{ .596,  .831,  .663} 9.41} &
  \multicolumn{1}{c}{35.29} &
  \multicolumn{1}{c|}{\cellcolor[rgb]{ .388,  .745,  .482} \textbf{16.46}} &
  \multicolumn{1}{c}{\cellcolor[rgb]{ .51,  .796,  .588} 10.83} &
  \multicolumn{1}{c}{39.90} &
  \multicolumn{1}{c|}{\cellcolor[rgb]{ .506,  .792,  .584} 17.04} &
  \multicolumn{1}{c}{\cellcolor[rgb]{ .906,  .957,  .929} 8.53} &
  \multicolumn{1}{c}{41.80} &
  \multicolumn{1}{c|}{\cellcolor[rgb]{ .529,  .804,  .604} 14.16} &
  \multicolumn{1}{c}{\cellcolor[rgb]{ .388,  .745,  .482} 11.11} &
  \multicolumn{1}{c}{31.96} &
  \multicolumn{1}{c|}{\cellcolor[rgb]{ .388,  .745,  .482} \textbf{16.49}} &
  \multicolumn{1}{c}{\cellcolor[rgb]{ .388,  .745,  .482} 10.42} &
  \multicolumn{1}{c}{38.47} &
  \multicolumn{1}{c|}{\cellcolor[rgb]{ .388,  .745,  .482} \textbf{16.40}} &
  \multicolumn{1}{c}{\cellcolor[rgb]{ .937,  .969,  .957} 7.46} &
  \multicolumn{1}{c}{34.01} &
  \multicolumn{1}{c|}{12.24} &
  
  \\
\multicolumn{1}{c|}{} &
  \multicolumn{1}{c|}{M + I + E} &
  \multicolumn{1}{c}{7.14} &
  \multicolumn{1}{c}{42.77} &
  \multicolumn{1}{c|}{\cellcolor[rgb]{ .604,  .835,  .671} 15.20} &
  \multicolumn{1}{c}{\cellcolor[rgb]{ .749,  .89,  .792} 9.25} &
  \multicolumn{1}{c}{53.42} &
  \multicolumn{1}{c|}{\cellcolor[rgb]{ .69,  .871,  .745} 15.77} &
  \multicolumn{1}{c}{\cellcolor[rgb]{ .961,  .976,  .976} 7.38} &
  \multicolumn{1}{c}{\cellcolor[rgb]{ .408,  .753,  .498} 79.61} &
  \multicolumn{1}{c|}{\cellcolor[rgb]{ .706,  .875,  .757} 13.50} &
  \multicolumn{1}{c}{7.45} &
  \multicolumn{1}{c}{54.85} &
  \multicolumn{1}{c|}{13.12} &
  \multicolumn{1}{c}{9.23} &
  \multicolumn{1}{c}{52.78} &
  \multicolumn{1}{c|}{15.71} &
  \multicolumn{1}{c}{\cellcolor[rgb]{ .976,  .984,  .992} 6.83} &
  \multicolumn{1}{c}{\cellcolor[rgb]{ .431,  .765,  .522} 78.59} &
  \multicolumn{1}{c|}{\cellcolor[rgb]{ .835,  .929,  .871} 12.57} &
  
  \\
\cline{2-20} &

  \\
\end{tabular}%
}
  \label{tab:mixed}%
\end{table}%

\begin{table}[htbp!]
  \centering
  \caption{Impact of using random oversampling for clean samples instead of adding additional clean samples on our strategies and the baselines. Starred rows indicate using oversampled cleans. Bold F1-scores show the best performance between the two settings for each augmentation strategy.}
  \label{tab:ros_clean}
  \resizebox{13.5cm}{!}{
% Table generated by Excel2LaTeX from sheet 'discussions - ros'ed cleans'
\begin{tabular}{rrrrrrrrrrrrrrrrrrrrrr}

  \\
 &
   &
   &
  \multicolumn{9}{c|}{\textit{\textbf{ChatGPT 3.5 Turbo}}} &
  \multicolumn{9}{c}{\textit{\textbf{CodeQwen1.5-7B-Chat}}} &
  
  \\
\cline{3-21} &
  \multicolumn{1}{r|}{} &
   \multicolumn{1}{c|}{\multirow{2}[4]{*}{{\raisebox{0.4cm}{\textbf{Strategy}}}}} &
  \multicolumn{3}{c|}{\textbf{Devign}} &
  \multicolumn{3}{c|}{\textbf{Reveal}} &
  \multicolumn{3}{c|}{\textbf{Linevul}} &
  \multicolumn{3}{c|}{\textbf{Devign}} &
  \multicolumn{3}{c|}{\textbf{Reveal}} &
  \multicolumn{3}{c|}{\textbf{Linevul}} &
  
  \\
\cline{4-21} &
  \multicolumn{1}{r|}{} &
  \multicolumn{1}{c|}{} &
  \multicolumn{1}{c}{P} &
  \multicolumn{1}{c}{R} &
  \multicolumn{1}{c|}{F1} &
  \multicolumn{1}{c}{P} &
  \multicolumn{1}{c}{R} &
  \multicolumn{1}{c|}{F1} &
  \multicolumn{1}{c}{P} &
  \multicolumn{1}{c}{R} &
  \multicolumn{1}{c|}{F1} &
  \multicolumn{1}{c}{P} &
  \multicolumn{1}{c}{R} &
  \multicolumn{1}{c|}{F1} &
  \multicolumn{1}{c}{P} &
  \multicolumn{1}{c}{R} &
  \multicolumn{1}{c|}{F1} &
  \multicolumn{1}{c}{P} &
  \multicolumn{1}{c}{R} &
  \multicolumn{1}{c|}{F1} &
  
  \\
  \cline{3-21}\multicolumn{1}{c}{\multirow{11}[2]{*}{\begin{sideways}\textbf{Reveal}\end{sideways}}} &
\multicolumn{1}{c|}{} &

% 
  % 13.28 &
  % 39.93 &
  % 19.93 &
  % 14.35 &
  % 54.52 &
  % 22.72 &
  % 9.88 &
  % 22.71 &
  % 13.77 &
  % 13.28 &
  % 39.93 &
  % 19.93 &
  % 14.35 &
  % 54.52 &
  % 22.72 &
  % 9.88 &
  % 22.71 &
  % 13.77

  \multicolumn{1}{c|}{ROS} &
  \multicolumn{1}{c}{13.28} &
  \multicolumn{1}{c}{39.93} &
  \multicolumn{1}{c|}{19.93} &
  \multicolumn{1}{c}{14.35} &
  \multicolumn{1}{c}{54.52} &
  \multicolumn{1}{c|}{22.72} &
  \multicolumn{1}{c}{9.88} &
  \multicolumn{1}{c}{22.71} &
  \multicolumn{1}{c|}{13.77} &
  \multicolumn{1}{c}{13.28} &
  \multicolumn{1}{c}{39.93} &
  \multicolumn{1}{c|}{19.93} &
  \multicolumn{1}{c}{14.35} &
  \multicolumn{1}{c}{54.52} &
  \multicolumn{1}{c|}{22.72} &
  \multicolumn{1}{c}{9.88} &
  \multicolumn{1}{c}{22.71} &
  \multicolumn{1}{c|}{13.77} &
  \\
  \cline{3-21}
 &
  \multicolumn{1}{c|}{} &
  \multicolumn{1}{c|}{VulGen} &
  \multicolumn{1}{c}{ 10.45} &
  \multicolumn{1}{c}{ 57.90} &
  \multicolumn{1}{c|}{ \textbf{17.70}} &
  \multicolumn{1}{c}{ 12.46} &
  \multicolumn{1}{c}{ 64.60} &
  \multicolumn{1}{c|}{ \textbf{20.90}} &
  \multicolumn{1}{c}{ 5.31} &
  \multicolumn{1}{c}{ 98.09} &
  \multicolumn{1}{c|}{ 10.08} &
  \multicolumn{1}{c}{ 10.45} &
  \multicolumn{1}{c}{ 57.90} &
  \multicolumn{1}{c|}{ \textbf{17.70}} &
  \multicolumn{1}{c}{ 12.46} &
  \multicolumn{1}{c}{ 64.60} &
  \multicolumn{1}{c|}{ \textbf{20.90}} &
  \multicolumn{1}{c}{ 5.31} &
  \multicolumn{1}{c}{ 98.09} &
  \multicolumn{1}{c|}{ 10.08} &
  
  \\
 &
  \multicolumn{1}{c|}{} &
  \multicolumn{1}{c|}{VulGen*} &
  \multicolumn{1}{c}{ 8.56} &
  \multicolumn{1}{c}{ 64.29} &
  \multicolumn{1}{c|}{ 15.10} &
  \multicolumn{1}{c}{ 5.64} &
  \multicolumn{1}{c}{ 78.38} &
  \multicolumn{1}{c|}{ 10.52} &
  \multicolumn{1}{c}{ 5.37} &
  \multicolumn{1}{c}{ 100.00} &
  \multicolumn{1}{c|}{ \textbf{10.20}} &
  \multicolumn{1}{c}{ 8.56} &
  \multicolumn{1}{c}{ 64.29} &
  \multicolumn{1}{c|}{ 15.10} &
  \multicolumn{1}{c}{ 5.64} &
  \multicolumn{1}{c}{ 78.38} &
  \multicolumn{1}{c|}{ 10.52} &
  \multicolumn{1}{c}{ 5.37} &
  \multicolumn{1}{c}{ 100.00} &
  \multicolumn{1}{c|}{ \textbf{10.20}} &
  
  \\
 &
  \multicolumn{1}{c|}{} &
  \multicolumn{1}{c|}{VGX} &
  \multicolumn{1}{c}{ 11.02} &
  \multicolumn{1}{c}{ 59.35} &
  \multicolumn{1}{c|}{ \textbf{18.58}} &
  \multicolumn{1}{c}{ 11.69} &
  \multicolumn{1}{c}{ 65.62} &
  \multicolumn{1}{c|}{ \textbf{19.84}} &
  \multicolumn{1}{c}{ 9.51} &
  \multicolumn{1}{c}{ 87.24} &
  \multicolumn{1}{c|}{ \textbf{11.47}} &
  \multicolumn{1}{c}{ 11.02} &
  \multicolumn{1}{c}{ 59.35} &
  \multicolumn{1}{c|}{ \textbf{18.58}} &
  \multicolumn{1}{c}{ 11.69} &
  \multicolumn{1}{c}{ 65.62} &
  \multicolumn{1}{c|}{ \textbf{19.84}} &
  \multicolumn{1}{c}{ 9.51} &
  \multicolumn{1}{c}{ 87.24} &
  \multicolumn{1}{c|}{ \textbf{11.47}} &
  
  \\
 &
  \multicolumn{1}{c|}{} &
  \multicolumn{1}{c|}{VGX*} &
  \multicolumn{1}{c}{ 8.86} &
  \multicolumn{1}{c}{ 65.98} &
  \multicolumn{1}{c|}{ 15.62} &
  \multicolumn{1}{c}{ 5.74} &
  \multicolumn{1}{c}{ 74.40} &
  \multicolumn{1}{c|}{ 10.66} &
  \multicolumn{1}{c}{ 5.38} &
  \multicolumn{1}{c}{ 99.91} &
  \multicolumn{1}{c|}{ 10.21} &
  \multicolumn{1}{c}{ 8.86} &
  \multicolumn{1}{c}{ 65.98} &
  \multicolumn{1}{c|}{ 15.62} &
  \multicolumn{1}{c}{ 5.74} &
  \multicolumn{1}{c}{ 74.40} &
  \multicolumn{1}{c|}{ 10.66} &
  \multicolumn{1}{c}{ 5.38} &
  \multicolumn{1}{c}{ 99.91} &
  \multicolumn{1}{c|}{ 10.21} &
  
  \\
 &
  \multicolumn{1}{c|}{} &
  \multicolumn{1}{c|}{Mutation} &
  \multicolumn{1}{c}{ 10.98} &
  \multicolumn{1}{c}{ 53.44} &
  \multicolumn{1}{c|}{ \textbf{18.22}} &
  \multicolumn{1}{c}{ 12.23} &
  \multicolumn{1}{c}{ 68.46} &
  \multicolumn{1}{c|}{ \textbf{20.75}} &
  \multicolumn{1}{c}{ 7.80} &
  \multicolumn{1}{c}{ 22.62} &
  \multicolumn{1}{c|}{ \textbf{11.60}} &
  \multicolumn{1}{c}{ 10.96} &
  \multicolumn{1}{c}{ 61.64} &
  \multicolumn{1}{c|}{ \textbf{18.60}} &
  \multicolumn{1}{c}{ 15.01} &
  \multicolumn{1}{c}{ 57.96} &
  \multicolumn{1}{c|}{ \textbf{23.85}} &
  \multicolumn{1}{c}{ 9.48} &
  \multicolumn{1}{c}{ 30.43} &
  \multicolumn{1}{c|}{ \textbf{14.45}} &
  
  \\
 &
  \multicolumn{1}{c|}{} &
  \multicolumn{1}{c|}{Mutation*} &
  \multicolumn{1}{c}{ 9.05} &
  \multicolumn{1}{c}{ 59.95} &
  \multicolumn{1}{c|}{ 15.72} &
  \multicolumn{1}{c}{ 9.90} &
  \multicolumn{1}{c}{ 73.34} &
  \multicolumn{1}{c|}{ 17.45} &
  \multicolumn{1}{c}{ 4.66} &
  \multicolumn{1}{c}{ 44.77} &
  \multicolumn{1}{c|}{ 8.44} &
  \multicolumn{1}{c}{ 9.07} &
  \multicolumn{1}{c}{ 66.77} &
  \multicolumn{1}{c|}{ 15.98} &
  \multicolumn{1}{c}{ 6.25} &
  \multicolumn{1}{c}{ 86.17} &
  \multicolumn{1}{c|}{ 11.65} &
  \multicolumn{1}{c}{ 5.66} &
  \multicolumn{1}{c}{ 70.12} &
  \multicolumn{1}{c|}{ 10.47} &
  
  \\
 &
  \multicolumn{1}{c|}{} &
  \multicolumn{1}{c|}{Injection} &
  \multicolumn{1}{c}{ 13.44} &
  \multicolumn{1}{c}{ 53.26} &
  \multicolumn{1}{c|}{ \textbf{21.46}} &
  \multicolumn{1}{c}{ 15.43} &
  \multicolumn{1}{c}{ 61.46} &
  \multicolumn{1}{c|}{ \textbf{24.66}} &
  \multicolumn{1}{c}{ 11.57} &
  \multicolumn{1}{c}{ 26.43} &
  \multicolumn{1}{c|}{ 16.10} &
  \multicolumn{1}{c}{ 13.67} &
  \multicolumn{1}{c}{ 54.28} &
  \multicolumn{1}{c|}{ \textbf{21.83}} &
  \multicolumn{1}{c}{ 14.23} &
  \multicolumn{1}{c}{ 64.48} &
  \multicolumn{1}{c|}{ \textbf{23.32}} &
  \multicolumn{1}{c}{ 9.99} &
  \multicolumn{1}{c}{ 25.79} &
  \multicolumn{1}{c|}{ 14.40} &
  
  \\
 &
  \multicolumn{1}{c|}{} &
  \multicolumn{1}{c|}{Injection*} &
  \multicolumn{1}{c}{ 11.16} &
  \multicolumn{1}{c}{ 63.21} &
  \multicolumn{1}{c|}{ 18.97} &
  \multicolumn{1}{c}{ 13.07} &
  \multicolumn{1}{c}{ 69.24} &
  \multicolumn{1}{c|}{ 21.99} &
  \multicolumn{1}{c}{ 11.69} &
  \multicolumn{1}{c}{ 27.25} &
  \multicolumn{1}{c|}{ \textbf{16.36}} &
  \multicolumn{1}{c}{ 10.33} &
  \multicolumn{1}{c}{ 48.01} &
  \multicolumn{1}{c|}{ 17.00} &
  \multicolumn{1}{c}{ 7.90} &
  \multicolumn{1}{c}{ 65.66} &
  \multicolumn{1}{c|}{ 14.10} &
  \multicolumn{1}{c}{ 11.35} &
  \multicolumn{1}{c}{ 22.34} &
  \multicolumn{1}{c|}{ \textbf{15.05}} &
  
  \\
 &
  \multicolumn{1}{c|}{} &
  \multicolumn{1}{c|}{Extension} &
  \multicolumn{1}{c}{ 14.97} &
  \multicolumn{1}{c}{ 37.21} &
  \multicolumn{1}{c|}{ \textbf{21.35}} &
  \multicolumn{1}{c}{ 14.76} &
  \multicolumn{1}{c}{ 50.00} &
  \multicolumn{1}{c|}{ \textbf{22.80}} &
  \multicolumn{1}{c}{ 18.01} &
  \multicolumn{1}{c}{ 12.81} &
  \multicolumn{1}{c|}{ 14.97} &
  \multicolumn{1}{c}{ 15.20} &
  \multicolumn{1}{c}{ 38.18} &
  \multicolumn{1}{c|}{ \textbf{21.75}} &
  \multicolumn{1}{c}{ 14.82} &
  \multicolumn{1}{c}{ 52.71} &
  \multicolumn{1}{c|}{ \textbf{23.13}} &
  \multicolumn{1}{c}{ 20.65} &
  \multicolumn{1}{c}{ 12.62} &
  \multicolumn{1}{c|}{ \textbf{15.67}} &
  
  \\
 &
  \multicolumn{1}{c|}{} &
  \multicolumn{1}{c|}{Extension*} &
  \multicolumn{1}{c}{ 10.51} &
  \multicolumn{1}{c}{ 56.82} &
  \multicolumn{1}{c|}{ 17.73} &
  \multicolumn{1}{c}{ 12.52} &
  \multicolumn{1}{c}{ 54.46} &
  \multicolumn{1}{c|}{ 20.36} &
  \multicolumn{1}{c}{ 12.33} &
  \multicolumn{1}{c}{ 30.43} &
  \multicolumn{1}{c|}{ \textbf{17.55}} &
  \multicolumn{1}{c}{ 10.24} &
  \multicolumn{1}{c}{ 46.62} &
  \multicolumn{1}{c|}{ 16.80} &
  \multicolumn{1}{c}{ 7.23} &
  \multicolumn{1}{c}{ 53.74} &
  \multicolumn{1}{c|}{ 12.74} &
  \multicolumn{1}{c}{ 6.83} &
  \multicolumn{1}{c}{ 60.58} &
  \multicolumn{1}{c|}{ 12.28} &
  
  \\
\cline{3-21}\multicolumn{1}{c}{\multirow{11}[2]{*}{\begin{sideways}\textbf{Bigvul}\end{sideways}}} &
\multicolumn{1}{c|}{} &
  \multicolumn{1}{c|}{ROS} &
  \multicolumn{1}{c}{7.54} &
  \multicolumn{1}{c}{25.60} &
  \multicolumn{1}{c|}{11.65} &
  \multicolumn{1}{c}{7.96} &
  \multicolumn{1}{c}{33.39} &
  \multicolumn{1}{c|}{12.86} &
  \multicolumn{1}{c}{10.91} &
  \multicolumn{1}{c}{13.16} &
  \multicolumn{1}{c|}{11.93} &
  \multicolumn{1}{c}{7.54} &
  \multicolumn{1}{c}{25.60} &
  \multicolumn{1}{c|}{11.65} &
  \multicolumn{1}{c}{7.96} &
  \multicolumn{1}{c}{33.39} &
  \multicolumn{1}{c|}{12.86} &
  \multicolumn{1}{c}{10.91} &
  \multicolumn{1}{c}{13.16} &
  \multicolumn{1}{c|}{11.93} &
  \\
  \cline{3-21}
 &
  \multicolumn{1}{c|}{} &
  \multicolumn{1}{c|}{VulGen} &
  \multicolumn{1}{c}{ 6.49} &
  \multicolumn{1}{c}{ 62.32} &
  \multicolumn{1}{c|}{ \textbf{11.75}} &
  \multicolumn{1}{c}{ 6.36} &
  \multicolumn{1}{c}{ 70.59} &
  \multicolumn{1}{c|}{ \textbf{11.66}} &
  \multicolumn{1}{c}{ 8.51} &
  \multicolumn{1}{c}{22.150 } &
  \multicolumn{1}{c|}{ \textbf{12.29}} &
  \multicolumn{1}{c}{ 6.49} &
  \multicolumn{1}{c}{ 62.32} &
  \multicolumn{1}{c|}{ \textbf{11.75}} &
  \multicolumn{1}{c}{ 6.36} &
  \multicolumn{1}{c}{ 70.59} &
  \multicolumn{1}{c|}{ \textbf{11.66}} &
  \multicolumn{1}{c}{ 8.51} &
  \multicolumn{1}{c}{22.15 } &
  \multicolumn{1}{c|}{ \textbf{12.29}} &
  
  \\
 &
  \multicolumn{1}{c|}{} &
  \multicolumn{1}{c|}{VulGen*} &
  \multicolumn{1}{c}{ 5.72} &
  \multicolumn{1}{c}{ 72.02} &
  \multicolumn{1}{c|}{ 10.60} &
  \multicolumn{1}{c}{ 5.77} &
  \multicolumn{1}{c}{ 85.21} &
  \multicolumn{1}{c|}{ 10.81} &
  \multicolumn{1}{c}{ 6.76} &
  \multicolumn{1}{c}{ 50.51} &
  \multicolumn{1}{c|}{ 11.92} &
  \multicolumn{1}{c}{ 5.72} &
  \multicolumn{1}{c}{ 72.02} &
  \multicolumn{1}{c|}{ 10.60} &
  \multicolumn{1}{c}{ 5.77} &
  \multicolumn{1}{c}{ 85.21} &
  \multicolumn{1}{c|}{ 10.81} &
  \multicolumn{1}{c}{ 6.76} &
  \multicolumn{1}{c}{ 50.51} &
  \multicolumn{1}{c|}{ 11.92} &
  
  \\
 &
  \multicolumn{1}{c|}{} &
  \multicolumn{1}{c|}{VGX} &
  \multicolumn{1}{c}{ 6.06} &
  \multicolumn{1}{c}{ 59.30} &
  \multicolumn{1}{c|}{ \textbf{10.99}} &
  \multicolumn{1}{c}{ 6.89} &
  \multicolumn{1}{c}{ 73.29} &
  \multicolumn{1}{c|}{ \textbf{12.60}} &
  \multicolumn{1}{c}{ 5.35} &
  \multicolumn{1}{c}{ 98.82} &
  \multicolumn{1}{c|}{ 10.15} &
  \multicolumn{1}{c}{ 6.06} &
  \multicolumn{1}{c}{ 59.30} &
  \multicolumn{1}{c|}{ \textbf{10.99}} &
  \multicolumn{1}{c}{ 6.89} &
  \multicolumn{1}{c}{ 73.29} &
  \multicolumn{1}{c|}{ \textbf{12.60}} &
  \multicolumn{1}{c}{ 5.35} &
  \multicolumn{1}{c}{ 98.82} &
  \multicolumn{1}{c|}{ 10.15} &
  
  \\
 &
  \multicolumn{1}{c|}{} &
  \multicolumn{1}{c|}{VGX*} &
  \multicolumn{1}{c}{ 5.82} &
  \multicolumn{1}{c}{ 69.16} &
  \multicolumn{1}{c|}{ 10.74} &
  \multicolumn{1}{c}{ 5.75} &
  \multicolumn{1}{c}{ 77.27} &
  \multicolumn{1}{c|}{ 10.71} &
  \multicolumn{1}{c}{ 6.98} &
  \multicolumn{1}{c}{ 49.03} &
  \multicolumn{1}{c|}{ \textbf{12.22}} &
  \multicolumn{1}{c}{ 5.82} &
  \multicolumn{1}{c}{ 69.16} &
  \multicolumn{1}{c|}{ 10.74} &
  \multicolumn{1}{c}{ 5.75} &
  \multicolumn{1}{c}{ 77.27} &
  \multicolumn{1}{c|}{ 10.71} &
  \multicolumn{1}{c}{ 6.98} &
  \multicolumn{1}{c}{ 49.03} &
  \multicolumn{1}{c|}{ \textbf{12.22}} &
  
  \\
 &
  \multicolumn{1}{c|}{} &
  \multicolumn{1}{c|}{Mutation} &
  \multicolumn{1}{c}{ 7.80} &
  \multicolumn{1}{c}{ 63.91} &
  \multicolumn{1}{c|}{ \textbf{13.90}} &
  \multicolumn{1}{c}{ 7.63} &
  \multicolumn{1}{c}{ 67.73} &
  \multicolumn{1}{c|}{ \textbf{13.72}} &
  \multicolumn{1}{c}{ 6.78} &
  \multicolumn{1}{c}{ 74.14} &
  \multicolumn{1}{c|}{ \textbf{12.42}} &
  \multicolumn{1}{c}{ 7.56} &
  \multicolumn{1}{c}{ 57.23} &
  \multicolumn{1}{c|}{ \textbf{13.36}} &
  \multicolumn{1}{c}{ 9.37} &
  \multicolumn{1}{c}{ 53.26} &
  \multicolumn{1}{c|}{ \textbf{15.94}} &
  \multicolumn{1}{c}{ 6.65} &
  \multicolumn{1}{c}{ 78.31} &
  \multicolumn{1}{c|}{ \textbf{12.25}} &
  
  \\
 &
  \multicolumn{1}{c|}{} &
  \multicolumn{1}{c|}{Mutation*} &
  \multicolumn{1}{c}{ 6.33} &
  \multicolumn{1}{c}{ 70.75} &
  \multicolumn{1}{c|}{ 11.63} &
  \multicolumn{1}{c}{ 6.39} &
  \multicolumn{1}{c}{ 78.22} &
  \multicolumn{1}{c|}{ 11.82} &
  \multicolumn{1}{c}{ 5.96} &
  \multicolumn{1}{c}{ 83.50} &
  \multicolumn{1}{c|}{ 11.12} &
  \multicolumn{1}{c}{ 6.20} &
  \multicolumn{1}{c}{ 82.51} &
  \multicolumn{1}{c|}{ 11.53} &
  \multicolumn{1}{c}{ 6.15} &
  \multicolumn{1}{c}{ 85.69} &
  \multicolumn{1}{c|}{ 11.48} &
  \multicolumn{1}{c}{ 6.08} &
  \multicolumn{1}{c}{ 92.22} &
  \multicolumn{1}{c|}{ 11.40} &
  
  \\
 &
  \multicolumn{1}{c|}{} &
  \multicolumn{1}{c|}{Injection} &
  \multicolumn{1}{c}{ 8.79} &
  \multicolumn{1}{c}{ 29.41} &
  \multicolumn{1}{c|}{ \textbf{13.53}} &
  \multicolumn{1}{c}{ 11.64} &
  \multicolumn{1}{c}{ 38.00} &
  \multicolumn{1}{c|}{ \textbf{17.82}} &
  \multicolumn{1}{c}{ 11.43} &
  \multicolumn{1}{c}{ 18.81} &
  \multicolumn{1}{c|}{ 14.22} &
  \multicolumn{1}{c}{ 10.33} &
  \multicolumn{1}{c}{ 31.80} &
  \multicolumn{1}{c|}{ \textbf{15.59}} &
  \multicolumn{1}{c}{ 9.91} &
  \multicolumn{1}{c}{ 38.16} &
  \multicolumn{1}{c|}{ \textbf{15.74}} &
  \multicolumn{1}{c}{ 7.85} &
  \multicolumn{1}{c}{ 30.49} &
  \multicolumn{1}{c|}{ \textbf{12.49}} &
  
  \\
 &
  \multicolumn{1}{c|}{} &
  \multicolumn{1}{c|}{Injection*} &
  \multicolumn{1}{c}{ 7.24} &
  \multicolumn{1}{c}{ 54.21} &
  \multicolumn{1}{c|}{ 12.77} &
  \multicolumn{1}{c}{ 8.86} &
  \multicolumn{1}{c}{ 59.46} &
  \multicolumn{1}{c|}{ 15.42} &
  \multicolumn{1}{c}{ 12.82} &
  \multicolumn{1}{c}{ 23.45} &
  \multicolumn{1}{c|}{ \textbf{16.57}} &
  \multicolumn{1}{c}{ 7.46} &
  \multicolumn{1}{c}{ 40.38} &
  \multicolumn{1}{c|}{ 12.59} &
  \multicolumn{1}{c}{ 7.90} &
  \multicolumn{1}{c}{ 71.54} &
  \multicolumn{1}{c|}{ 14.23} &
  \multicolumn{1}{c}{ 8.61} &
  \multicolumn{1}{c}{ 24.37} &
  \multicolumn{1}{c|}{ 12.72} &
  
  \\
 &
  \multicolumn{1}{c|}{} &
  \multicolumn{1}{c|}{Extension} &
  \multicolumn{1}{c}{ 10.60} &
  \multicolumn{1}{c}{ 17.33} &
  \multicolumn{1}{c|}{ 13.16} &
  \multicolumn{1}{c}{ 11.13} &
  \multicolumn{1}{c}{ 34.02} &
  \multicolumn{1}{c|}{ \textbf{16.77}} &
  \multicolumn{1}{c}{ 19.03} &
  \multicolumn{1}{c}{ 11.96} &
  \multicolumn{1}{c|}{ 14.68} &
  \multicolumn{1}{c}{ 9.89} &
  \multicolumn{1}{c}{ 31.48} &
  \multicolumn{1}{c|}{ \textbf{15.05}} &
  \multicolumn{1}{c}{ 10.22} &
  \multicolumn{1}{c}{ 33.86} &
  \multicolumn{1}{c|}{ \textbf{15.70}} &
  \multicolumn{1}{c}{ 15.74} &
  \multicolumn{1}{c}{ 11.86} &
  \multicolumn{1}{c|}{ \textbf{13.53}} &
  
  \\
 &
  \multicolumn{1}{c|}{} &
  \multicolumn{1}{c|}{Extension*} &
  \multicolumn{1}{c}{ 7.76} &
  \multicolumn{1}{c}{ 47.85} &
  \multicolumn{1}{c|}{ \textbf{13.35}} &
  \multicolumn{1}{c}{ 7.58} &
  \multicolumn{1}{c}{ 51.67} &
  \multicolumn{1}{c|}{ 13.22} &
  \multicolumn{1}{c}{ 11.75} &
  \multicolumn{1}{c}{ 24.28} &
  \multicolumn{1}{c|}{ \textbf{15.84}} &
  \multicolumn{1}{c}{ 8.48} &
  \multicolumn{1}{c}{ 41.18} &
  \multicolumn{1}{c|}{ 14.06} &
  \multicolumn{1}{c}{ 6.86} &
  \multicolumn{1}{c}{ 52.15} &
  \multicolumn{1}{c|}{ 12.12} &
  \multicolumn{1}{c}{ 6.67} &
  \multicolumn{1}{c}{ 71.73} &
  \multicolumn{1}{c|}{ 12.21} &
  
  \\
\cline{3-21} &

  \\
\end{tabular}%
}
\end{table}%

\subsection{Ensemble of Strategies}
\rb{Since our strategies appeared to be useful, we decided to investigate whether a mix of strategies would perform better than using a single strategy. For such a comparison, we used the results of RQ1, and added three settings for using two strategies, with each contributing 2.5K samples, amounting to the same 5K extra samples, and an all-strategy setting where one third of the samples came from each of the strategies. Table~\ref{tab:mixed} shows the result of this experiment. As observed, the combination does not always outperform the single strategy. In half (6 out of 12) of the settings, a mixed strategy works worse than a single strategy. Specifically, a mix of all three (M+I+E) and a combination of Mutation and Extension always fail to beat all of the single strategies. The only promising combination is the mix of Injection and Extension (I+E), which achieves the best performance in 4 out of 12 settings.
}
\rqboxc{The best ensemble setting appears to be a combination of Injection and Extension settings, yet this setting appears to be less useful than only using Injection. Yet, since Extension appears to beat Injection in some settings (e.g. RQ4), we recommend using Injection and Extension strategies and a mix of these to get to achieve a higher performance.}

\subsection{Impact Analysis of the Extra Clean Samples}

\rb{Following VGX and Vulgen, a proportional number of extra clean samples are added to the dataset to keep the ratio of vulnerable and clean samples in RQs 1-3. To examine the effect of the extra clean samples, we did an additional study alongside the results of RQ1, where instead of adding extra clean samples, we randomly oversampled the original clean samples to keep the ratio. Aside from our strategies, we also included VGX and Vulgen to see if we get a different result. As observed in Table~\ref{tab:ros_clean}, VGX and Vulgen get better performance in 10 out of 12 settings, while this is in all cases of Mutation, and 9 out of 12 settings for Injection and Extension. Specifically, Vulgen, VGX, Mutation, Injection, and Extension gain 22.75\%, 20.07\%, 28.4\%, 12.76\%, and 18.48\% on average F1-score, respectively. Hence, Mutation appears to be the most dependent on the addition of extra clean samples, and Injection appears to be the least dependent method. \\ \\
Compared to ROS (of RQ1), where only extra clean samples are added and vulnerable samples are oversampled, Vulgen, VGX, and Mutation drop 21.46\%, 20.42\%, and 17.21\%, on average F1-score, respectively. Injection and Extension gain an improvement of 5.12\% and 0.27\% in terms of the average F1-score. This shows that the effect of vulnerability augmentation is more important for achieving a higher performance than adding extra clean samples for Injection and Extension, while the extra clean samples play a bigger role for Vulgen, VGX, and Mutation strategies.}
% \\ \\
% It is important to note that for vulnerability augmentation, clean samples are available widely online, and according to the results, the addition of extra clean samples for maintaining the ratio is crucial. Our examination also shows that some methods rely on the clean samples’ diversity more than others. In other words, selecting better clean samples affects methods differently, and our Injection and Extension methods are less sensitive to that data. In fact, without adding extra cleans, we still beat the baselines, VGX, and Vulgen. (ROS’s equivalent for this study would be the same as the baseline, since applying ROS on both sides will lead to a dataset with duplicates)

\subsection{Cost Analysis} 
Based on our experiments, generating every 1,000 vulnerable samples with one prompt at a time and no concurrency, takes 3.4 hours and costs about \$1.88  with GPT3.5-Turbo and about 9 GPU hours with CodeQwen1.5-7B-Chat on two RTX3090s. This shows the feasibility of \ourTool as it takes less than \$19 with ChatGPT to augment a dataset like Devign to twice its size and improves the performance of a DLVD model up to 128.95\%. 
\subsection{Quality Analysis of Augmented Vulnerable Samples}
\rb{We focus on data augmentation and not generation in this study, and introducing noise is inevitable. This is common in the application of other data augmentation techniques in the fields of natural language processing and computer vision~\cite{mixup, SenMixup, cutout,olsson2021classmix, yun2019cutmix}. It is crucial to realize that data augmentation is used as a means to improve the generalizability of the models. A reasonable amount of noise can help with the regularization of the model. Otherwise, performance degradation will be seen. In addition, based on the results of recent studies~\cite{ExploringReprLevelAug, BoostSourceCodeAug} show that code data augmentation techniques that slightly break the syntax can help the training of the models. However, we recognize the importance of evaluating the quality of augmented vulnerable samples to provide practical insights for practitioners when using our approach. 
\\
To this end, we conducted a manual assessment of the generated code. Specifically, we randomly selected 50 samples per augmentation strategy (i.e., Mutation, Injection, and Extension), produced using ChatGPT-3.5 in RQ1 (150 samples in total). We manually inspected each sample to determine whether the LLM followed our instructions and successfully generated a vulnerable example that preserved the intended vulnerability pattern. Our analysis revealed that 72\%, 82\%, and 90\% of the samples generated via Mutation, Injection, and Extension, respectively, were correctly implemented and truly vulnerable. These results indicate an acceptable level of quality. The higher success rates for Injection and Extension may help explain their superior performance compared to Mutation. This also suggests that LLMs may find it easier to perform injection and extension tasks than mutation. We also study the failure cases, and we observe certain patterns. For instance, the common case for the Mutation was changing the code in a way different from the methods mentioned, such that the output is significantly different from the instruction. For Injection, wrong injections or half-complete injections were more common. For Extension, this had to do with the removal of the input code instead of extending it.}

\subsection{Threats to Validity}

\noindent\textbf{Internal Validity}
A common threat to the validity of our study is our hyperparameter settings for DLVD models and LLMs. For DLVD, hyperparameter tuning is extremely expensive given the scale of our study; as such, we followed the setting of previous studies~\cite{VGX, VULGEN}. In addition, when comparing different data augmentation approaches, we used the same settings for DLVD models, which ensures our comparisons are fair. For LLMs, we mostly used the defaults, and only set the number of new tokens to ChatGPT's maximum, and lowered ChatGPT's temperature to 0.5. Better results might be achievable by fine-tuning these hyperparameters. \\ \\
\rb{Label noise in the datasets can pose another threat to the validity of our experiments. Particularly, Devign and BigVul, even though they are commonly used in previous studies, are known to be noisy datasets~\cite{Croft2023VulDataQuality, Risse2024VDLimits, ding2024primevul}, meaning that a considerable number of vulnerable samples are incorrectly labeled, which can affect the results of our experiments. To ensure these results can be trusted, we expanded our study by dedicating an additional RQ to using PrimeVul, which is a more recent dataset with considerably less label noise, and showed that our results still hold. \\ \\
Another threat is that LLMs might not always follow the prompts as instructed and could produce hallucinations. To alleviate this threat, we add a verifier component to filter out low-quality code, and we use a retry mechanism that prompts the LLM up to three times in case of errors (including an empty code response). In addition, we study the quality of augmented vulnerable samples, and found that the quality is at an acceptable level. We encourage future research to integrate more advanced approaches to reduce hallucinations. More importantly, our goal in this study is to augment existing vulnerable datasets to help models capture the vulnerability patterns and generalize better, rather than generating high-quality vulnerable samples. Slightly breaking the syntax and not having completely correct code remains beneficial for training of the models~\cite{wang2024natural, ExploringReprLevelAug, BoostSourceCodeAug}, and data with subtle noise helps with the generalization of models~\cite{NoisyDataTikhonovRegularization,ClassificationWithNoisyLabels, NoiseModelling1996}. We acknowledge that more sophisticated methods can be used for the verification phase, but starting from a simple module (i.e., a parser) was necessary for evaluating whether using LLMs for vulnerability detection is even feasible.\\} 

\noindent\textbf{External Validity}
relates to the generalizability of our findings. Even though we used two different LLMs (ChatGPT 3.5 Turbo and CodeQwen1.5-7B-Chat), evaluated on three commonly used datasets, and covered three SOTA DLVD models from two different families, our findings may not generalize well to real-world scenarios where different models, LLMs, and datasets are used. To encourage future research to investigate more LLMs (including encoder-decoder-based models), datasets, and DLVD models.

\section{Conclusion}\label{sec:conclusion}

In this paper, we propose \ourTool, a novel yet effective framework for augmenting vulnerable samples that uses a customized RAG mechanism to formulate carefully designed prompt templates which are then used for prompting the LLMs. We also use a fuzzy parser as a verification method, to make sure that the generated code doesn't have severe syntax problems. Our evaluation results show that our method significantly outperforms SOTA methods, and is suitable for large-scale vulnerability augmentation with low cost.
% In this paper, we propose a simple yet effective framework to identify commit-level vulnerability fixes by enhancing code change representation learning to improve the capacity of our model for capturing small code changes, and simplifying the training process to enable us to train the embedding models and final classification model jointly. Our evaluation shows that \ourTool outperforms SOTA baselines and achieves an F1 of 0.33, a \costFive of 0.632, and a \costTwe of 0.76, with an improvement of 77.4\%, 8.4\%, and 5.1\% over the best baselines, respectively. Our ablation analysis shows that our \codeDelta plays a critical role in \ourTool, and is typically effective in capturing small
% code changes. Evaluation of the temporal dataset demonstrates the superiority of \ourTool over baselines with a large margin improvement in real-world situation. 

\section{Data Availability}
We made our replication package publicly available~\cite{datarepo} to encourage future research on enhancing the performance of LLM-generated vulnerable code and more advanced filtering mechanisms.

%\begin{acks}
%To Robert, for the bagels and explaining CMYK and color spaces.
%\end{acks}

\bibliographystyle{ACM-Reference-Format}
\bibliography{sample-base}

% \appendix









\end{document}